\documentclass[twocolumn,prd,nofootinbib,showpacs]{revtex4}
\usepackage{graphicx}
\usepackage{color}
\usepackage{amsmath,amssymb,amsfonts,dsfont}
\usepackage{multirow}
\usepackage{sistyle}
\usepackage{ulem}

\pdfoutput=1 

\newcommand{\lsim}{\mathrel{\mathop{\kern 0pt \rlap
  {\raise.2ex\hbox{$<$}}}
  \lower.9ex\hbox{\kern-.190em $\sim$}}}
\newcommand{\gsim}{\mathrel{\mathop{\kern 0pt \rlap
  {\raise.2ex\hbox{$>$}}}
  \lower.9ex\hbox{\kern-.190em $\sim$}}}

\def \aap  {A\&A}

\def \apjs  {ApJS}
\def \apjl  {ApJL}

\begin{document}

\title{Detection of a $\gamma$-ray halo around Geminga with the {\it Fermi}-LAT and implications for the positron flux}

\author{Mattia Di Mauro,}\email{mdimauro@slac.stanford.edu
}
\affiliation{NASA Goddard Space Flight Center, Greenbelt, MD 20771, USA}
\affiliation{Catholic University of America, Department of Physics, Washington DC 20064, USA}
\author{Silvia Manconi}\email{manconi@to.infn.it}
\affiliation{Dipartimento di Fisica, Universit\`a di Torino, via P. Giuria 1, 10125 Torino, Italy}
\affiliation{Istituto Nazionale di Fisica Nucleare, Sezione di Torino, Via P. Giuria 1, 10125 Torino, Italy}
\author{Fiorenza Donato}\email{donato@to.infn.it}
\affiliation{Dipartimento di Fisica, Universit\`a di Torino, via P. Giuria 1, 10125 Torino, Italy}
\affiliation{Istituto Nazionale di Fisica Nucleare, Sezione di Torino, Via P. Giuria 1, 10125 Torino, Italy}

\begin{abstract}
The HAWC Collaboration has discovered a $\gamma$-ray emission extended about 2 degrees around the Geminga and Monogem pulsar wind nebulae (PWNe) at $\gamma$-ray energies $E_\gamma >5$ TeV. 
We analyze, for the first time, almost 10 years of $\gamma$-ray data obtained with the {\it Fermi} Large Area Telescope at $E_\gamma >$ 8 GeV in the direction of Geminga and Monogem.
Since these two pulsars are close the Galactic plane we run our analysis with 10 different interstellar emission models (IEMs) to study the systematics due to the modeling of this component. 
We detect a $\gamma$-ray halo around Geminga with a significance in the range $7.8-11.8\sigma$ depending on the IEM considered.
This measurement is compatible with $e^+$ and $e^-$ emitted by the PWN, which inverse-Compton scatter (ICS) with photon fields located within a distance of about 100 pc from the pulsar, where the diffusion coefficient is estimated to be around $1.1 \times 10^{27}$ cm$^2$/s at 100 GeV.
We include in our analysis the proper motion of the Geminga pulsar which is relevant for $\gamma$ rays produced for ICS in the {\it Fermi}-LAT energy range.
We find that an efficiency of about 1\% for the conversion of the spin-down energy of the pulsar into $e^+$ and $e^-$ is required to be consistent with $\gamma$-ray data from {\it Fermi}-LAT and HAWC.
The inferred contribution of Geminga to the $e^+$ flux is at most 20\% at the highest energy AMS-02 data. 
Our results are compatible with the interpretation that the cumulative emission from Galactic pulsars explains the positron excess. 
\end{abstract}

\maketitle

\section{Introduction}
\label{sec:intro}

High precision data of cosmic electron ($e^-$) and positron ($e^+$) fluxes are currently available over 4 decades in energy. 
In particular, the positron fraction ($e^+/(e^++e^-)$) has been measured by the Pamela \citep{2009Natur.458..607A}, {\it Fermi}-LAT \citep{2012PhRvL.108a1103A} and 
AMS-02 \citep{PhysRevLett.110.141102,PhysRevLett.113.121101} Collaborations. 
AMS-02 on board the International Space Station has measured with unprecedented precision the positron flux up to 1 TeV 
\citep{2014PhRvL.113l1101A,AMS_2018,PhysRevLett.122.041102}. 
The interpretation of $e^+$ (and, almost equivalently,  the $e^+/(e^++e^-)$ ratio) data is still under debate.
Below 10 GeV the  data are described well as the secondary production given by spallation reactions of primary 
cosmic rays (CRs) with the atoms of the interstellar medium (ISM) (see, e.g., \cite{2014JCAP...04..006D,Manconi:2016byt,Manconi:2018azw}).
On the other hand, the $e^+$ flux above a few tens of GeV strongly exceeds the predicted secondary component  \cite{2014JCAP...04..006D}. 
In order to explain the $e^+$ excess, the annihilation or  decay of dark matter particles, the emission from pulsar wind nebulae (PWNe) (see, e.g., \cite{2010JCAP...01..009I,Hooper:2008kg,Pato:2010im,Linden:2013mqa,2013PhRvD..88b3013C,2014JCAP...04..006D, Boudaud:2014dta, 2016JCAP...05..031D}) or supernova remnants (SNRs) (see, e.g., \cite{2009PhRvL.103e1104B,Ahlers:2009ae,Mertsch:2014poa,Tomassetti:2015cva, 2014PhRvD..89d3013C}) have been invoked.

The Milagro Collaboration has reported the detection of $\gamma$-ray emission from 1-100 TeV from the direction of Geminga with an extension of $2.6^{\circ}$ \cite{2009ApJ...700L.127A}. 
This observation has been confirmed  by the HAWC Collaboration with a $\gamma$-ray spectrum measured  from 5 TeV to 40 TeV \cite{Abeysekara:2017science} (hereafter HAWC2017) and an extension of about $2^{\circ}$. The HAWC observatory has also detected very high-energy emission in the direction of the pulsar B0656+14 (also known as Monogem), with a similar spatial extension.

These measurements play a crucial role in the understanding of the $e^\pm$ acceleration from pulsars and their PWNe, and can be used to estimate the contribution of these sources to the $e^+$ flux \cite{2009PhRvL.103e1101Y,Abeysekara:2017science,Hooper:2017gtd,Profumo:2018fmz}.
Indeed, the extended TeV $\gamma$-ray emission detected around PWNe can be interpreted as inverse Compton scattering (ICS) emission of $e^\pm$ accelerated and released by these sources off ambient photons of the interstellar radiation field (ISRF). The ISRF, composed of the cosmic microwave background (CMB), infrared (IR) and starlight (SL), is then scattered up to $\gamma$-ray energies.
In particular, given the extension of the detected TeV $\gamma$-ray emission  and the age of the sources, ICS photons must be generated by $e^\pm$ escaped from the PWN. 
The pairs produced in PWNe may be effectively released in the ISM when the system leaves the parent SNR, as extensively discussed in \cite{Bykov:2017xpo,2011ASSP...21..624B} for nebulae surrounding high-speed pulsars, classified as bow shock PWNe.
These particles then propagate in the Galaxy and can be detected at Earth.
In addition to $\gamma$-ray data, radio and X-ray data are available for many PWNe, even if they typically probe structures on much smaller scales, such as the jet and torii seen for the Geminga PWN \cite{2017ApJ...835...66P}.
Photons at these lower energies are produced by $e^\pm$ through Synchrotron and Bremsstrahlung radiation, which are predominantly trapped inside the PWN. 
Therefore, the spectral energy distribution (SED) from radio to $\gamma$-ray energies provides valuable information about the population of $e^\pm$ produced by these sources (see, e.g., \cite{Bykov:2017xpo} for a recent review).

The HAWC experiment measures $\gamma$ rays between $5-40$ TeV. 
These photons can be produced via ICS off the ISRF by $e^\pm$ at average energies of at least tens of TeV. 
Since the $e^+$ AMS-02 excess is between a few tens up to hundreds of GeV, the HAWC data cannot test directly the origin of this excess.
The use of HAWC $\gamma$-ray data in order to predict the $e^+$ flux at AMS-02 energies is indeed an extrapolation, which can vary significantly depending on the assumptions made.

In this paper, we analyze, for the first time, {\it Fermi}-LAT data from 8 GeV up to TeV energies in the direction of the Geminga and Monogem PWNe to search for an extended emission that can be attributed to the interaction of the accelerated $e^\pm$ with the ISRF. 
Our analysis is unique in $\gamma$-ray astronomy because we will include the proper motion of Geminga pulsar \cite{Faherty:2007} because, as we will show, this effect is relevant for the spatial morphology of the ICS $\gamma$-ray halo.
We will show that {\it Fermi}-LAT data are ideal, together with HAWC measurements, to constrain the $e^+$ flux from these two sources.

The paper is organized as follows.
In Sec.~\ref{sec:mw}  we explain our model for the photon emission from ICS and Synchrotron radiation, 
and for the emission and propagation of $e^+$ from PWNe.
In Sec.~\ref{sec:hawcresults} we present the predictions for the contribution of Geminga and Monogem to the positron flux using HAWC data. We show that it is not possible to provide a precise prediction for the contribution of these two sources to the $e^+$ excess at lower energies using only these data.
In Sec.~\ref{sec:fermianalysis} we analyze 10 years of {\it Fermi}-LAT data above 8 GeV to search for a halo emission around these two sources, and we calculate their contribution to the $e^+$ flux.  
In Sec.~\ref{sec:conclusions} we draw our conclusions.

\bigskip

\section{$\gamma$-ray and $e^{\pm}$ emission from Pulsar Wind Nebulae}
\label{sec:mw} 

The photon emission observed in the direction of PWNe covers a wide range of energies (see, e.g., \cite{2017hsn..book.2159S} 
for a recent review).
From radio to X-ray energies, photons are produced by $e^+$ and $e^-$  through synchrotron radiation caused by the magnetic field present in the ISM.
On the other hand, at higher energies $\gamma$ rays are produced via ICS of very-high energy $e^+$ and $e^-$ escaped from the PWN off the ISRF.
In what follows we describe the models we employed for the flux of photons emitted from ICS and synchrotron
radiation, and for the  $e^+$ source spectrum. 
We note that we are interested in the extended (few degrees at TeV energies) halo emission around PWNe, which could be attributed to the $e^\pm$  pairs accelerated and escaped from the PWN, and not to the small-scale (few arcseconds to arcminutes) structures observed in the nebula, as for example jets and torii (see e.g. \cite{2017ApJ...835...66P}). 
Nevertheless, the following equations hold for any  $e^+$ and $e^-$ input spectrum, target photon fields and for the synchrotron and ICS emission mechanisms.

In general, the photon flux emitted for the ICS or synchrotron mechanism by a source, at an energy $E_{\gamma}$ and for a solid angle $\Delta \Omega$, can be written as \cite{1970RvMP...42..237B,Cirelli:2010xx}:
\begin{equation} \label{eq:phflux}
 \phi^{\rm IC, Sync} (E_{\gamma}, \Delta \Omega)= \int_{m_e c^2}^{\infty} dE \mathcal{M}(E, \Delta \Omega) \mathcal{P}^{\rm IC, Sync}(E, E_{\gamma})\,.
\end{equation}
The term $\mathcal{M}(E,\Delta \Omega)$ represents the spectrum of $e^+$ and $e^-$  of energy $E$ propagating in the Galaxy and from a solid angle $\Delta \Omega$:
\begin{equation}
\mathcal{M}(E,\Delta \Omega)  = \int_{\Delta \Omega} d\Omega \int_0^{\infty} d s \, \mathcal{N}_e (E,s).
 \label{eq:M}
\end{equation}
 $\mathcal{N}_e (E,s)$ is the energy spectrum of $e^\pm$  of energy $E$ emitted by the source,  $s$ is the line of sight,
while $\mathcal{P}^{\rm IC, Sync}(E,E_{\gamma})$ is the power of photons emitted by a single $e^\pm$ for ICS or synchrotron emissions, as detailed in the next sub-sections. 
The solid angle $\Delta \Omega$ is parametrized using the angular separation between the line of sight and the direction of the source $\theta$.

\subsection{$\gamma$ rays from Inverse Compton Scattering}
\label{sec:ICS}
The $e^\pm$ propagating  in the Galaxy produce $\gamma$ rays through ICS with the Galactic ISRF.
The Galactic ISRF is composed of the CMB, described by a blackbody energy density at temperature $T_{{\rm CMB}} = 2.753$ K, the IR light with the peak of the spectrum at $T_{{\rm IR}} = 3.5 \cdot 10^{-3}$ eV and by the SL with $T_{{\rm SL}} = 0.3$ eV \cite{Vernetto:2016alq,2006ApJ...648L..29P,2017MNRAS.470.2539P}.
We define the ICS power of photons of energy $E_{\gamma}$ produced by electrons of energy $E$ as in \cite{1970RvMP...42..237B,2010A&A...524A..51D}: 
\begin{eqnarray}
&&\mathcal{P}^{IC}(E,E_{\gamma}) = \frac{3 \sigma_T c \, m^2_e c^4}{4 {E}^2} \int_{\frac{m_ec^2}{4E}}^{1} dq \frac{d\mathcal{N}}{d\epsilon}(\epsilon(q)) \times  \\
&& \times \left( 1 - \frac{m^2_e c^4}{4 q E^2 (1-\tilde {\epsilon})} \right) \nonumber 
 \left[ 2 q \log{q}+q+1 -2 q^2 + \frac{\tilde {\epsilon}(1-q)}{2-2 \tilde{\epsilon}}  \right],
 \label{eq:P}
\end{eqnarray}
where $\epsilon$ is the ISRF photon energy, $\frac{d\mathcal{N}}{d\epsilon}(\epsilon(q))$ is the energy spectrum of the ISRF, and 
\begin{equation}
q = \frac{\tilde {\epsilon}}{\Gamma_{\epsilon}(1-\tilde {\epsilon})} {\rm \:\: , \:\:}  \Gamma_{\epsilon} = \frac{4 \epsilon E}{m^2_e c^4} 
{\rm \:\: , \:\:}  \tilde {\epsilon} = \frac{E_{\gamma}}{E}\,.
 \label{eq:ICdefpar}
\end{equation}
 \\
Models for the  local ISRF are provided in \cite{Vernetto:2016alq,2006ApJ...648L..29P,2017MNRAS.470.2539P}, which all contain a careful description of the photons at all  frequencies. 
These ISRF models are based on the estimate for the interactions between SL and the interstellar matter, which take into account an accurate knowledge of the stars, gas, and dust in the Galaxy, and the IR emissivities per dust grain.
Our results are obtained for the ISRF energy density in the local Galaxy reported in \cite{Vernetto:2016alq}.
We do not consider any spatial variation in the model. We have explicitly checked that our results do not get modified by using the model in Ref.~\cite{2006ApJ...648L..29P}. We do not expect significant changes 
with the model in Ref.~\cite{2017MNRAS.470.2539P}, since it is very similar to the other ones in the local Galaxy, which is the relevant scale for our analysis.\footnote{The distances to Monogem and Geminga are 0.250~kpc and 0.288~kpc respectively.} 

\subsection{Synchrotron Radiation}
\label{sec:sync} 
The $e^\pm$  produce photons from radio to X-ray energies through synchrotron radiation due to their interaction with the Galactic magnetic field.
The flux of synchrotron photons has the same expression as in Eq.~\ref{eq:phflux}, where the synchrotron power  $\mathcal{P}^{\rm Sync}(E,E_{\gamma})$ is now given by \cite{2010PhRvD..82d3002A}:
\begin{equation}\label{eq:syncspectrum}
 \mathcal{P}^{\rm Sync}= \frac{dN_{\rm Sync}}{dE_{\gamma} dt}\,. 
\end{equation}
The quantity defined in Eq.~\eqref{eq:syncspectrum} is connected to the energy emitted by one lepton per unit frequency and unit time,  $\frac{dE_{\rm sync}}{d\nu dt}$, as:
\begin{equation}
 \frac{dN_{\rm Sync}}{dE_{\gamma} dt}= \frac{1}{h E_{\gamma}} \frac{dE_{\rm sync}}{d\nu dt}
\end{equation}
since $N_{\rm sync} E_{\gamma}= E_{\rm Sync}$.
To obtain the emissivity function in a random magnetic field one should average out the standard synchrotron formula (see \cite{1970RvMP...42..237B}) over the directions of the magnetic field. 
For $e^\pm$ with arbitrary pitch angle, the emitted energy per unit frequency and time is thus given by (see \cite{2010PhRvD..82d3002A}):
\begin{equation}
 \frac{dE_{\rm sync}}{d\nu dt} = \frac{\sqrt{3} e^3 B}{m_e c^2} G(x)
\end{equation}
where $e$ and $  m_e$ are the electron charge and mass, $B$ is the magnetic field and $c$ is the speed of light. 
The function $G(x)$ is an analytical approximation for the dimensionless synchrotron integral as defined in \cite{2010PhRvD..82d3002A} (Eq.~D7), where $x=\nu/\nu_c$ and $\nu=E_{\gamma}/h$ and
\begin{equation}\label{eq:nnuc}
 \nu_c= \nu_c(E)= \frac{3e B E^2}{4\pi m_e^3 c^5}\,. 
\end{equation}


\subsection{Cosmic $e^+$ and $e^-$ emission}
\label{sec:e+e-}
PWNe are among the major accelerators of $e^+$ and $e^-$ in the Galaxy.  
Under the influence of winds and shocks, $e^\pm$ can detach from the surface of the neutron star and initiate cascade processes that lead to the production of a cloud of charged particles that surrounds the pulsar, which is called a PWN (see, e.g., \cite{Amato:2013fua}).
Within the nebula, the $e^+$ and $e^-$ are believed to be accelerated to very high energies at the termination shock and then injected into the ISM after a few tens of kyr \citep{1996ApJ...459L..83C, 2011ASSP...21..624B}.

Two different assumptions are usually made for the emission mechanism of $e^\pm$  from PWNe.
In the burst-like injection scenario all the particles are emitted from the sources at a time equal to the age of the source ($t^{\star}$).
Therefore, the time-dependence is a delta function $\delta(t-t^{\star})$.
On the other hand, in the continuous injection scenario the particles are emitted with a rate that follows the pulsar spin-down energy.

The injection spectrum of $e^\pm$ emitted by a PWN in the burst-like injection scenario can be described as: \cite{2010A&A...524A..51D}:  
 \begin{equation}
 Q(E)= Q_{0} \left( \frac{E}{E_0}\right)^{- \gamma_e} \exp \left(-\frac{E}{E_c} \right),
 \label{eq:Q_E}
\end{equation}
where $Q_{0}$ is in units of GeV$^{-1}$ and $E_c$ is a cutoff energy. 
If not stated differently, we adopt $E_c= 10^3$ TeV. 
We stress that a value of $E_c$ well above 10 TeV is necessary to produce $\gamma$ rays through ICS at the energies measured around the Geminga PWN with HAWC and Milagro.
The normalization of the power law is fixed to $E_0= 1$~GeV.
Given the injection spectrum in Eq.~\ref{eq:Q_E}, the total energy emitted in $e^-$ and $e^+$ in units of GeV can be obtained through (see \cite{2010A&A...524A..51D}):
\begin{equation}
 E_{\rm tot} = \int _{E_1} ^\infty dE \, E \,Q(E) \,,
 \label{eq:Etot}
\end{equation}
where we fix $E_1 = 0.1$ GeV. This is the typical value considered for the minimum energy of non-thermal electrons \cite{Buesching:2008hr,Sushch:2013tna}.
The normalization $Q_{0}$ for a single PWN is obtained assuming that a fraction $\eta$ of the total spin-down energy $W_0$ emitted by the pulsar is released in form of $e^\pm$ pairs, i.e.:
\begin{equation}
E_{\rm tot} = \eta W_0. 
\label{eq:EtotPWN}
\end{equation}
The value of $W_0$ can be computed starting from the age of the pulsar $t^{\star}$, 
the typical pulsar decay time $\tau_0$, and the spin-down luminosity $\dot{E}$:
\begin{equation}
 W_0 = \tau_0 \dot{E} \left( 1+ \frac{t^{\star}}{\tau_0} \right)^2\,.
 \label{eq:W0PWN}
\end{equation}
The spin-down luminosity $\dot{E}$, the \textit{observed} age $t_{\rm obs}$ (where $t^{\star} = t_{\rm obs} + d/c$ is the \textit{actual} age) and the distance $d$ for the pulsars are taken from the ATNF catalog \cite{2005AJ....129.1993M},  while $\tau_0$ is the characteristic pulsar spin-down timescale. We assume that the magnetic braking index is $k=3$.
We use $\tau_0=12$ kyr if not stated otherwise, following HAWC2017.  
Moreover, we assume $d = $250 pc, $ t_{\rm obs} = $ 370 kyr and $\dot{E} = 3.2\cdot 10^{34}$ erg/s for  Geminga and $d = $ 288 pc, $ t_{\rm obs}$ =110 kyr and $\dot{E} = 3.8\cdot 10^{34}$  erg/s for Monogem \cite{2005AJ....129.1993M}. 
Only middle-aged pulsars, with an observed age $50$~kyr$<t_{\rm obs}<10000$~kyr, are supposed to emit $e^\pm$. In younger pulsars  $e^\pm$ are believed to be confined until the expanding medium merges with the ISM,
which should occur at least $40-50$~kyr after the pulsar formation \citep{1996ApJ...459L..83C, 2011ASSP...21..624B}. 

In the burst-like injection scenario the flux $\mathcal{N}(E,\mathbf{r})$ of $e^\pm$ at a position $\mathbf{r}$ (in Galactic coordinates) and energy $E$  considering an infinite diffusion halo is given by  (see, e.g., \cite{2014JCAP...04..006D}):
 \begin{equation}\label{eq:singlesourcesolution}
  \mathcal{N}(E,\mathbf{r}) = \frac{b(E_s)}{b(E)} \frac{1}{(\pi \lambda^2)^{\frac{3}{2}}} \exp\left({-\frac{|\mathbf{r} -\mathbf{r_{s}} |^2}{ \lambda^2}}\right)Q(E_s)
\end{equation}
where $b(E)$ is the energy loss function, $\mathbf{r_{s}}$ indicates the source position, and $\lambda$ is the typical propagation scale length:
\begin{equation}
\label{eq:lambda}
 \lambda^2= \lambda^2 (E, E_s) \equiv 4\int _{E} ^{E_s} dE' \frac{D(E')}{b(E')},
\end{equation} 
with the diffusion coefficient $D(E)$ given by $D(E)= D_0(E/1{\rm \,GeV})^{\delta}$.
$E_s$ is the initial energy of $e^\pm$ that cool down to $E$ in a {\rm loss time} $\Delta \tau$:
\begin{equation}
 \Delta \tau (E, E_s) \equiv \int_{E} ^{E_s} \frac{dE'}{b(E')} = t-t_{{\rm obs}} .
\end{equation}
\\
In the continuous injection scenario and with a homogeneous diffusion in the Galaxy, the flux $\mathcal{N}_e(E,\mathbf{r},t)$ of $e^\pm$ at an  energy $E$, a position $\mathbf{r}$, and  time $t$ is given by the following equation \cite{Yuksel:2008rf}:
 \begin{eqnarray}
\label{eq:N_cont}
 && \mathcal{N}_e(E,\mathbf{r},t) = \int_0^{t} dt_0 \frac{b(E_s(t_0))}{b(E)} \frac{1}{(\pi \lambda^2(t_0,t,E))^{\frac{3}{2}}} \times  \nonumber\\
 && \times \exp\left({-\frac{|\mathbf{r} -\mathbf{r_{s}} |^2}{ \lambda(t_0,t,E)^2}}\right)Q(E_s(t_0)),
\end{eqnarray}
where the integration over $t_0$ is included since the PWN releases $e^\pm$ continuously in time.
The expression for the injection spectrum is now time-dependent:
 \begin{equation}
 Q(E, t)= L(t) \left( \frac{E}{E_0}\right)^{- \gamma} \exp \left(-\frac{E}{E_c} \right) ,
 \label{eq:Q_E_cont}
\end{equation}
and the total energy emitted by the source is given by:
 \begin{equation}
 E_{{\rm tot}} = \int_0^{T} dt \int_{E_{1}}^{\infty} dE E Q(E,t) = 
\int_0^{T} dt L(t).
 \label{eq:Q_E_cont}
\end{equation}
$L(t)$ is the magnetic dipole braking:
 \begin{equation}
 L(t) = \frac{L_0}{\left( 1+ \frac{t}{\tau_0} \right)^{2} }.
 \label{eq:Lt}
\end{equation}

The HAWC2017 data suggest that the diffusion coefficient in the vicinity of the source may be much smaller than the one usually derived for the average of the Galaxy. 
A possible phenomenological remedy of this discrepancy is the implementation of a two-zone diffusion model, where the region of inefficient diffusion is contained around the source, and delimited by an empirical radius \cite{Profumo:2018fmz,Tang:2018wyr}. 
The inhibition of diffusion near pulsars has been recently discussed in \cite{2018arXiv180709263E}, where a possible theoretical interpretation is provided. 
We implement here the following diffusion coefficient \cite{Tang:2018wyr}:
\begin{eqnarray}
\label{eq:conddm}
D(r) =  
\left\{
\begin{array}{rl}
& D_0 (E/1{\rm \,GeV})^\delta {\rm \;for\;} 0 < r < r_b, \\
& D_2 (E/1{\rm \,GeV})^\delta {\rm \;for\;} r \geq r_b,
\end{array}
\right.
\label{eq:Diff}
\end{eqnarray}
where $r_b$ is the boundary between the low-diffusion and high-diffusion zones.
The $e^{\pm}$ density in Eq.~\ref{eq:N_cont} takes the form:
 \begin{equation}
\label{eq:N_cont_2z}
  \mathcal{N}_e(E,\mathbf{r},t) = \int_0^{t} dt_0 \frac{b(E(t_0))}{b(E)} Q(E(t_0)) \mathcal{H}(\mathbf{r},E),
\end{equation}
where $\mathcal{H}(\mathbf{r},E)$ is:
\begin{eqnarray}
&& \mathcal{H}(\mathbf{r},E) =  \frac{\xi(\xi+1)}{(\pi \lambda_0^2)^{\frac{3}{2}} [2 \xi^2 {\rm erf}(\epsilon) - \xi(\xi-1) {\rm erf}(2\epsilon) + 2 {\rm erfc}(\epsilon)]} \nonumber \\
&&  \left\{
\begin{array}{rl}
 e^{({-\frac{\Delta r^2}{ \lambda_0^2}})} + \left(\frac{\xi-1}{\xi+1}\right) \left(\frac{2 r_b}{r}-1\right) e^{({-\frac{(\Delta r - 2 r_b)^2}{ \lambda_0^2}})},  0 < r < r_b \\
 \left( \frac{2\xi}{\xi+1} \right) \left[ \frac{r_b}{r} +\xi\left( 1 - \frac{r_b}{r} \right) \right] e^{( -[{\frac{(\Delta r - r_b)}{ \lambda_2}} + \frac{r_b}{\lambda_0} ]^2 )}, r \geq r_b,
\end{array}
\right.
\label{eq:Diff}
\end{eqnarray}
where $\Delta r = |\bf{r}-\bf{r_s}|$, $\xi$ is defined as $\xi=\sqrt{D_0/D_2}$, $\lambda_0$ and $\lambda_2$ are the typical propagation lengths for $D_0$ and $D_2$ (see Eq.~\ref{eq:lambda}) and $\epsilon = r_b/\lambda_0$.
In the case of $D_0=D_2$  or assuming $r_b \gg r$,  Eqs.~\ref{eq:N_cont_2z} for the two-zone diffusion model becomes Eq.~\ref{eq:N_cont}, which is valid for a one-zone model.
\\

In our model, the parameters that account for the two diffusion zones are $D_0$, $D_2$ and $r_b$. Different combinations of these parameters could generate very similar morphologies of the $\gamma$-ray ICS halo. This is particularly true for a very extended halo as the one we will search for Geminga and Monogem. In addition to this, the transition between the low and high-diffusion zones can be parametrized a priori with any arbitrary function of the distance, i.e.~an Heaviside or a smoother transition with an exponential, a power-law or a logarithmic function. All these additional effects, if included in our model, would make our analysis of $\gamma$-ray data extremely challenging. 
Moreover, these are second order effects that, if calibrated on the same $\gamma$-ray ICS halo morphology and flux, are not going to change significantly the results on the maximum contribution of these sources to the $e^+$ flux at Earth at very-high energies.
Therefore, we decide to assume the simplest approach for the $\gamma$-ray ICS halo that includes the one-zone model. Then, we calculate the $e^+$ flux at Earth considering the more complex two-zone diffusion model. In particular we will provide the results for the $e^+$ flux for different values of $r_b$ compatible with the $\gamma$-ray observations. This method will provide predictions for the $e^+$ flux at Earth that partialy incorporates the uncertainty in the spatial distribution of the diffusion reported above.

The Geminga pulsar has a proper motion of $178.2 \pm 1.8$ mas/year that corresponds to a transverse velocity of  $v_T \approx 211 (d/250\rm{pc})$ km s$^{-1}$ \cite{Faherty:2007}. 
On the other hand, the line of sight velocity is negligible, and thus it is not considered in this paper.
The transverse velocity affects significantly the morphology of the $\gamma$-ray emission from Geminga for energies smaller than about 100 GeV. A photon with an energy of 10 GeV is produced by an electron with an average energy of 100 GeV. These electrons propagate in the Galaxy for Myr timescales while loosing most of their energy. 
On this timescale, the Geminga pulsar travels many tens of parsecs.
The proper motion of Monogem is $44$ mas/year \cite{Hobbs:2005yx}, and does not affect the ICS $\gamma$-ray morphology.
We include the proper motion for Geminga in our calculation by replacing its position $\mathbf{r_s}$ in Eq.~\ref{eq:N_cont} with $\mathbf{r_s}+\mathbf{v}_T t_0$ where $\mathbf{v}_T$ is the pulsar transverse velocity which is a vector because we must specify the direction of motion.

\begin{figure*}[t]
\centering\includegraphics[width=0.49\textwidth]{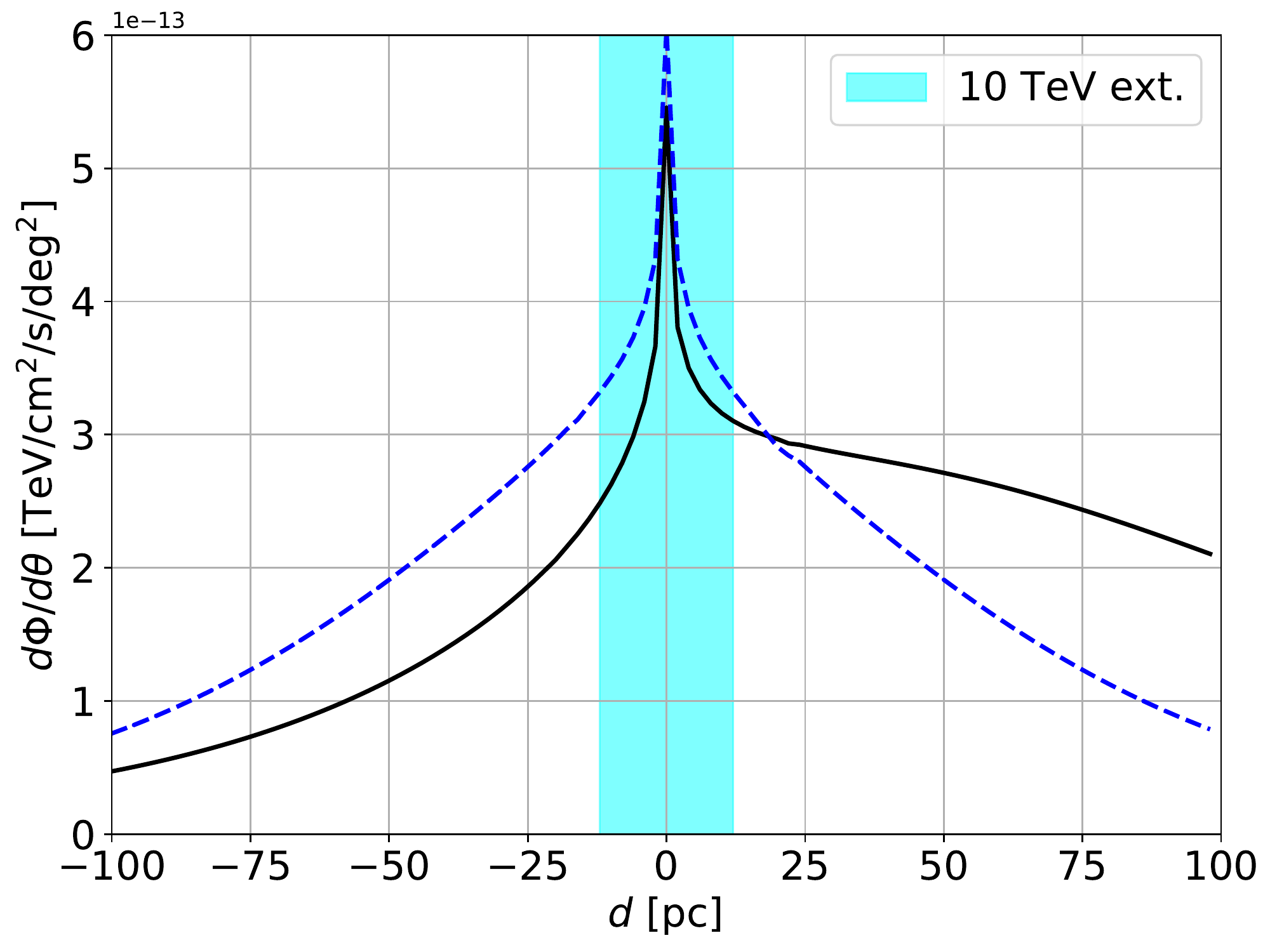}
\centering\includegraphics[width=0.49\textwidth]{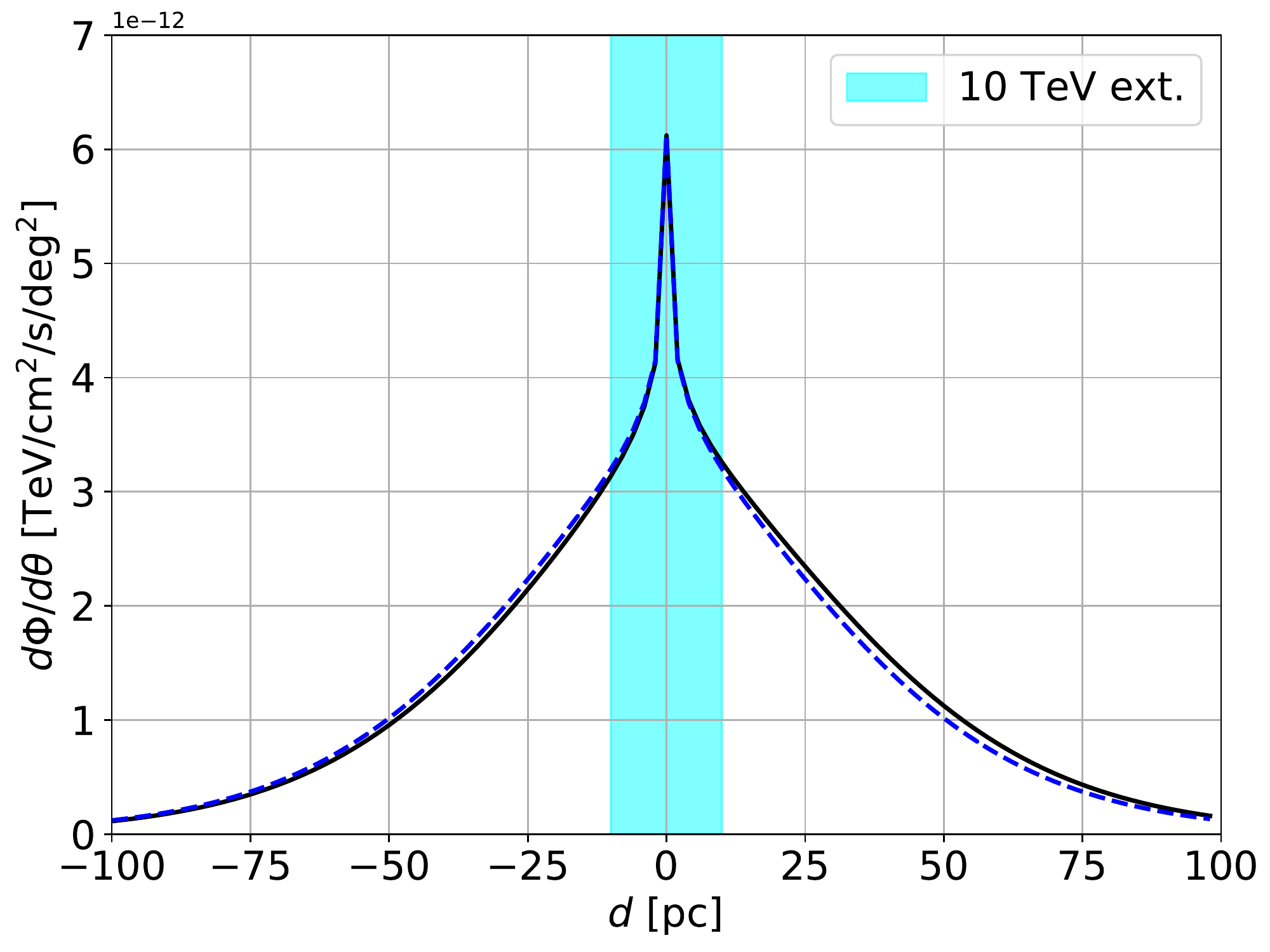}
\caption{$\gamma$-ray surface brightness $d\Phi_{\gamma}/d\theta$ calculated for Geminga (left panel) and Monogem (right panel) pulsars at $30$ GeV as a function of the transverse distance from the source $d_T$. We use here $D_0=2\cdot 10^{26}$ cm$^2$/s and $\delta=1/3$. The black (blue) lines are with (without) the proper motion for the pulsar. The direction of motion considered here is aligned to the pulsar proper motion.}  
\label{fig:spatialdistr}
\end{figure*}

In Fig.~\ref{fig:spatialdistr} we show the surface brightness $d\Phi_{\gamma}/d\theta$ as computed for Geminga and Monogem pulsars as a function of the transverse distance from the source ($d_T$) for a $\gamma$-ray energy of 30 GeV. 
$\theta$ is the angular separation between the line of sight and the direction of the source.
The direction of motion considered in this figure is aligned to the position vector from which the distance $d_T$ is calculated. Therefore, this represents the maximum effect that the pulsar velocity produces in the $\gamma$-ray morphology.
The $d\Phi_{\gamma}/d\theta$ is calculated with Eq.~\ref{eq:phflux} and \ref{eq:M} without integrating over the solid angle $\Delta \Omega$.
The proper motion of Geminga has a large impact on the ICS $\gamma$-ray emission which is larger by a factor of a few for positive $d_T$ that represents the position of the pulsar in the past.
The effect of Monogem proper motion is negligible as shown in Fig.~\ref{fig:spatialdistr} (right panel).

In either of the models the flux of $e^\pm$ at Earth is given by:
\begin{equation}
 \Phi_{e^+}(E) = \frac{c}{4\pi} \mathcal{N}_e(E,r=d,t=t^{\ast}),
 \label{eq:flux}
\end{equation}
where $d$ is the distance to the source. 
As a side note, we stress that $\gamma$ rays are produced by ICS from either $e^+$ and $e^-$ emitted by a PWN.
This implies that there is factor of 0.5 between the normalization of the injection spectrum derived from $\gamma$-ray data and the one that is then used to predict the $e^+$ flux at Earth.

\begin{figure*}[t]
\centering\includegraphics[width=0.49\textwidth]{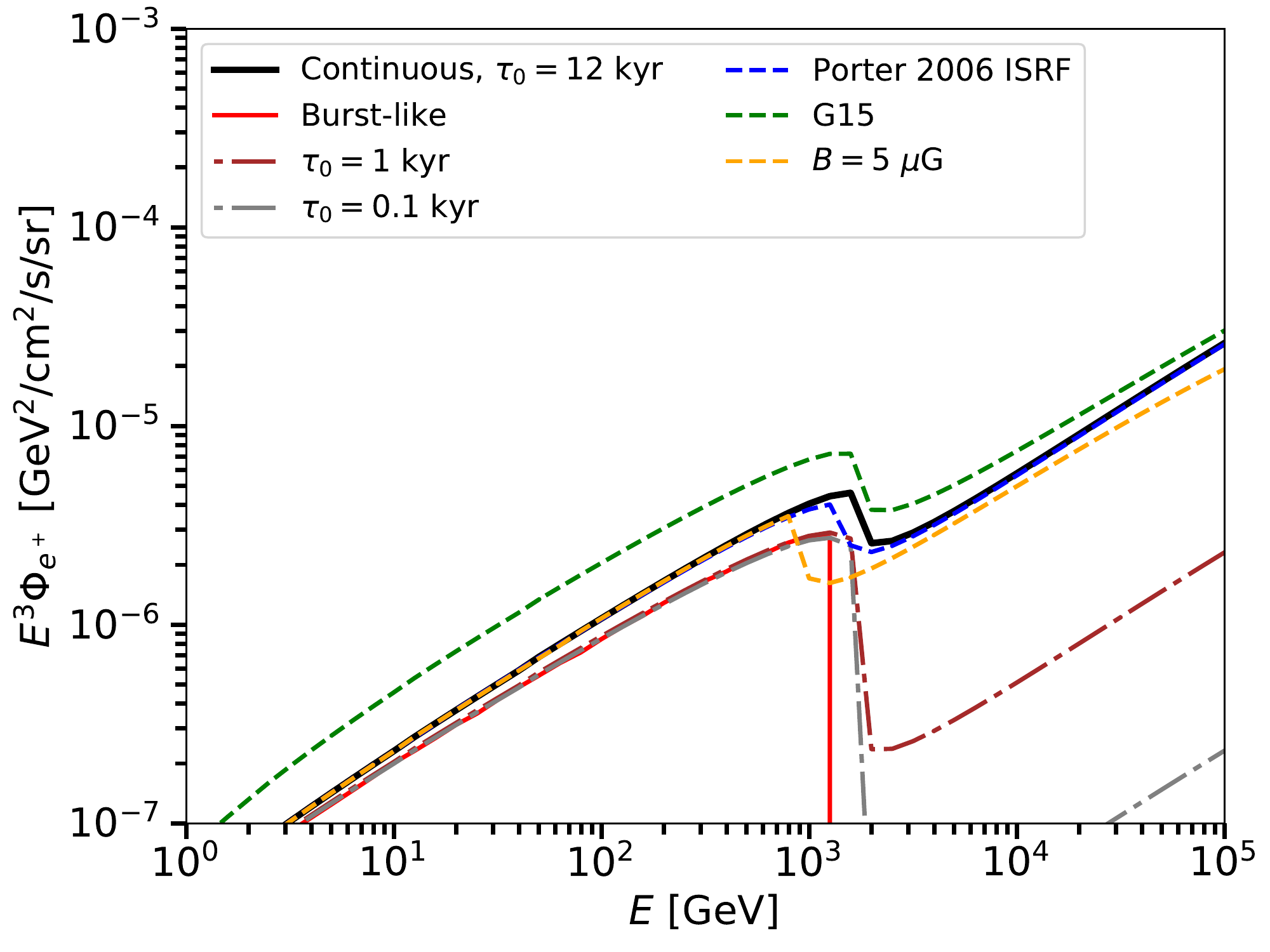}
\centering\includegraphics[width=0.49\textwidth]{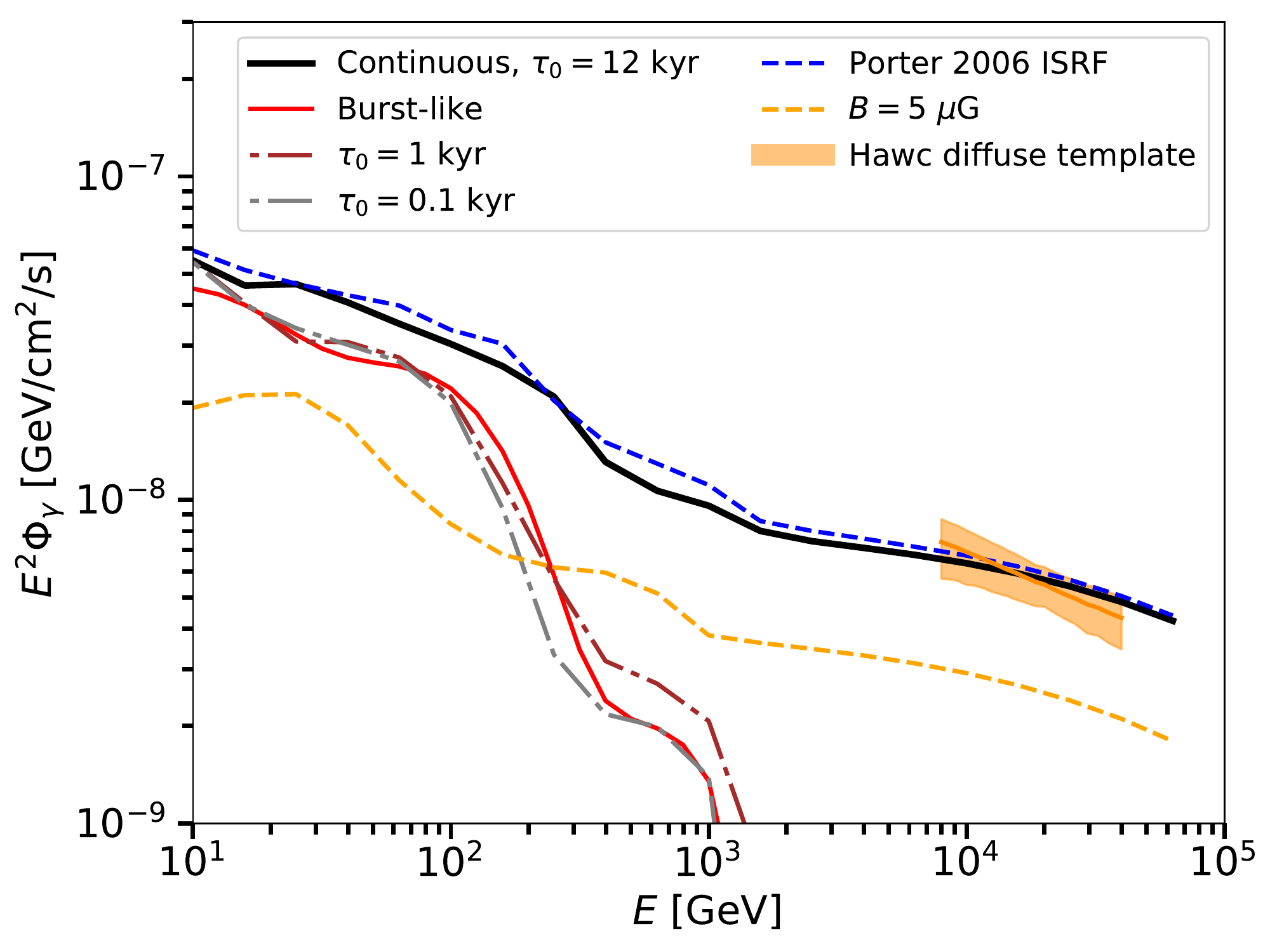}
\caption{
The $e^+$ flux (left panel) and the $\gamma$-ray flux from ICS (right panel) computed within the benchmark case as defined in Sec.~\ref{sec:e+e-} using a continuous injection and $\tau_0=$ 12 kyr (black solid line) or a burst-like injection (red solid line) for Geminga with a total energy emitted into $e^+$ and $e^-$ of $\eta W_0=1.5\times 10^{47}$ erg. 
We assume $\gamma_e=1.8$, $B=3 \mu$G, the ISRF model in the local Galaxy as in \cite{Vernetto:2016alq}. 
For the propagation model we assume K15 for the $e^+$ flux and $\delta=1/3$ and $D_0=10^{26}$ cm$^2$/s for the ICS $\gamma$-ray flux. 
We also display the effect on the $e^+$ and $\gamma$-ray fluxes of changing one at a time the parameters of the above model: $\tau_0= 1$ kyr (brown dot-dashed) and $\tau_0= 0.1$ kyr (gray dot-dashed), \cite{2006ApJ...648L..29P} ISRF (blue dashed line) and $B=5$ $\mu$G (orange dashed line). For the $e^+$ flux we also show the result derived with the G15 propagation model.}  
\label{fig:contburst}
\end{figure*}

We now implement our model to predict the flux of $e^{\pm}$ (we will then be interested specifically in $e^+$) and ICS $\gamma$ rays at Earth. 
In Fig.~\ref{fig:contburst} we show the predictions for the $e^+$ flux at Earth (left panel) and for the ICS $\gamma$-ray flux (right panel) for the burst-like and continuous injection, for Geminga. 
We assume a benchmark case with a spectral slope for the $e^+$ injection spectrum of $\gamma_e=1.8$, $\tau_0=$ 12 kyr, $\eta W_0=1.5\times 10^{47}$ erg, a strength of the Galactic magnetic field of $B = 3 \mu$G, the ISRF model of \cite{Vernetto:2016alq} in the local Galaxy and the propagation model in \cite{Kappl:2015bqa} (hereafter K15).
The K15 model as well as the one in \cite{Genolini:2015cta} (hereafter G15) have been tuned according to fits to CR data performed within a semi-analytical diffusion model.
In K15 the diffusion is modeled as $D_0=0.0967$ kpc$^2$/Myr and $\delta=0.408$, while for the G15 model with $D_0=0.05$ kpc$^2$/Myr and $\delta=0.445$.
The values found in these two papers are also compatible with the ones derived in \cite{Johannesson:2016rlh,Korsmeier:2016kha}.

This benchmark case has been inspired by the $\gamma$-ray flux observed with HAWC, but no fit has been performed.  
We also test the effect of the choice of CR $e^+$ injection spectrum, the CR propagation model, and the pulsar parameters on the $e^+$ flux and ICS $\gamma$-ray flux at Earth in the one-zone diffusion. 
Here we use the one-zone diffusion model but similar modifications are expected for the two-zone diffusion model.
We assume K15 propagation model for the e+ flux and $\delta = 1/3$ and $D_0 = 10^{26}$ cm$^2$/s for the ICS $\gamma$-ray flux.
As for the $e^+$ flux, the burst-like case shows a sharp cut-off at TeV energies because very-high energy $e^+$, injected from the source at a time equal to the source age, lose most of their energy during their propagation in the Galaxy for synchrotron and ICS cooling. 
On the other hand, the continuous injection scenario produces higher $e^+$ and ICS $\gamma$-ray fluxes at TeV energies for increasing values of $\tau_0$.
Indeed, when the release time is longer, a part of the most energetic $e^+$ is released much later than the time of the pulsar birth. 
Therefore, these very high-energy particles are characterized by a larger propagation length $\lambda$, which permits them to reach the Earth and to generate TeV $\gamma$ rays.
On the other hand, if $\tau_0$ is very small  ($\tau_0 < 1$ kyr) the predictions for the continuous injection asymptotically tends to the burst-like scenario.
We also test the effect on the $e^+$ and ICS $\gamma$-ray flux for different choices of the ISRF, propagation model, and Galactic magnetic field strength. 
A change from the ISRF model in \cite{Vernetto:2016alq} to the one in \cite{2006ApJ...648L..29P}  has a negligible effect for $e^+$ flux and of about $10\%$ in the case of the ICS $\gamma$-ray flux. 
The difference in the $e^+$ flux as propagated using the K15 model and the one in G15 is a normalization factor of about 1.5.
Finally, changing the Galactic magnetic field from 3 to 5 $\mu$G has an effect in the $e^+$ flux only at around 1 TeV.
Indeed at TeV energies the synchrotron radiation mechanism of $e^{\pm}$ energy losses, that at lower energy is subdominant, is of the same order of the one due to ICS.
On the other hand, the ICS $\gamma$-ray flux changes by an overall factor of about 2 if we increase the magnetic field strength from 3 to 5 $\mu$G because we are using 
for this case a low-diffusion propagation model with $D_0 = 10^{26}$ cm$^2$/s.
It is visible from the figure that a pulsar spin-down timescale $\tau_0 \gsim $ 10 kyr is needed in order to be compatible with the HAWC2017 observations. 
In the following, we will fix $B= 3\mu$G,  the \cite{Vernetto:2016alq} ISRF model in the local Galaxy as well as  $\tau_0 =$ 12 kyr.

\section{$e^+$ flux from Geminga and Monogem PWNe derived with HAWC data}
\label{sec:hawcresults}

\begin{figure*}[t]
\centering\includegraphics[width=0.49\textwidth]{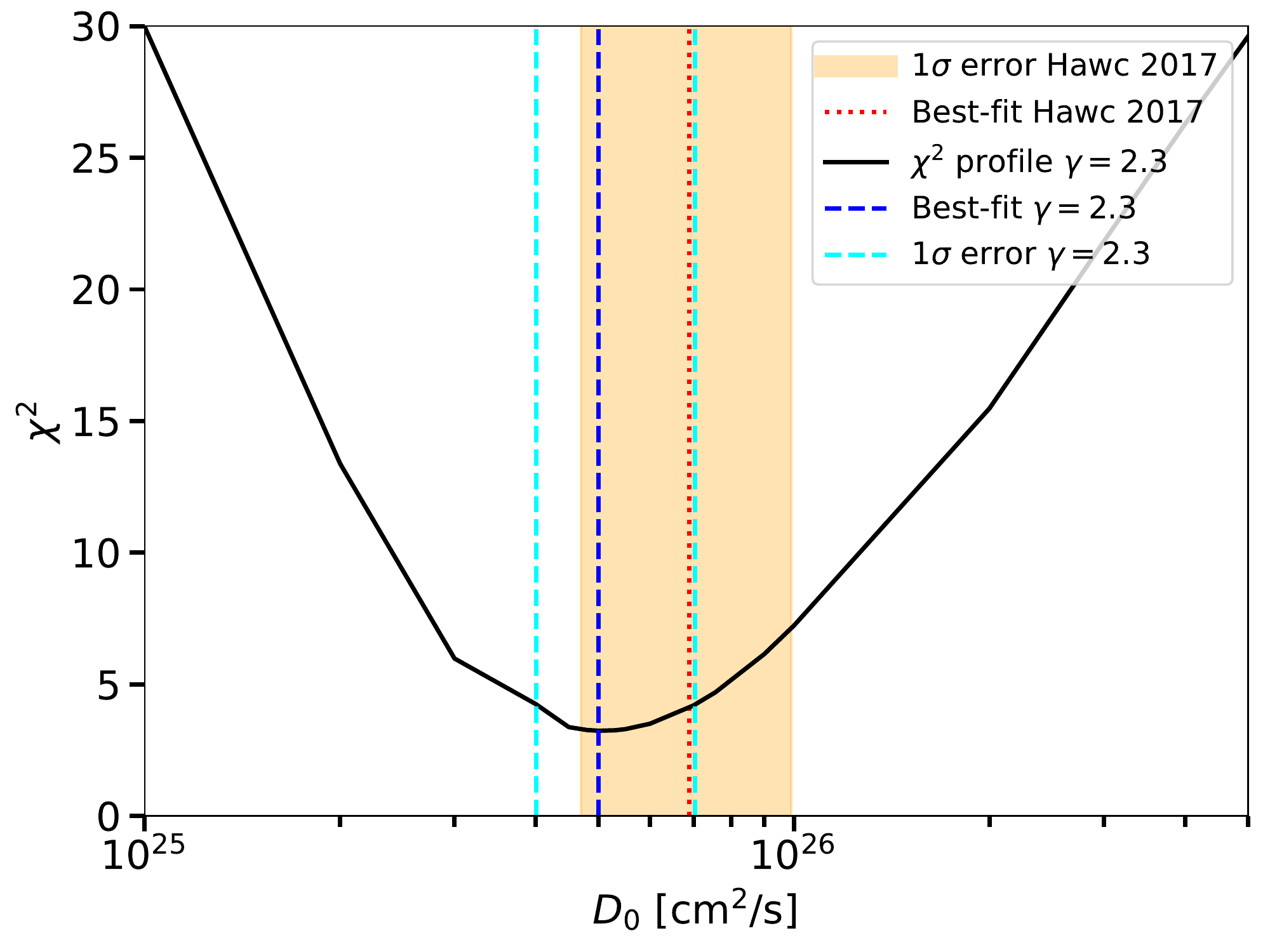}
\centering\includegraphics[width=0.49\textwidth]{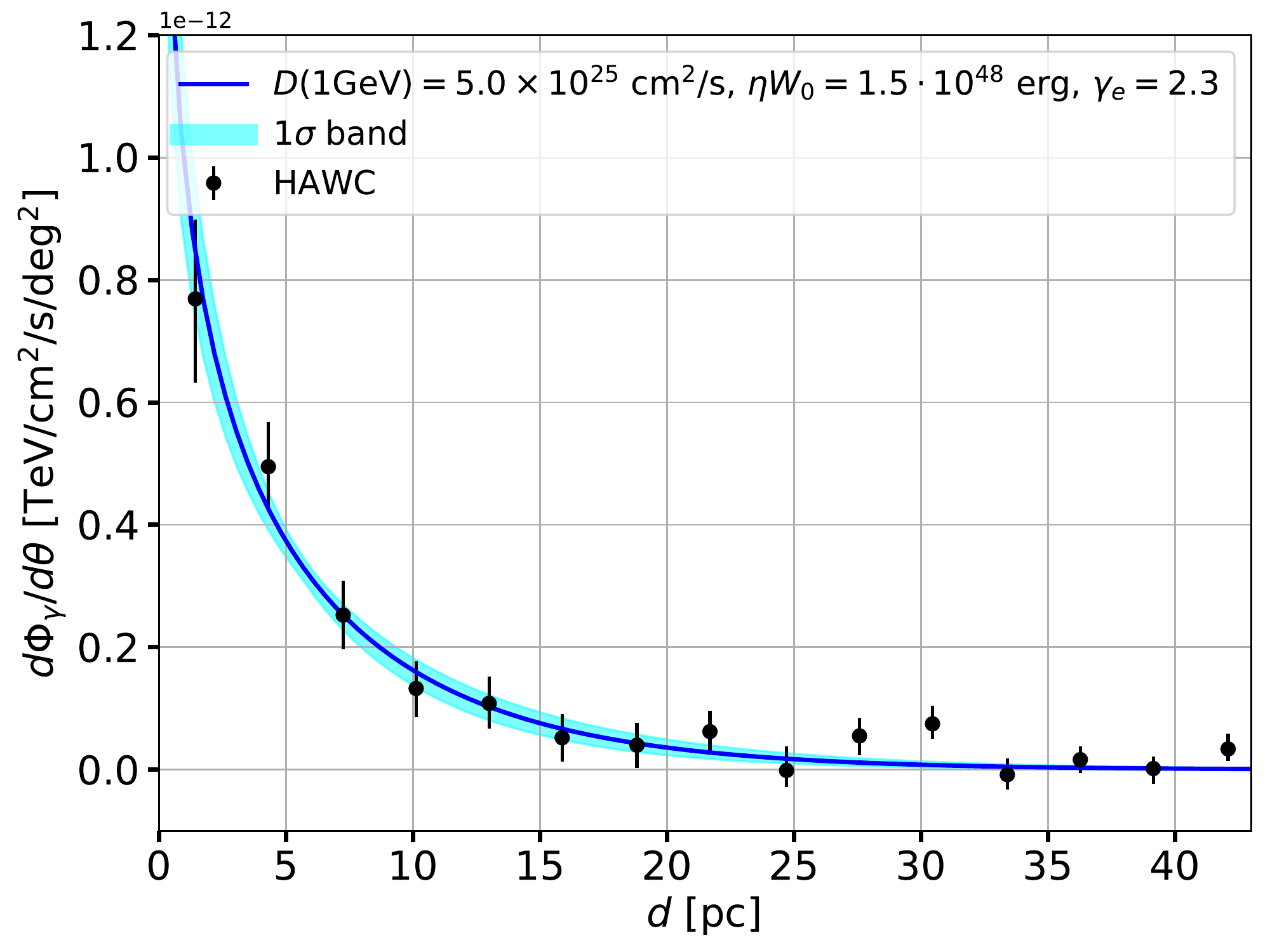}
\caption{Fit to the Geminga surface brightness. Left panel: $\chi^2$ as a function of $D_0$, with $\delta=1/3$ and $\gamma_e=2.3$. The black line is the $\chi^2$ profile, the blue dashed line is the best fit value and the cyan lines represent the $1\sigma$ errors. The orange band and red dotted line report  the $1\sigma$ band and the best fit value, respectively, as derived in HAWC2017. Right panel: Surface brightness profile between $5-50$ TeV. The blue solid line and the cyan bands are the best fit and $1\sigma$ uncertainty band derived in our analysis. We also display the HAWC data (HAWC2017).}  
\label{fig:SB_initial_Geminga_2p3}
\end{figure*}

\begin{figure*}[t]
\centering\includegraphics[width=0.49\textwidth]{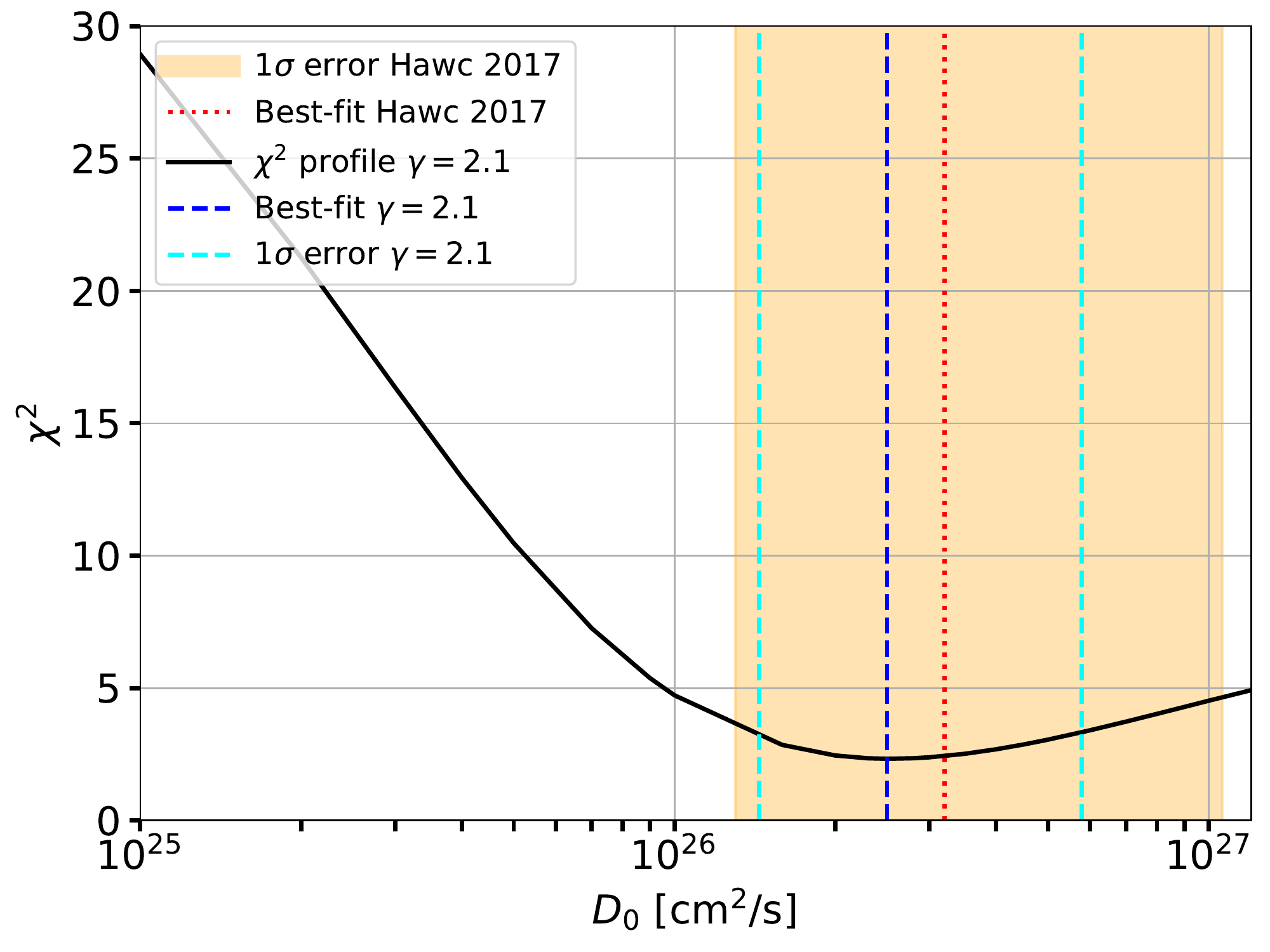}
\centering\includegraphics[width=0.49\textwidth]{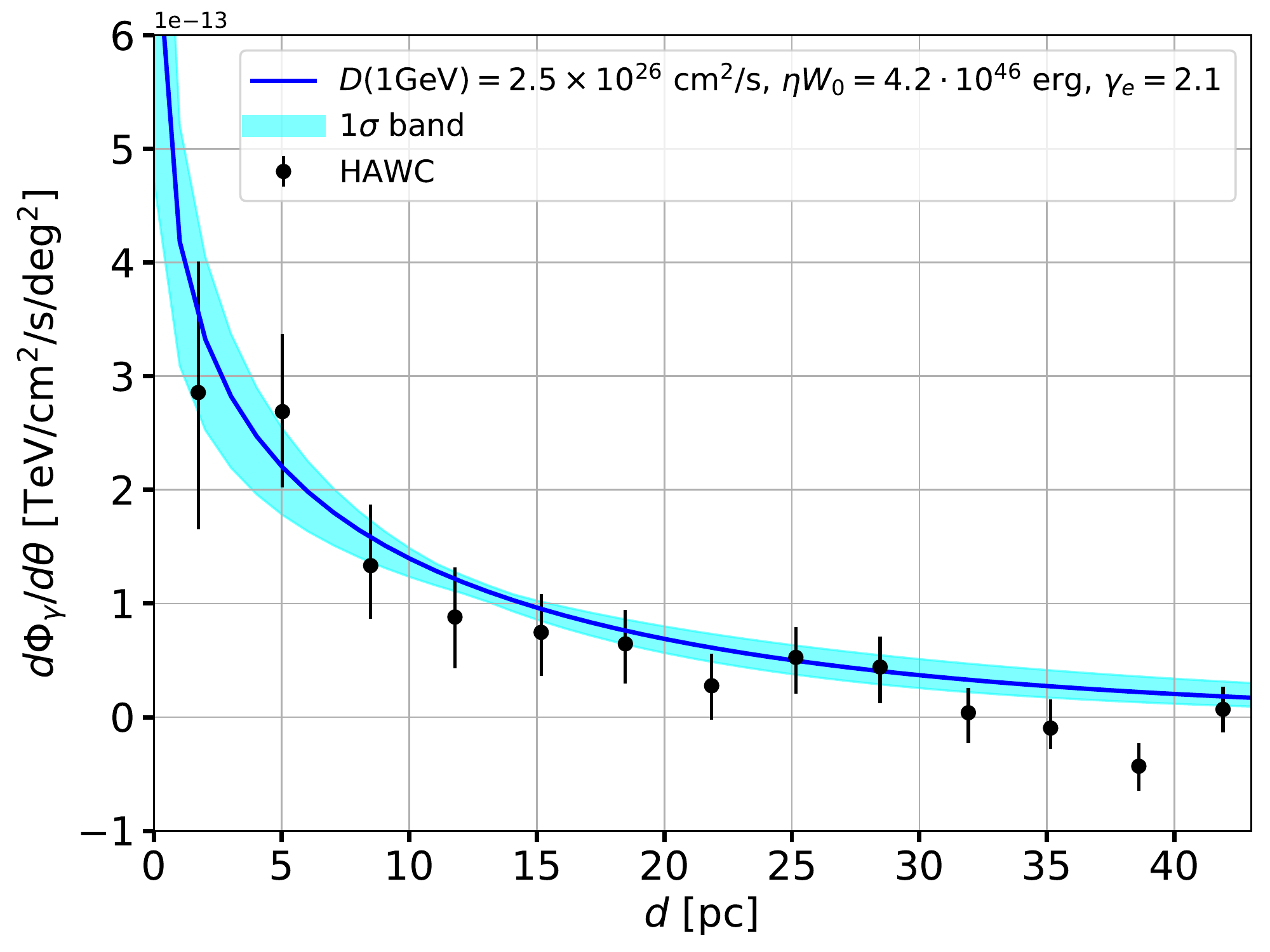}
\caption{Same as Fig.~\ref{fig:SB_initial_Geminga_2p3} for Monogem PWN and with $\gamma_e=2.1$.}  
\label{fig:SB_initial_Monogem_2p1}
\end{figure*}

\begin{figure*}[t]
\centering\includegraphics[width=0.49\textwidth]{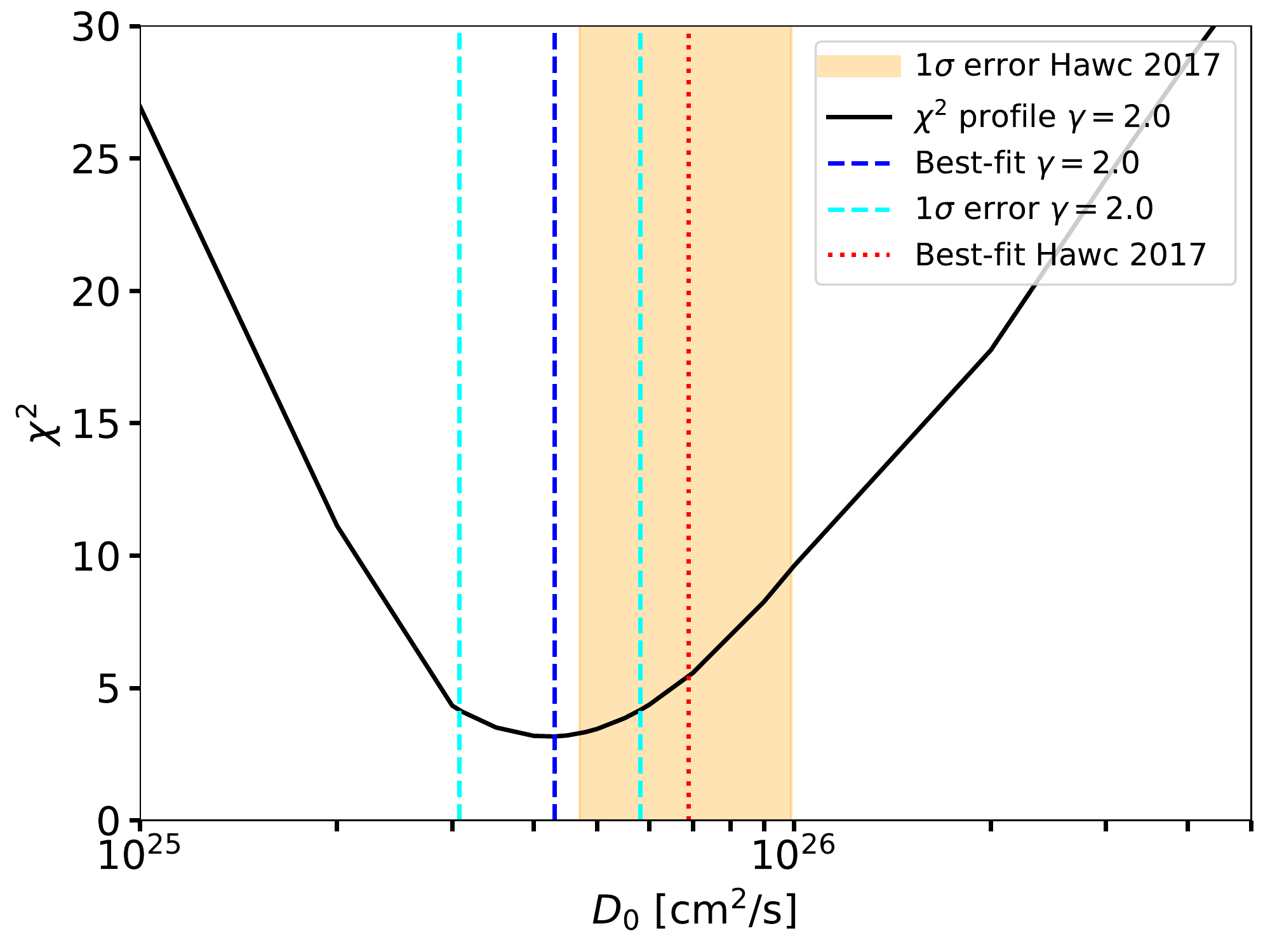}
\centering\includegraphics[width=0.49\textwidth]{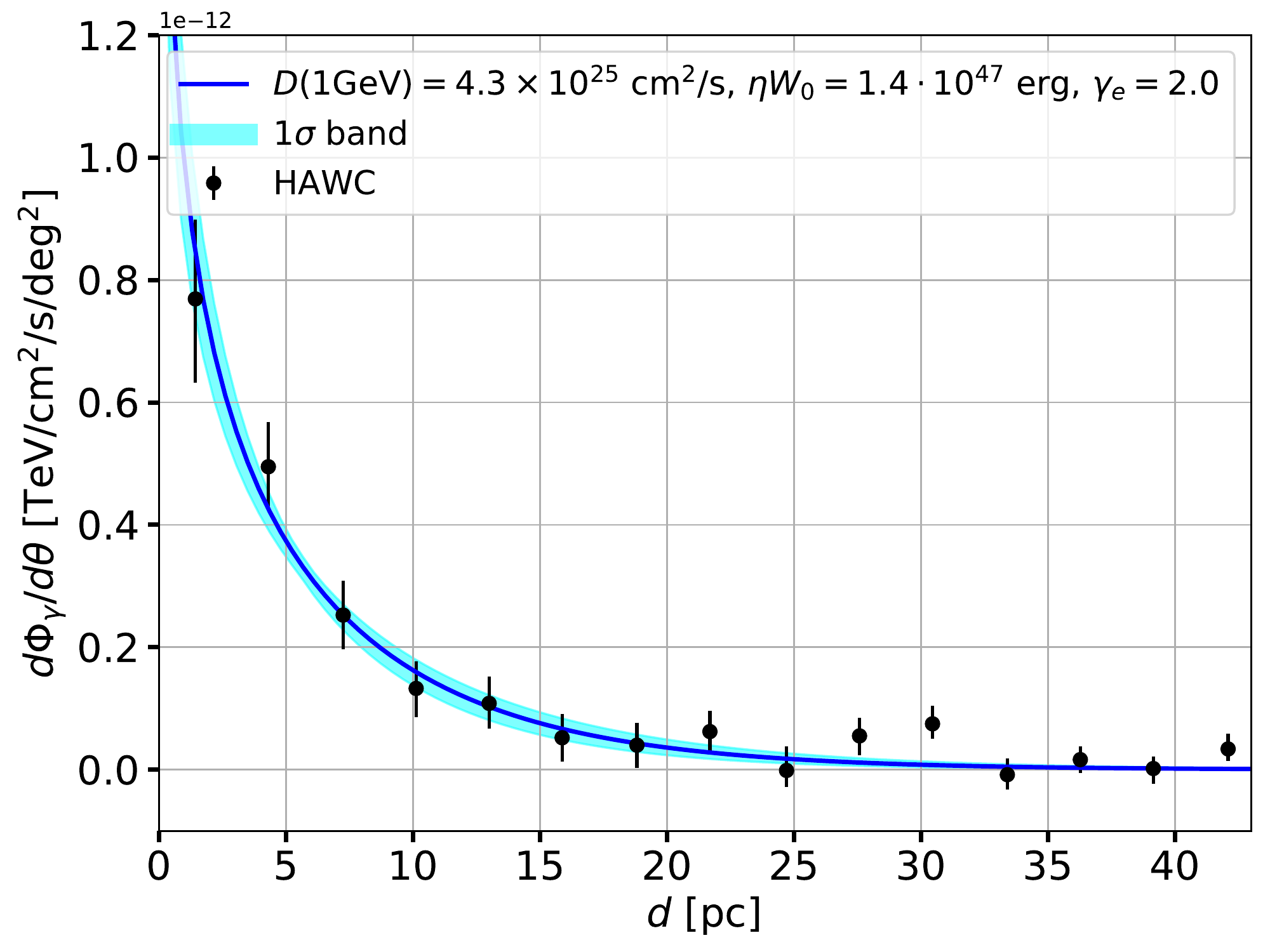}
\caption{Same as Fig.~\ref{fig:SB_initial_Geminga_2p3} for Geminga PWN with $\gamma_e=2.0$.}  
\label{fig:SB_initial_Geminga_2p0}
\end{figure*}

In this section we fit the surface brightness measured with HAWC to predict the $e^+$ flux at Earth produced by the Geminga and Monogem PWNe.
The surface brightness $d\Phi_{\gamma}/d\theta$ for ICS is computed from Eq.~\ref{eq:phflux} without integrating over $\theta$.
As in HAWC2017, we assume that the diffusion of $e^+$ and $e^-$ in the vicinity of the PWN ($D_0$) is different with respect to the average of the Galaxy. 
Therefore, we implement a one-zone diffusion model assuming $D_0$ in the entire region observed by HAWC.
We also proceed as in HAWC2017 by fixing $\delta=1/3$ and varying $D_0$ and $\eta W_0$ in the fit.
This value for $\delta$ is motivated by the Kolmogorov turbulence model \cite{1941DoSSR..30..301K}.
We first fix $\gamma_e = 2.30$ for Geminga and $\gamma_e = 2.10$ for Monogem as done by the HAWC Collaboration in their analysis.
The best fit gives $D_0 = 5.0^{+2.0}_{-1.0} \cdot 10^{25}$ cm$^2$/s and $\eta W_0 = 1.5 \cdot 10^{48}$ erg for Geminga and $D_0 = 2.5^{+3.3}_{-2.1} \cdot 10^{26}$ cm$^2$/s and  $\eta W_0 = 4.2 \cdot 10^{46}$ erg 
for Monogem.
Determining $W_0$ from Eq.~\ref{eq:W0PWN}, we derive $\eta= 0.12\pm0.02$ for Geminga and $\eta=0.03\pm0.01$ for Monogem.
The results of the fit to the HAWC surface brightness are shown in Figs.~\ref{fig:SB_initial_Geminga_2p3} and \ref{fig:SB_initial_Monogem_2p1} for Geminga and for Monogem, respectively.
The left panels show the chi-square ($\chi^2$) profile as a function of  $D_0$. The comparison  with the HAWC results demonstrates that we find compatible results.
The values  derived by the HAWC Collaboration are $D_0= 6.9^{+3.0}_{-2.2} \cdot 10^{25}$ cm$^2$/s for Geminga and $D_0= 3.2^{+10.6}_{-2.0} \cdot 10^{26}$ cm$^2$/s for Monogem\footnote{We find these values rescaling from 100 TeV to 1 GeV the results reported in HAWC2017 for the diffusion coefficient and assuming $\delta=1/3$.}.
We also try the same fit for a harder $e^\pm$ injection spectrum index of $\gamma_e = 2.00$ for Geminga. As reported in Fig.~\ref{fig:SB_initial_Geminga_2p0}, we find 
that the best fit value is $D_0 = 4.3^{+1.5}_{-1.2} \cdot 10^{25}$ cm$^2$/s and $\eta=0.011$, still falling in the HAWC uncertainty band.  
This test indicates that the harder injection spectrum is still giving a good fit to the Geminga surface brightness.

\begin{figure}[t]
\centering\includegraphics[width=0.49\textwidth]{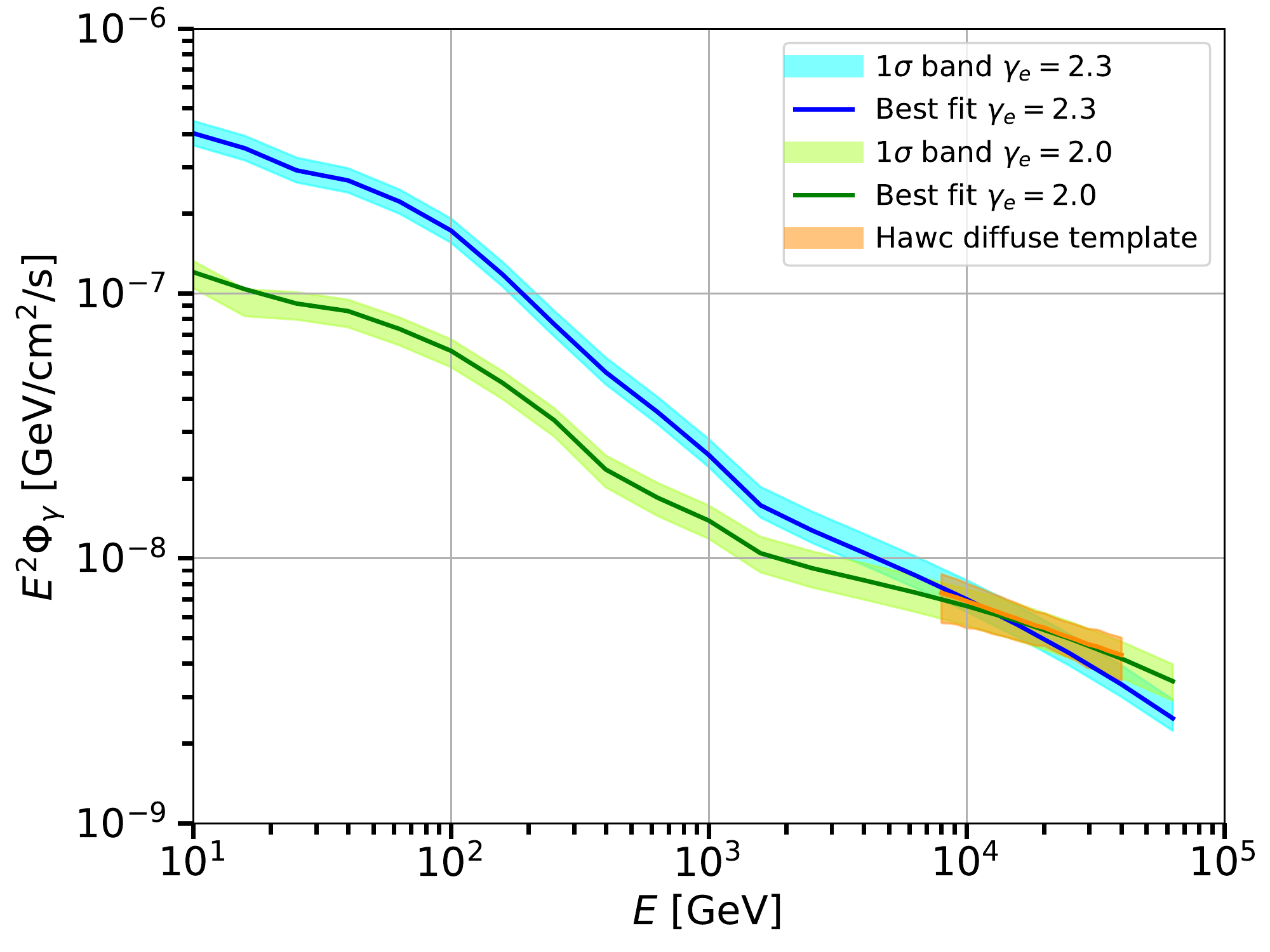}
\caption{$\gamma$-ray flux for ICS from Geminga for $\gamma_e=2.3$ (blue) and 2.0 (green). The solid line represents the best fit while the bands are the $1\sigma$ uncertainty derived from a fit to the HAWC surface brightness. Together with the theoretical predictions we report also the flux measured by HAWC using a diffuse template compatible with ICS $\gamma$ rays.}  
\label{fig:ICGeminga_2p0_2p3}
\end{figure}

\begin{figure*}[t]
\centering\includegraphics[width=0.49\textwidth]{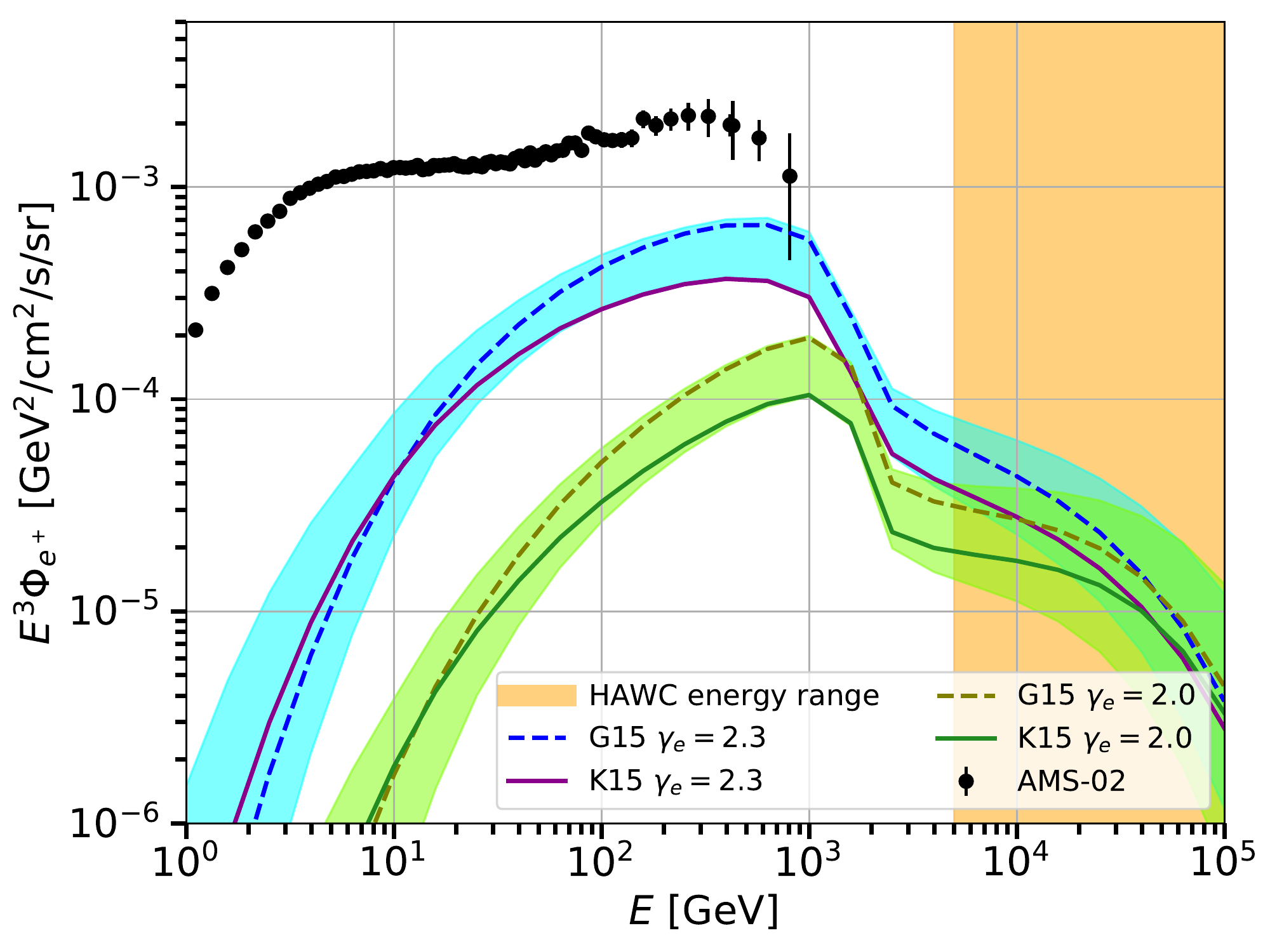}
\centering\includegraphics[width=0.49\textwidth]{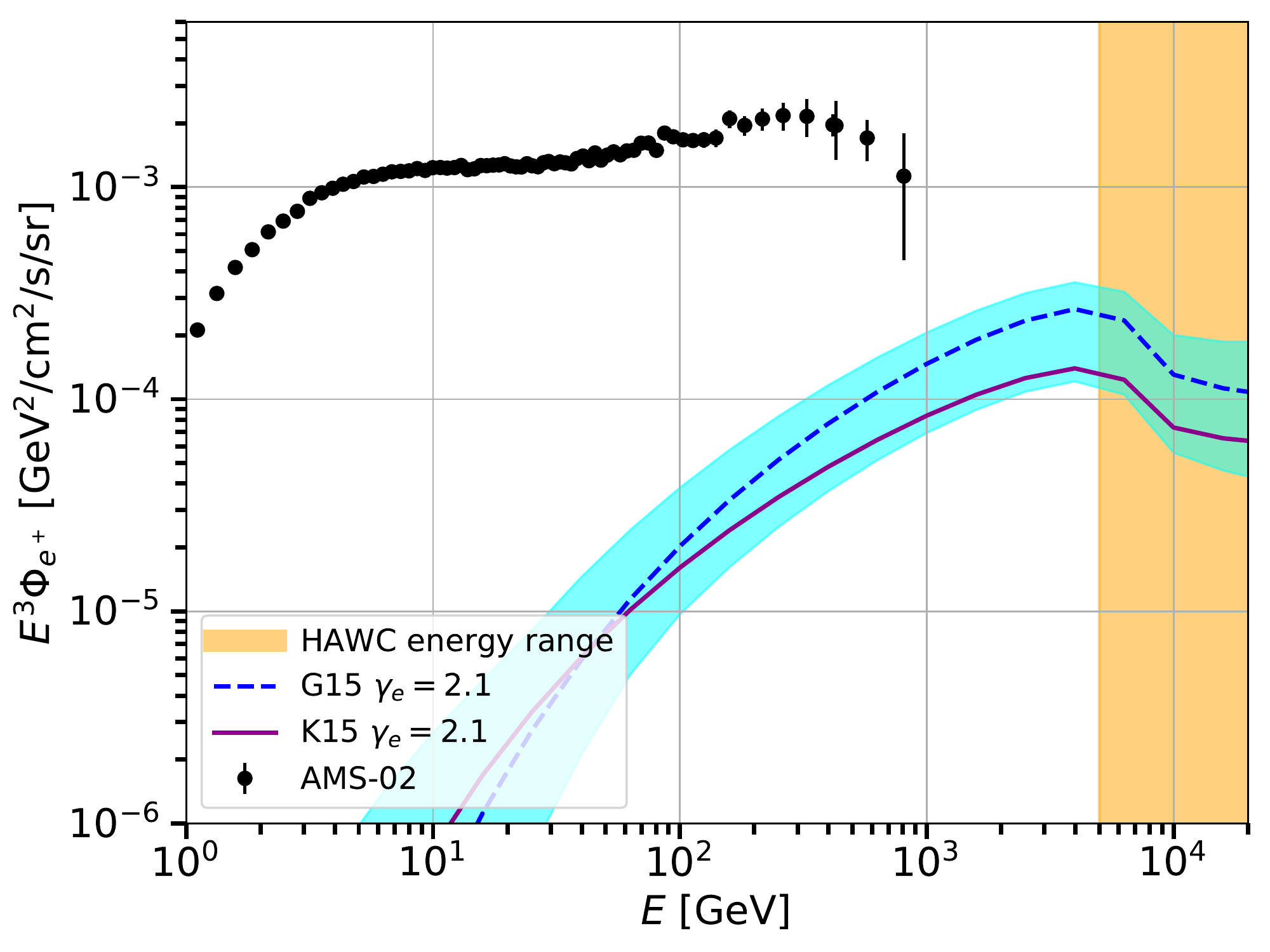}
\caption{Flux at Earth of $e^+$ for Geminga (left) and Monogem (right). Solid and dashed lines are for K15 and G15 propagation parameters. Upper (lower) curves  are for $\gamma_e=2.3$ ($\gamma_2=2.0$). 
The bands include  the uncertainty in $\eta W_0$ and changing the propagation model (see text for details). The orange band outlines the energy range covered by HAWC.
For Monogem $\gamma_e=2.1$.}  
\label{fig:Jinitial}
\end{figure*}

We now use the results obtained from HAWC data on the surface brightness to predict the flux of photons due to ICS between 10 GeV to 10 TeV.
In Fig.~\ref{fig:ICGeminga_2p0_2p3} we show the $\gamma$-ray flux for ICS from Geminga, implementing  our results for both $\gamma_e=2.3$ and 2.0. We also provide a  comparison to the HAWC flux data as measured in 
 HAWC2017 for a diffuse template compatible with the ICS process.
We can clearly see that the predicted flux for the two $\gamma_e$ are very similar in the HAWC energy range, but spread  by about a factor of 4 in the {\it Fermi}-LAT energy range between $10-100$ GeV.
This indicates that the extrapolated fluxes can in principle be tested  with  {\it Fermi}-LAT data. 

A similar consideration holds for $e^+$ flux at Earth. 
While the HAWC surface brightness profile is well fitted with different $\gamma_e$  and $\eta$ values, the corresponding $e^+$ flux at AMS-02 energies can be significantly different. 
In Fig.~\ref{fig:Jinitial} we quantify this consideration for Geminga (left panel) and Monogem (right panel) together with AMS-02 data.
In order to show the effect of the propagation from the source to the Earth, we additionally calculate the flux for the two different sets of  propagation parameters K15 and G15. 
Since, the size of the $\gamma$-ray ICS halo measured in HAWC2017 for Geminga is of the order of $20$ pc (see, e.g., Fig.~\ref{fig:SB_initial_Geminga_2p3}) and this represents a negligible volume for the propagation of $e^{\pm}$ from the pulsar to the Earth, we assume here K15 and G15 models which parametrize the average propagation of CRs in the Galaxy.
The bands (the blue and green one) are by themselves a convolution of the uncertainties brought by $\eta W_0$ and the two propagation setups. 
Considering the variation for $\gamma_e$, Geminga can contribute to the $e^+$ AMS-02 high-energy data at most $30\%$ for $\gamma_e=2.3$, and $10\%$ with $\gamma_e=2.0$, while Monogem can contribute at most $5\%$.
The main consequence of this analysis is that with HAWC data alone it is not possible to constrain the contribution of Geminga and Monogem PWNe to the AMS-02 $e^+$ excess. 
Indeed, even changing the value of Geminga $\gamma_e=2.3$ by only 0.3, its contribution  can vary by at a factor of 5 at the highest AMS-02 energies. 
This result is not unexpected, since the $e^+$ flux is tuned to HAWC data at $E_\gamma>5$ TeV. 
For the physics involved in the ICS, HAWC data are able to constrain only $e^+$ at tens of TeV energies. 
In the Thomson regime, a $\gamma$ ray detected at 10 TeV is on average produced by a $e^+$ or an $e^-$ at energies around 60 TeV energy through ICS with the CMB. During the propagation from the source to Earth $e^+$ lose energy through synchrotron radiation and ICS and $e^+$ are detected with an energy of about 2 TeV\footnote{We assume for this simple calculation that the energy of a photon produced by a positron with a Lorentz factor of $\gamma$ for ICS on a photon field given by a black body distribution with a characteristic temperature of $T_{{\rm BB}}$ is $3.60 \gamma^2 k_B T_{{\rm BB}}$. Moreover we assumed energy losses given by $5\cdot 10^{-17} E^2$ GeV/s that are compatible with a Galactic magnetic field with $3\mu$G and the ISRF of \cite{Vernetto:2016alq,2006ApJ...648L..29P}.}.
In Fig.~\ref{fig:Jinitial} we see that the $e^+$ flux is similar for the two $e^+$ spectral slopes ($\gamma_e=2.3, 2$) only for $E \geq 5$ TeV.
Therefore, $\gamma$-ray data between $10-1000$ GeV would be highly desirable in order to constrain more precisely the Monogem and Geminga contribution to the $e^+$ excess.
Indeed, ICS photons in this energy range are produced by $e^{\pm}$ detected at Earth with average energies in the range $350-1500$ GeV rendering {\it Fermi}-LAT perfectly suited for their detection, as we will discuss in the next Sections.

As we partially discussed above, our results are only marginally comparable with the HAWC2017 ones, whose analysis is based on a strong assumption.
The same diffusion coefficient that explains the spatial morphology of the $\gamma$-ray emission near the PWNe halos 
($D_0 = 7\cdot 10^{25} {\rm cm}^2/s$, $\delta=1/3$) is assumed for the propagation of $e^+$ in the Galaxy. 
This value of $D_0$ is a factor of about 500 smaller than the average value assumed for the Galaxy. 
The size of the observed ICS emission is about 20 pc and the distance of these sources is about 250 pc, thus meaning that the propagation volume in the ICS region is only about 1\% of the volume traveled by $e^+$ from the source to the Earth. 

The efficiencies that we find are lower than the values derived by the HAWC Collaboration, namely  $\eta=0.40$ for Geminga and $\eta=0.04$ for Monogem. 
We have identified five main differences between our analysis and the one presented in HAWC2017.

In HAWC2017, the ISRF is described by three 
separate black body distributions, while we consider the model \cite{Vernetto:2016alq} without any approximation. 
The black body approximation gives an ISRF spectrum that predicts an emission lower by a factor of about $25\%$ than the full model, in particular for frequencies in the range $10^{13}-10^{14}$ Hz. 
Therefore, by using the black body approximation higher efficiency values are found.

The second difference is due to the minimum energy of $e^{\pm}$ (see Eq.~\ref{eq:Etot}) that in HAWC2017 is fixed to 1 GeV while we consider 0.1 GeV. This is particularly relevant for values of $\gamma_e$ much softer or harder than $2.0$.
For example, taking the efficiency derived in HAWC2017 for $>1$ GeV, rescaling it to $E_e>0.1$ GeV and assuming $\gamma_e=2.3$, we obtain $\eta=0.9$, almost equal to the total spin-down luminosity.
Using for the minimum $e^{\pm}$ energy 1 GeV reduces the efficiency by about a factor of 2.

In HAWC2017 they follow the treatment of the energy losses presented in \cite{Lopez-Coto:2017uku}. There is a discrepancy with our model of about a factor of 2 in the intensity of the energy losses at 10 TeV. This difference is reasonable given the current accuracy for the knowledge of the ISRF spectrum and magnetic field strength in the local Galaxy and considering that the location near the PWN could have different values of these quantities.
The different treatment of the energy losses gives a $40\%$ lower ICS $\gamma$-ray flux.

In HAWC2017 they use the diffusion coefficient that is found with the combined analysis of Geminga and Monogem, namely $D(100\rm{\,TeV})=4.5\cdot 10^{27}$ cm$^2$/s and an energy dependece $D(E)=(1+E/3\rm{\,GeV})^{1/3}$.
These assumptions make the ICS $\gamma$-ray flux smaller by about a factor of $25\%$ with respect to the ICS flux found using the value of $D_0$ determined for Geminga only.

Finally, they use for the total energy emitted by Geminga $W_0=1.1\cdot 10^{49}$ erg while we consider $W_0=1.3\cdot 10^{49}$ erg. On the other hand we assume the same value for Monogem. The discrepancy in the Geminga $W_0$ is probably due to a different spin-down energy used for this source. We take in our analysis the value reported in the ATNF catalog. 

Following the same prescriptions used in HAWC2017 for the quantities listed above and performing again a fit to the HAWC surface brightness, we find $\eta = 0.26\pm 0.06$ for Geminga and $\eta = 0.03\pm0.01$ for Monogem. The value of $\eta$ we find for Monogem is compatible with HAWC2017 result within $1\sigma$ error. 
On the other hand, for Geminga they are compatible at $2\sigma$ error. 
We finally note that there are further differences such as the implementation of the calculation of the $\gamma$-ray ICS emission or in the procedure used to fit the surface brightness data that are not identical and can have some role in the values of $\eta$.

Other studies recently considered the ICS flux and consequent $e^+$ flux from Geminga and Monogem in light of the recent HAWC measurements \cite{Hooper:2017gtd,Profumo:2018fmz,Shao-Qiang:2018zla,Tang:2018wyr,Fang:2018qco,Johannesson:2019jlk}.
Before explaining the differences and similarities with those papers we note that the value of the efficiency strongly depends on the shape of the $e^{\pm}$ spectrum and the energy range considered for the injection of these particles.
For example we have shown earlier in this Section that with $\gamma_e=2.3$ an efficiency of $0.12$ is required to fit the HAWC surface brightness while with $\gamma_e=2.0$, the efficiency is $\eta=0.011$. Indeed, the case with a softer $\gamma_e$ contributes fewer $e^{\pm}$ at TeV energies so a higher $\eta$ is necessary to fit the HAWC data.

The authors of \cite{Hooper:2017gtd} find that the $e^+$ excess can be entirely explained by PWNe, with Geminga contributing more than 10\% to the flux at AMS-02 energies.
However, the surface brightness from HAWC was not used because these data were not available at the time. They considered the flux data from the 2HWC HAWC catalog \cite{Abeysekara:2017hyn}, where a disk template was assumed for all sources. This profile provides a flux that is lower by about a factor of 4 with respect to the ICS template (see Fig. S2 in the supplementary materials of HAWC2017). The ICS template has been used by the HAWC Collaboration and found  to be preferred at more than $5\sigma$ significance with respect to the disk morphology for Geminga\footnote{Private communication with contact authors of HAWC2017.}
Finally, Ref.~\cite{Hooper:2017gtd} also uses the Milagro data point at 35 TeV \cite{2009ApJ...700L.127A}, where a template of a single point source has been considered. 
Therefore, rescaling their results (i.e., with the same $\gamma_e$ they use in the paper) by the difference between the HAWC data in the disk and diffuse assumptions (i.e., by multiplying them by a factor of about 4), Geminga would exceed the AMS-02 positron flux above 10 GeV by about a factor of two. This estimate is affected by an extrapolation below the energy range covered by HAWC. Indeed, HAWC $\gamma$-ray data above 10 TeV constrain the $e^+$ population above TeV energies for the flux at Earth. Therefore, the discrepancy reported above between the Geminga $e^+$ flux and AMS-02 data can be reconciled with harder indexes for $\gamma_e$.
In \cite{Profumo:2018fmz} the authors assumed a spectral slope of $\gamma_e=2.3$ for $e^+$ which, as we will show in this paper, is not compatible with the spectrum we find with {\it Fermi}-LAT data for the $\gamma$-ray emission from Geminga and Monogem PWNe. 
We also note that they work under the hypothesis of a burst-like injection spectrum for Geminga. As illustrated in Sec.~\ref{sec:e+e-}, the burst-like injection is not compatible with the HAWC observation of TeV 
$\gamma$ rays when interpreted as being produced from ICS. 
Finally, there is a missing factor of 2 in the definition of the diffusion length $r_{{\rm diff}}$ (Eq.~4) that likely affects their results\footnote{The authors of \cite{Profumo:2018fmz} confirmed this missing factor in the definition of $r_{{\rm diff}}$}.


Recently, Ref.~\cite{Johannesson:2019jlk} used the GALPROP code to inspect the propagation of $e^{\pm}$ in light of HAWC2017 data. They find that an efficiency of $0.26$ explains the data for Geminga. The comparison with their efficiency value is difficult since they use an injection spectrum given by a smoothly broken power-law which is very different from our model. In addition they calculate the ICS flux integrating a region $10^{\circ}$ wide around Geminga pulsar so their prediction for this quantity is underestimated at low $\gamma$-ray energy. Indeed, around 10 GeV the ICS flux is very extended (see Fig.~\ref{fig:mapcube_Geminga}). 
Nevertheless, their results (see Fig 2 and 3) are similar to ours for the positron flux when they calibrate the efficiency on the HAWC data.

Ref.~\cite{Shao-Qiang:2018zla} analyzed {\it Fermi}-LAT data to search for an extended emission from Geminga and Monogem PWNe. They do not find any significant emission so they derive upper limits in the energy range $10-500$ GeV. We will report in Sec.~\ref{sec:results} our explanation for their non-detection. They find that an efficiency of about $0.3$ is required to fit HAWC data assuming a slope of the $e^{\pm}$ spectrum of 2.25. 
The value of the energy range for $e^{\pm}$ considered in this paper is not clearly stated so it is difficult to compare their results for $\eta$ with ours. Nevertheless, the efficiencies they report are not far from the results we present in this section.

Ref.~\cite{Tang:2018wyr} reports an efficiency of 0.4 for $\gamma_e=2.34$, the one-zone diffusion model and for electrons of energy $0.1$ GeV.  Rescaling their efficiency for electron energies $>0.1$ GeV, as we are assuming in our paper, would make the efficiency of the order of 0.80 that is much larger than the value we report in our analysis.
Nevertheless, their predictions for the ICS flux for this case (see Fig.~2) is similar to the ones we present in this section. We find with the one-diffusion zone and with a similar value for $\gamma_e$ that the efficiency is $\eta=0.12$. Some of the reasons for the discrepancy in the $\eta$ value might be a different treatment of the energy losses and/or the ISRF model. 
We refer to the previous discussion of HAWC2017 results for a more quantitative comparison of the value of $\eta$ found using different assumptions for the model (e.g., ISRF model, energy losses, $e^{\pm}$ spectral index and minimum energy).

The predictions reported in Ref.~\cite{Fang:2018qco} are similar to the one presented in this section. Indeed, if they assume an $e^{\pm}$ spectral index consistent with HAWC2017, $r_b=100$ pc, which is the closest case to our model, and an efficiency of $1.0$, they find a flux with a similar shape as the one in Fig.~\ref{fig:Jinitial} and with a value of about $1.6\cdot 10^{-3}$ GeV$^2$/cm$^2$/s/sr at 1 TeV. Calculating the ICS flux with the same assumptions, we find a range of $2-4 \cdot 10^{-3}$ GeV$^2$/cm$^2$/s/sr assuming K15 or G15 propagation models for $r_b>100$ pc.

\section{{\it Fermi}-LAT analysis}
\label{sec:fermianalysis}
The Monogem and Geminga pulsars have been detected with {\it Fermi}-LAT and included in the source catalogs since one year after the beginning of the mission \cite{Ackermann:2013fwa}. The signal detected so far has a point-like morphology and is associated to the pulsed emission from the two sources. An extended emission of multiple degree size has never been claimed. We proceed here with a dedicated analysis to search for such a signal.

\subsection{Template for the ICS $\gamma$-ray emission}
\label{sec:template}
\begin{figure}[t]
\centering\includegraphics[width=0.49\textwidth]{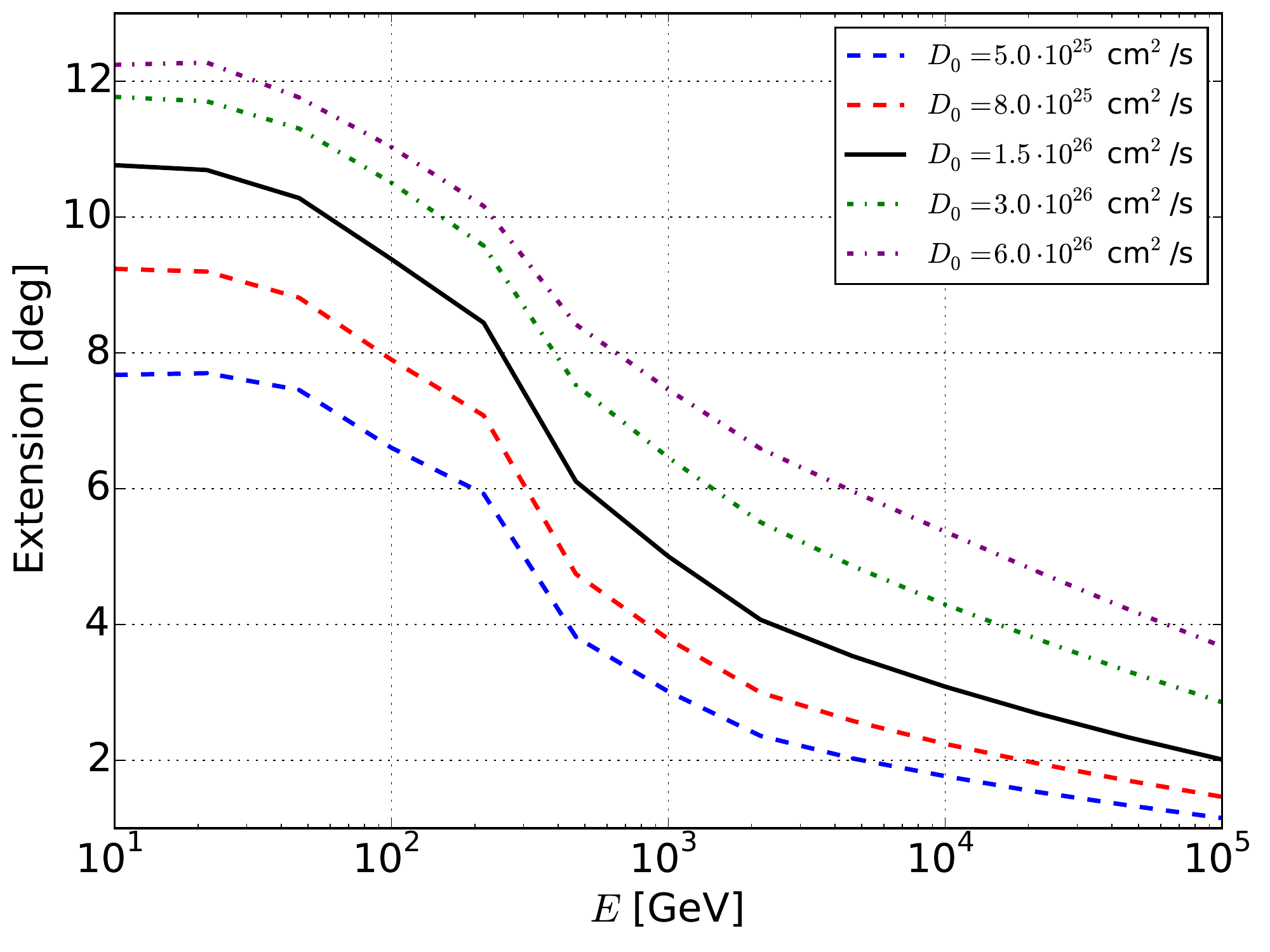}
\caption{Calculated angular extension for the ICS $\gamma$-ray emission from Geminga using $D_0 = [0.5,0.8,1.5,3.0,5.0] \cdot 10^{26}$ cm$^2$/s  and  $\gamma_e=1.8$. We do not consider the pulsar proper motion and the extension is defined as the angle that contains the 68\% of the total flux.}  
\label{fig:extension}
\end{figure}

We analyze {\it Fermi}-LAT data above 8 GeV to search for an extended emission around Geminga and Monogem interpreted as ICS $\gamma$ rays from $e^+$ and $e^-$ emitted by these sources and released in the ISM.
The morphology of ICS emission is energy dependent.
This is illustrated in Fig.~\ref{fig:extension}, where we report the computed angular extension of Geminga as a function of energy, for $D_0 =  [0.5,0.8,1.5,3.0,5.0] \cdot 10^{26}$ cm$^2$/s  and  $\gamma_e=1.8$. 
We define the angular extension as the angle that contains the 68\% of the total flux.
The larger the value of $D_0$ the more extended is the ICS emission. There is an increase of about a factor of 50\% in the angular extension by increasing  $D_0$ by a factor of 10 from $D_0 =  0.5 \cdot 10^{26}$ cm$^2$/s to $D_0 =  5 \cdot 10^{26}$ cm$^2$/s.
The results in Fig.~\ref{fig:extension} at around 10 TeV are comparable with the ones reported in Fig.~\ref{fig:SB_initial_Geminga_2p3}-\ref{fig:SB_initial_Geminga_2p0} for the surface brightness of Geminga above 5 TeV. Indeed, for $D_0 =  8 \cdot 10^{26}$ cm$^2$/s we found an extension of about 10 pc that represents about $2.2^{\circ}$ angular extension, similar to what is shown in Fig.~\ref{fig:extension}.
Focusing on $D_0 = 1.5 \cdot 10^{26}$ cm$^2$/s, for $E_\gamma< 40$ GeV the ICS $\gamma$-ray emission is spread out by $\theta\sim 10^{\circ}$, while the extension decreases significantly for higher energies reaching about $3^{\circ}$ in the HAWC energy range.
The trend of the angular extension is almost constant for $E_{\gamma}<50$ GeV because these photons are mainly produced by $e^{\pm}$ at energies of hundreds of GeV. The latter can travel relatively large distances from the source before losing their energy, which implies that the $\gamma$-ray emission from ICS is relatively extended.
On the other hand, TeV photons are produced by $e^{\pm}$ with much higher energies. They suffer intense energy losses; this implies a much smaller extension for the ICS $\gamma$-ray region.
We have not included the proper motion of the Geminga pulsar in this calculation since the goal is to give a rough estimate of the extension as a function of energy.
The size of the extension below 100 GeV is even larger when including the Geminga proper motion as we will show later in the paper when we will include this effect in the calculation.

In order to account for the energy dependence of the spatial morphology for the ICS emission we create a  \textit{mapcube} template.
This is a three dimensional table that, for each energy bin, gives the map of the $\gamma$-ray intensity in Galactic longitude and latitude.
This format matches the one used in the {\it Fermi}-LAT Science Tools to account for a component with a spatial morphology that changes with energy.
We fix $\gamma_e=1.8$ for both Monogem and Geminga. 
We stress that the choice of the positron spectral index does not influence significantly the spatial morphology of the ICS template and, more importantly, the results on the flux data. 
We have checked that by changing $\gamma_e$ from 1.8 to 2.3 the size of extension, calculated as the 68\% containment radius, changes by 1 per mille.
Indeed, in calculating the flux as a function of energy the ICS halo template is fitted to the data independently for each energy bin where inside each energy bin we assume a power-law shape. Therefore, given a free normalization factor for each energy bin, the initial value of $\gamma_e$ does not affect the result for the flux data.

We also remind the reader that we assume for simplicity a one-zone diffusion model for the $\gamma$-ray ICS halo. This is a reasonable choice since for the energies considered in our analysis the low-diffusion zone dominates our ROI. In Sec.~\ref{sec:results} then we will calculate the $e^+$ flux at Earth assuming the two-zone diffusion model which is more appropriate for the propagation of these particles travelling a significant path in the high-diffusion zone. We will choose different values of $r_b$, compatible with the $\gamma$-ray observations of the ICS halo, in order to account for the uncertainties in the modeling of the diffusion from the pulsars to the Earth.

\begin{figure*}[t]
\centering\includegraphics[width=0.49\textwidth]{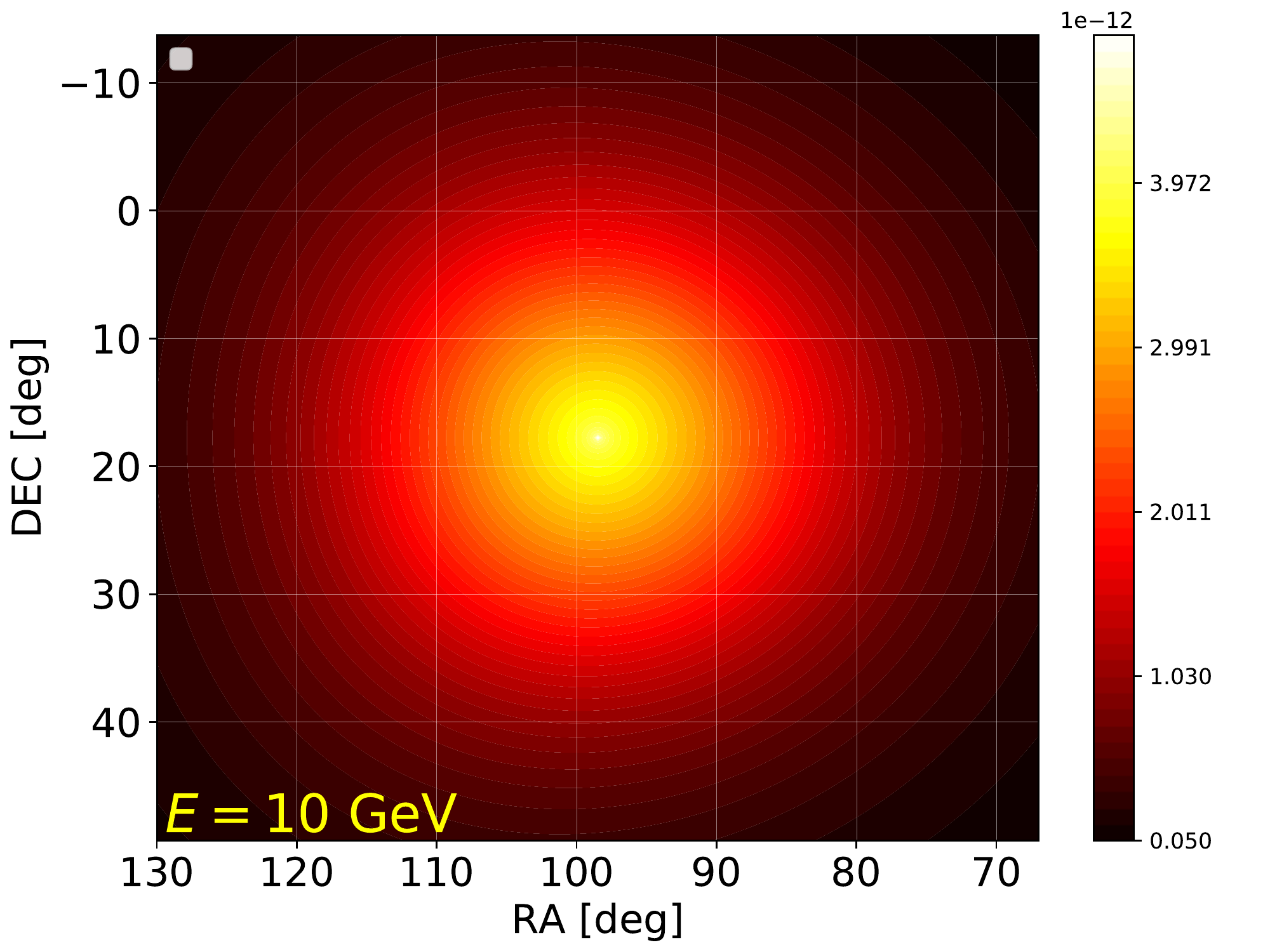}
\centering\includegraphics[width=0.49\textwidth]{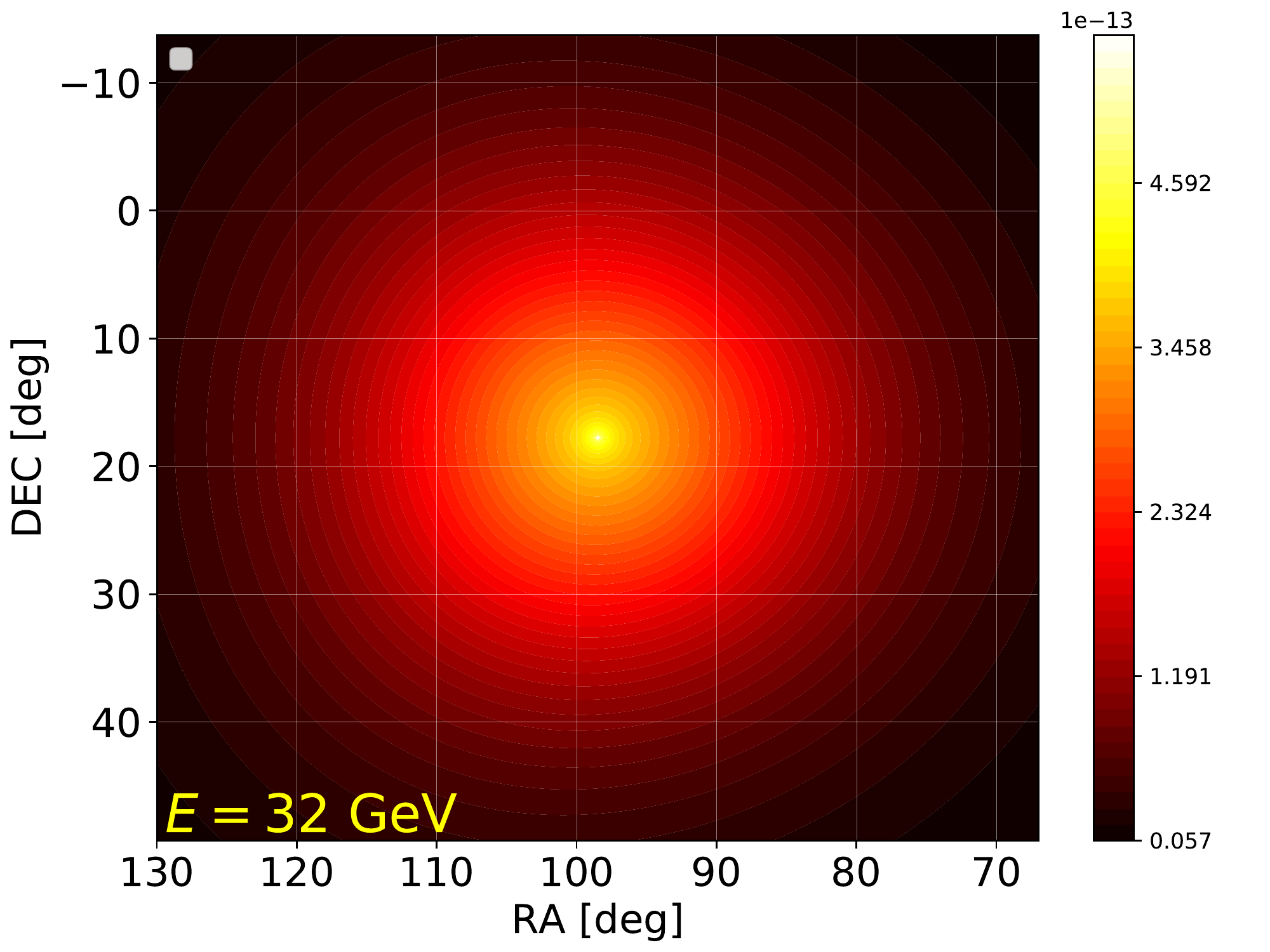}
\centering\includegraphics[width=0.49\textwidth]{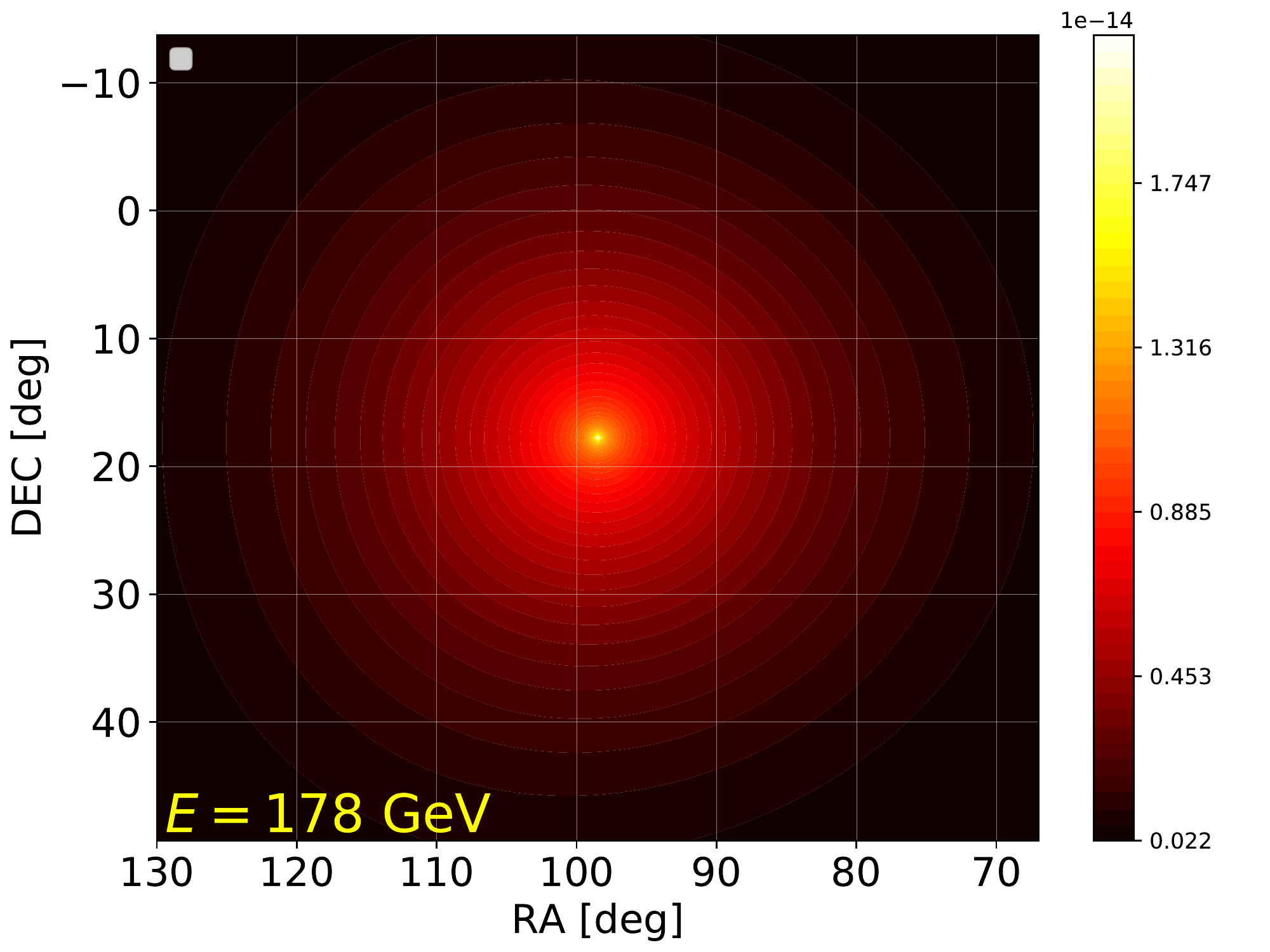}
\centering\includegraphics[width=0.49\textwidth]{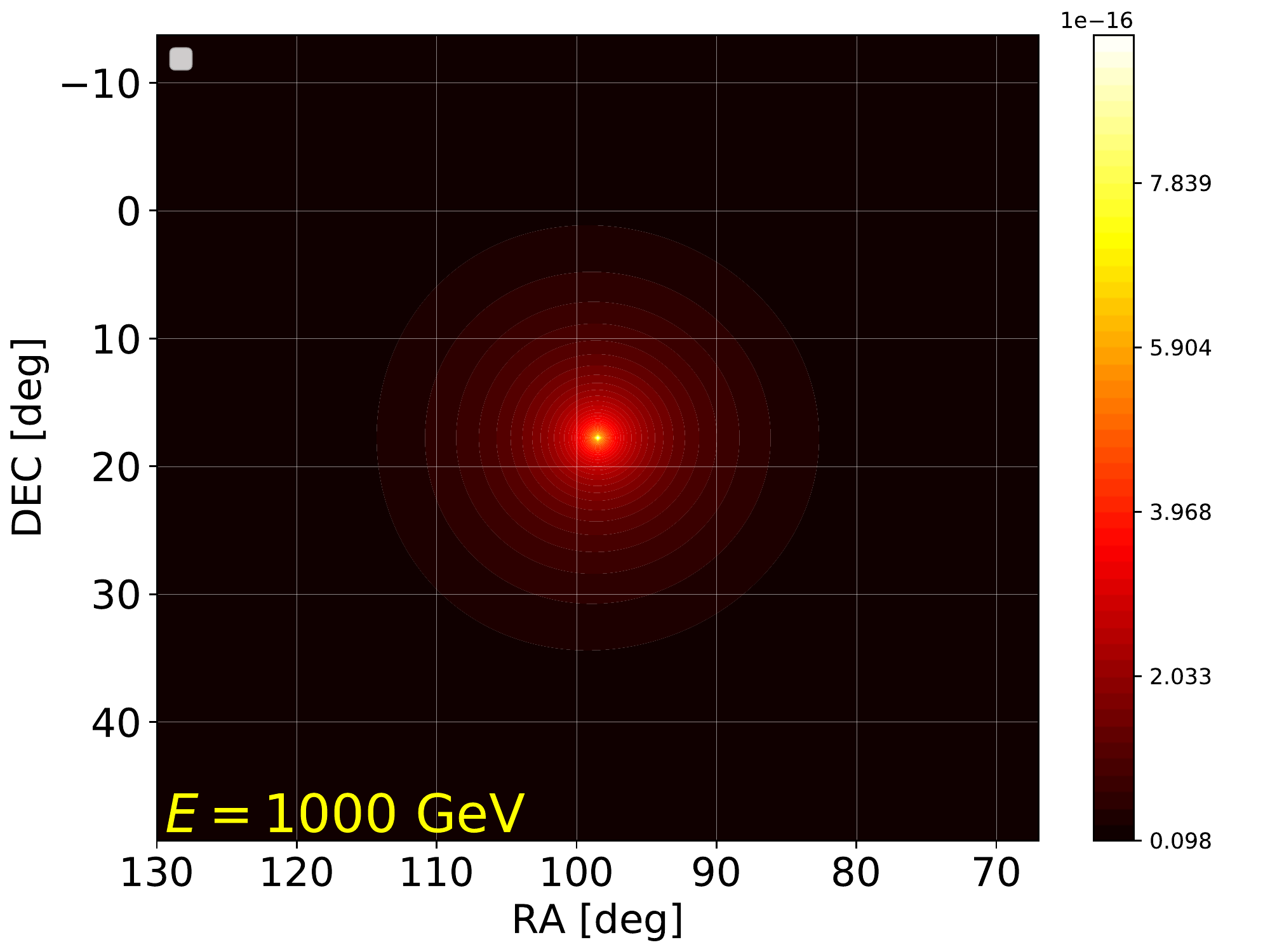}
\caption{Calculated intensity maps for Geminga ICS $\gamma$-ray emission at four energies without considering the pulsar proper motion. The color bar is in units of MeV$^{-1}$ cm$^{-2}$ s$^{-1}$ sr$^{-1}$. We use $\gamma_e=1.8$ and $D_0=2.0\cdot 10^{26}$ cm$^2$/s. The color bar minimum is a factor of 100 lower than the maximum.}  
\label{fig:mapcube_Geminga}
\end{figure*}

\begin{figure*}[t]
\centering\includegraphics[width=0.49\textwidth]{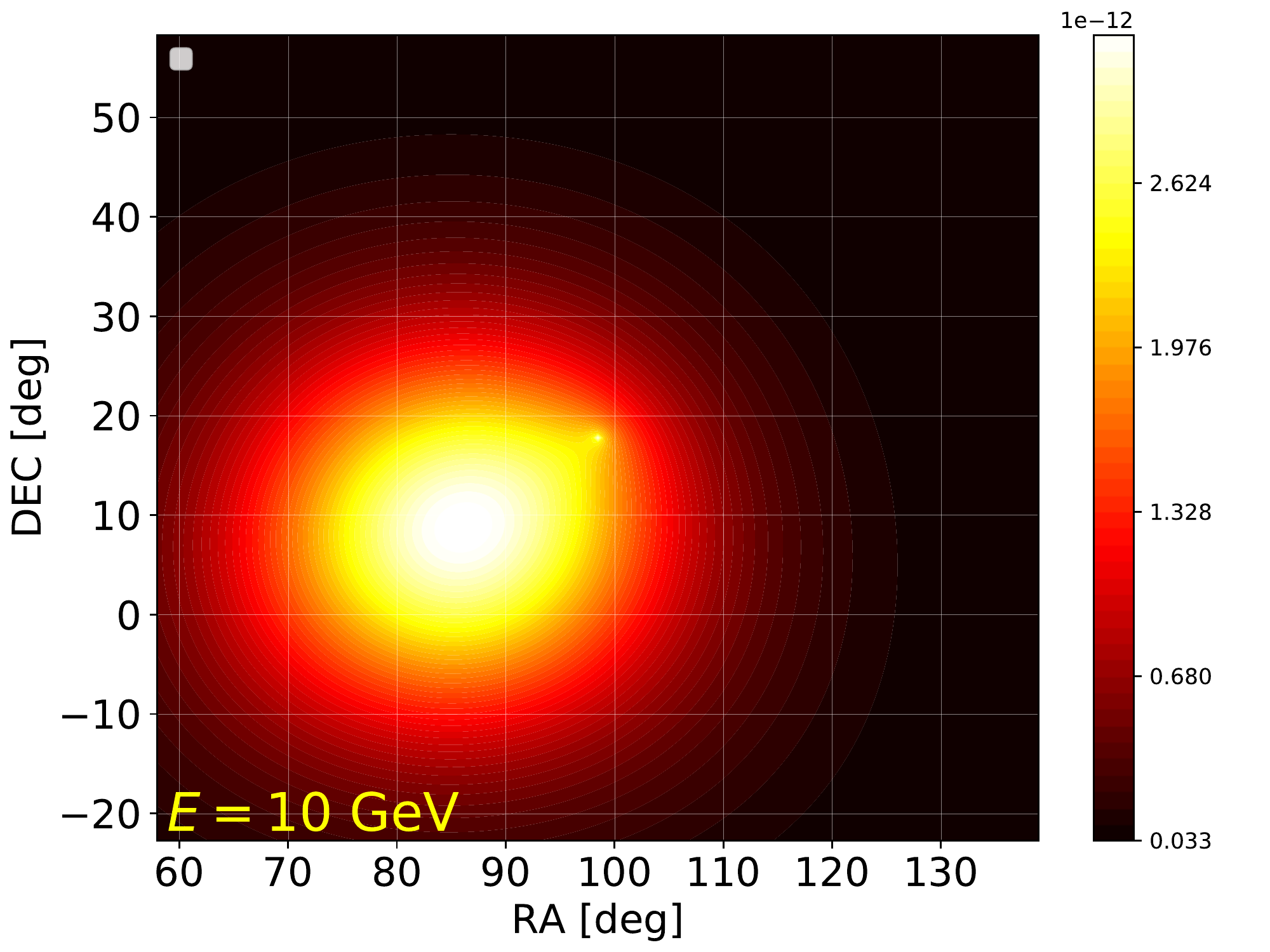}
\centering\includegraphics[width=0.49\textwidth]{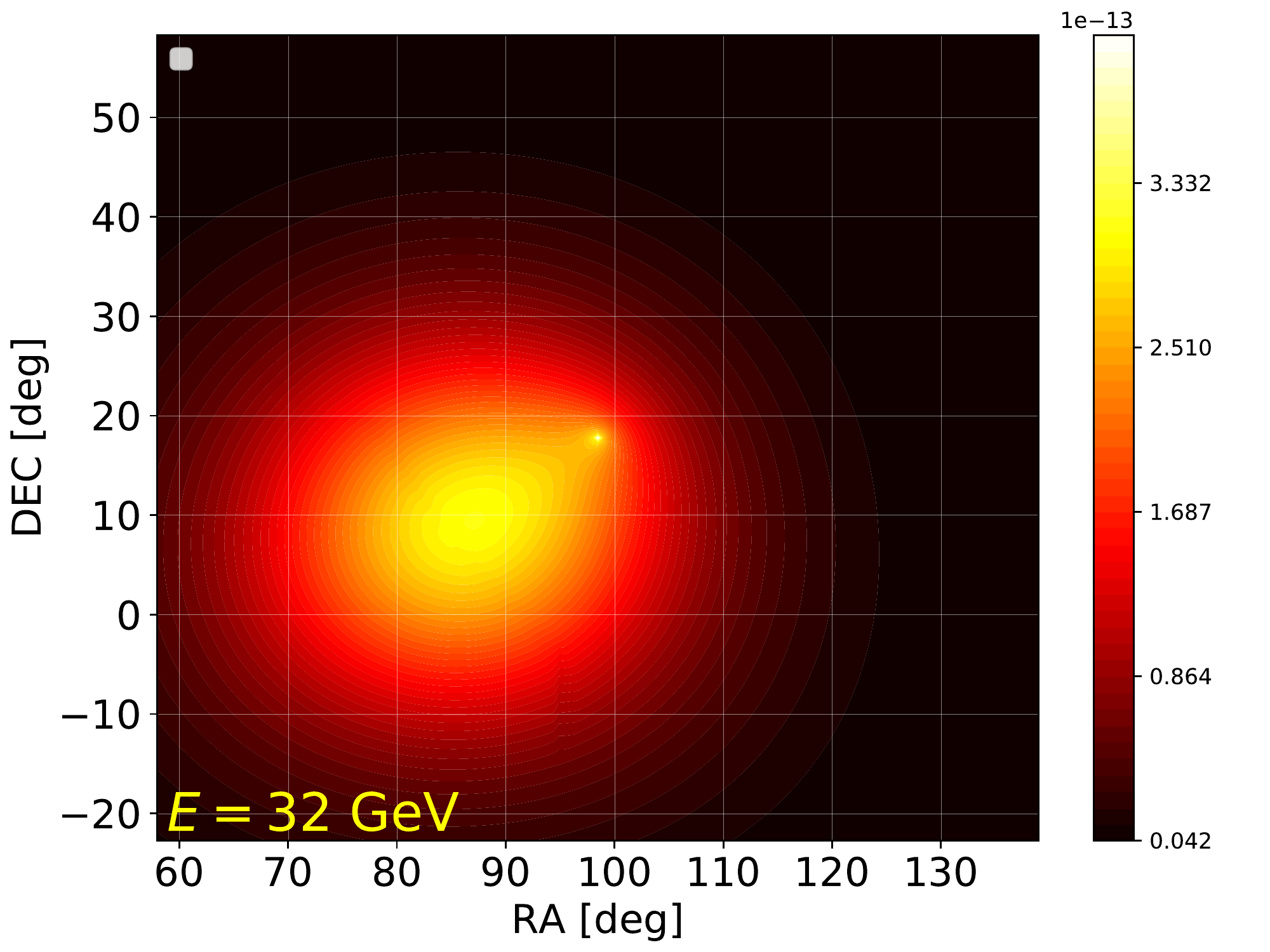}
\centering\includegraphics[width=0.49\textwidth]{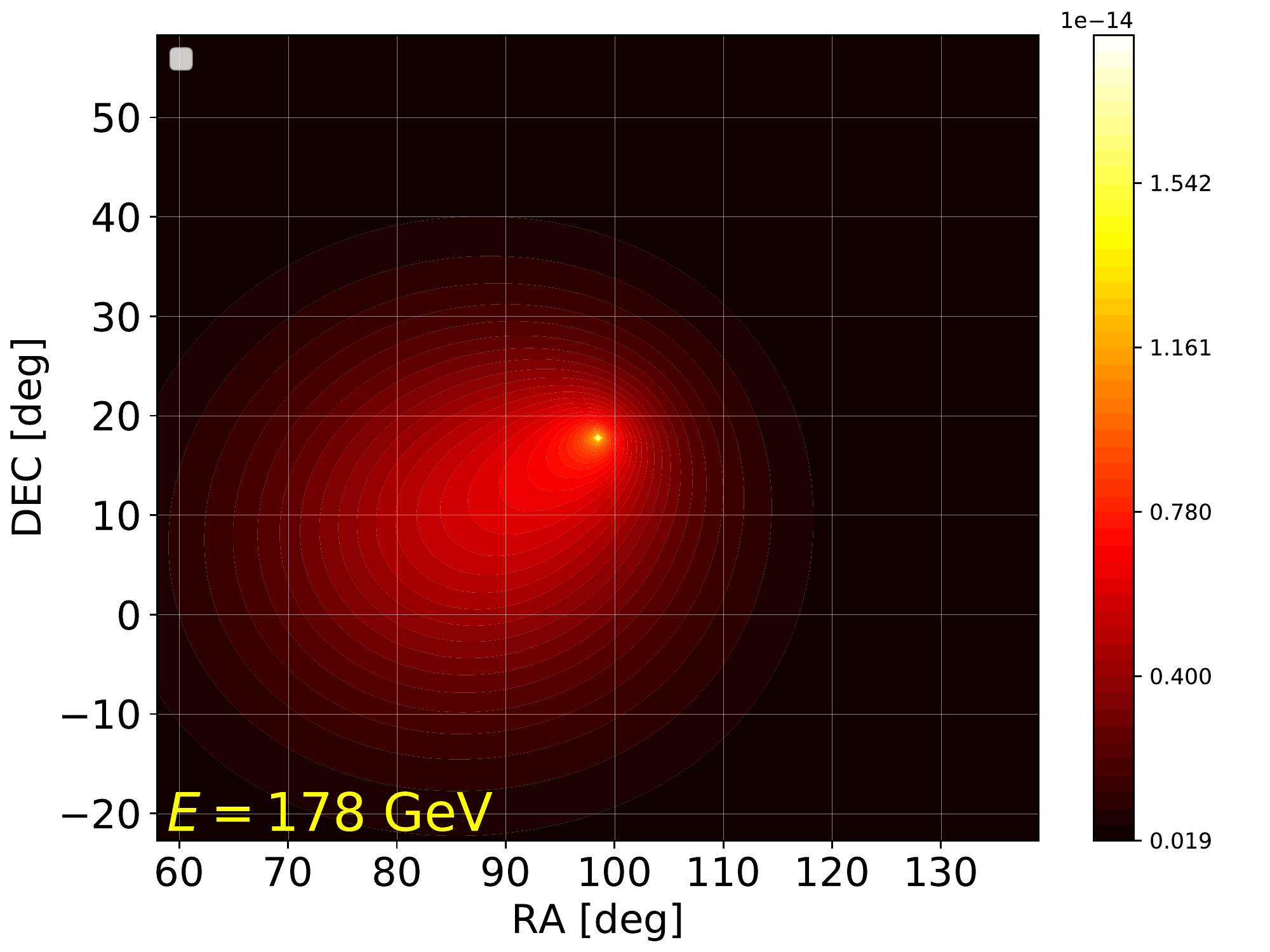}
\centering\includegraphics[width=0.49\textwidth]{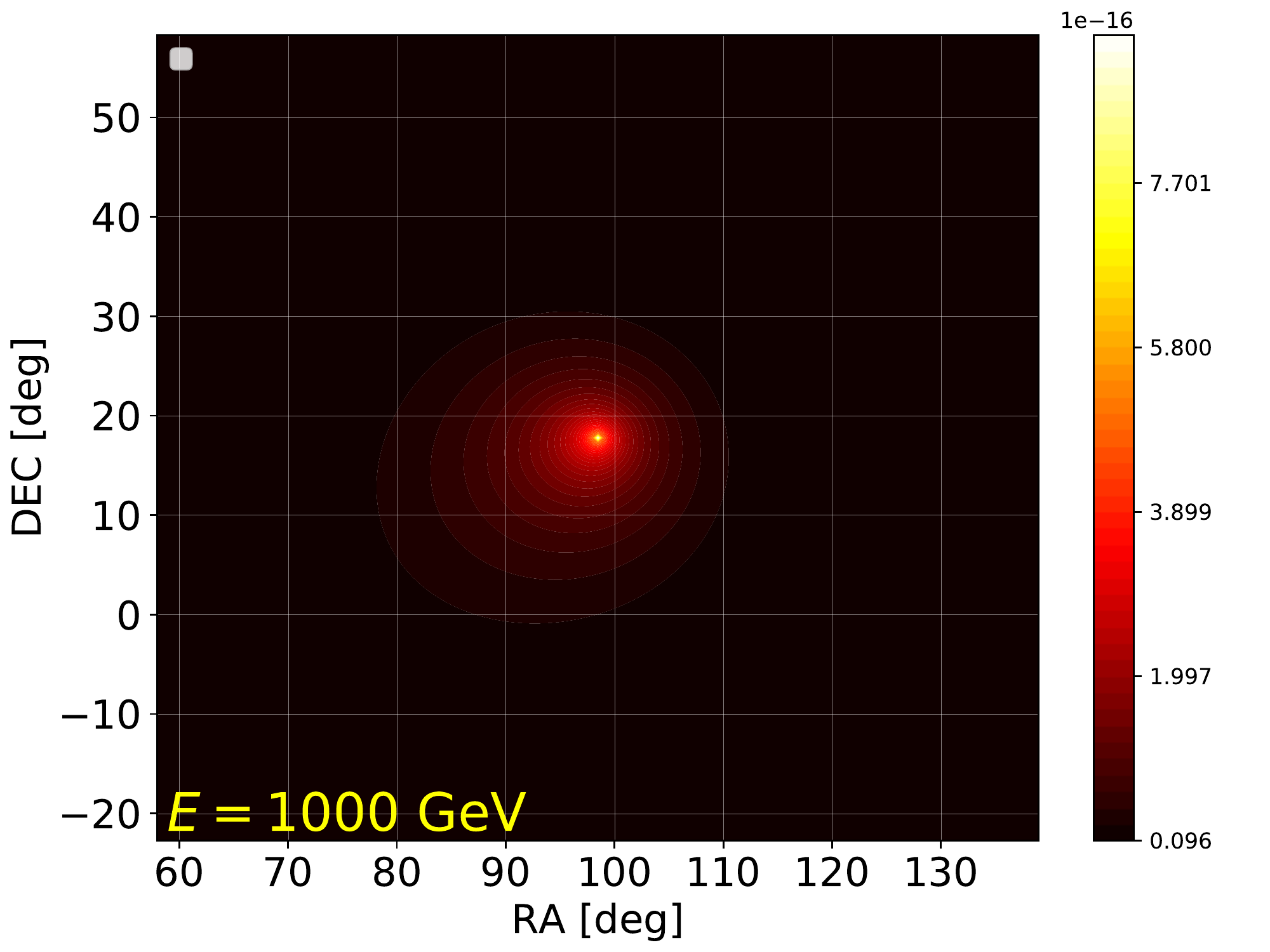}
\caption{Same as Fig.~\ref{fig:mapcube_Geminga} but considering the proper motion of Geminga pulsar.}  
\label{fig:mapcube_motionGeminga}
\end{figure*}

In Fig.~\ref{fig:mapcube_Geminga} (Fig.~\ref{fig:mapcube_motionGeminga}) we display the ICS template for Geminga assuming $D_0 = 2.0 \cdot 10^{26}$ cm$^2$/s and not considering (considering) the pulsar proper motion.
In the figures it is visible that the angular extension decreases significantly when moving to higher energies, as already noted for Fig.~\ref{fig:extension}.
In particular, at 10 GeV the emission is very extended and concentrated within about $10^{\circ}$ from the center of the source, while at TeV energies it is concentrated within $2^{\circ}$.
Fig.~\ref{fig:mapcube_motionGeminga} shows that the effect of the proper motion is mostly significant at low energies.
Indeed, at 10 GeV the peak of the $\gamma$-ray emission is located about $10^{\circ}$ away from the current location of the pulsar. Moreover, by increasing the photon energy the flux concentrates at the actual location of the pulsar becoming almost spherically symmetric above a few hundred GeV.

\subsection{Analysis setup}
\label{sec:setup}
We perform an analysis of 115 months of {\it Fermi}-LAT Pass 8 data, from 2008 August 4 to 2018 March 4.
We select $\gamma$-ray events in the energy range $E=[8,1000]$ GeV, passing standard data quality selection criteria\footnote{\url{https://fermi.gsfc.nasa.gov/ssc/data/analysis/documentation/Cicerone/Cicerone_Data_Exploration/Data_preparation.html}}.
We choose photon energies above 8 GeV because at lower energies the interstellar emission model (IEM) and the pulsed emission from the pulsar dominates the $\gamma$-ray data. 
Moreover, running the analysis above 10 GeV gives a slightly worse significance for the detection of Geminga while selecting $E>6$ GeV does not imply any significant improvement in the results of our analysis.
We consider events belonging to the Pass~8 {\tt SOURCE} event class, and use the corresponding instrument response functions {\tt P8R3\_SOURCE\_V2}, specific for source detection. 

The ROI we consider in our analysis is dominated by the IEM and it is thus central to test different models for this component.
We run the analysis using 10 different models in order to derive the systematics in the results given by the choice of the IEM.
We employ the IEM released with Pass 8 data \citep{Acero:2016qlg} (i.e., {\tt gll\_iem\_v06.fits}), routinely used in Pass~8 analyses and we refer to it as the {\it official} model (Off.).
This model is derived by performing a template fitting to {\it Fermi}-LAT $\gamma$-ray data. 
It is thus based on the spatial correlations between $\gamma$-ray data and a linear combination of gas and ICS maps in order to model the diffuse background for source studies and estimate the $\gamma$-ray emissivity of the gas in different regions across the Galaxy. 
This model contains a patch component to account for extended excess emission regions of unknown origin. However, in the Geminga region there is not any of this patch component.
We include the standard template for the isotropic emission ({\tt iso\_P8R3\_SOURCE\_V2\_v06.txt})\footnote{For descriptions of these templates, see
  \url{http://fermi.gsfc.nasa.gov/ssc/data/access/lat/BackgroundModels.html}.} associated to the Off. IEM.

Then, we run the analysis for 8 other IEMs and the correspondent isotropic emission models used in the first {\it Fermi}-LAT SNR catalog \cite{Acero:2015prw}.
They all assume the same underlying data for the HI and CO emission and they implement the inverse Compton component derived from GALPROP\footnote{https://galprop.stanford.edu}.
These models have been generated by varying the CR source distribution, height of the CR propagation halo, and HI spin temperature in order to test the effect of the choice of the IEM in the flux and spatial distribution of SNRs. 
Finally, the models are all adjusted to match the data from the LAT in order to provide a good representation of the $\gamma$-ray sky.
These 8 models have been used in the SNR catalog to explore the systematic effects on SNRs fitted properties, including the size and morphology of the extension, caused by IEM modeling. We will label these models as alternatives (Alt.).


The final IEM that we consider is the sample model used in the {\it Fermi}-LAT analysis of the $\gamma$-ray excess in the direction of the Galactic center \cite{TheFermi-LAT:2017vmf}.
This model is based on Galprop and it has been tuned on 6.5 years of {\it Fermi}-LAT of Pass 8 data.
We will label this model as Galactic Center IEM (GC).

We have implemented an analysis pipeline using {\tt FermiPy}, a Python package that automates analyses with the {\it Fermi} Science Tools \citep{2017arXiv170709551W}\footnote{See \url{http://fermipy.readthedocs.io/en/latest/}.}.
{\tt FermiPy} includes tools that 1) generate simulations of the $\gamma$-ray sky, 2) detect sources, and 3) calculate the characteristics of their SED.   
For more details on {\tt FermiPy} we refer to the Appendices of \cite{Fermi-LAT:2017yoi}. 

We consider a region of interest (ROI) of $70^{\circ}\times70^{\circ}$ centered at RAJ2000$=95^{\circ}$ and DEJ2000$=13^{\circ}$.
Geminga and Monogem are separated only by about $7^{\circ}$ and the ICS $\gamma$-ray emission is expected to be very extended in {\it Fermi}-LAT data as we have seen in Figs.~\ref{fig:mapcube_Geminga} and \ref{fig:mapcube_motionGeminga}.
This is why we choose to consider this position and width of our ROI.
We bin the data with a pixel size of $0.06^{\circ}$ and 6 bins per energy decade.
Our model includes the IEM, the isotropic template and cataloged sources from the preliminary 8 years list\footnote{\url{https://fermi.gsfc.nasa.gov/ssc/data/access/lat/fl8y/gll_psc_8year_v5.fit}}.
In the analysis the normalization and the spectral shape parameters of the point sources and of the IEM are free to vary while for the isotropic template only the normalization is a free parameter.
For the templates of the two source halos, we vary $D_0$ in the range $10^{25}-10^{29}$ cm$^2$/s and their spectral slope.

\subsection{Analysis results}           
\label{sec:results}

We start our analysis with a fit to the ROI, where we include Geminga and Monogem both as pulsar point sources and their ICS halos. 
We include the proper motion of Geminga in our analysis when we calculate the ICS $\gamma$-ray flux.
The flux of the Geminga pulsed emission is particularly relevant between $8-20$ GeV, while Monogem is very faint with a $TS\approx0$\footnote{The Test Statistic ($TS$) is defined as twice the difference in maximum log-likelihood 
between the null hypothesis (i.e., no source present)
  and the test hypothesis: $TS = 2 ( \log\mathcal{L}_{\rm test} -
  \log\mathcal{L}_{\rm null} )$~\cite{1996ApJ...461..396M}.}.
 Indeed, these sources have an energy cutoff in their energy spectrum of about 700 MeV for Geminga and 400 MeV for Monogem \cite{Acero:2015hja}.
We re-localize all the sources in the ROI and then search for new point sources with $TS >25$.
We perform this analysis for different values of $D_0$ for Monogem and Geminga ICS halos.

\begin{table*}[t]
\begin{center}
\begin{tabular}{|c|c|c|c|c|c|}
\hline
IEM & $TS^{\rm{Geminga}}$ & $D^{\rm{Geminga}}_0$ & $TS^{\rm{motion}}$  & $TS^{\rm{Monogem}}$ & $D^{\rm{Monogem}}_0$   \\ 
 &  & [$10^{26}$ cm$^2$/s] &  &  & [$10^{26}$ cm$^2$/s]   \\ 
\hline
Off.   & 65  & $2.1^{+1.0}_{-0.7}$ & 28 & 25 & $>2$   \\ 
\hline
Alt. 1   & 104  & $2.6^{+1.4}_{-0.8}$ &30 & 3 & $>1$   \\ 
\hline
Alt. 2   & 92  & $2.6^{+1.2}_{-0.8}$ & 22 & 14 & $>3$   \\ 
\hline
Alt. 3  & 87  & $3.3^{+1.6}_{-1.1}$ & 24 & 16 & $>4$   \\ 
\hline
Alt. 4  & 102  & $3.5^{+1.8}_{-1.1}$ & 20 & 26 & $>3$   \\ 
\hline
Alt. 5   & 111  & $2.4^{+1.0}_{-0.6}$ & 51 & 12 & $>2$   \\ 
\hline
Alt. 6   &143  & $2.6^{+1.2}_{-0.8}$ & 43 & 10 & $>3$   \\ 
\hline
Alt. 7   & 128  & $2.8^{+1.3}_{-0.9}$ & 41 & 12 & $>10$   \\ 
\hline
Alt. 8   & 134  & $3.1^{+1.3}_{-0.9}$ & 39 & 25 & $>8$   \\ 
\hline
GC   & 71  & $1.6^{+0.6}_{-0.4}$ & 35 & 8 & $>1$   \\ 
\hline
\end{tabular}
\caption{Results for the Monogem and Geminga ICS halos derived using each of the 10 IEMs considered in our analysis. We report the $TS$ for the detection of the ICS halo from Geminga ($TS^{\rm{Geminga}}$) with the correspondent value for the diffusion coefficient ($D^{\rm{Geminga}}_0$), the $TS$ for the presence of the proper motion for Geminga ($TS^{\rm{motion}}$, see the text for further details). In the last two columns we display the $TS$ for the detection of the ICS halo from Monogem ($TS^{\rm{Monogem}}$) with the $95\%$ CL lower limit for the diffusion coefficient ($D^{\rm{Monogem}}_0$).}
\label{tab:altmodel}
\end{center}
\end{table*}

\begin{figure*}[t]
\centering\includegraphics[width=0.60\textwidth]{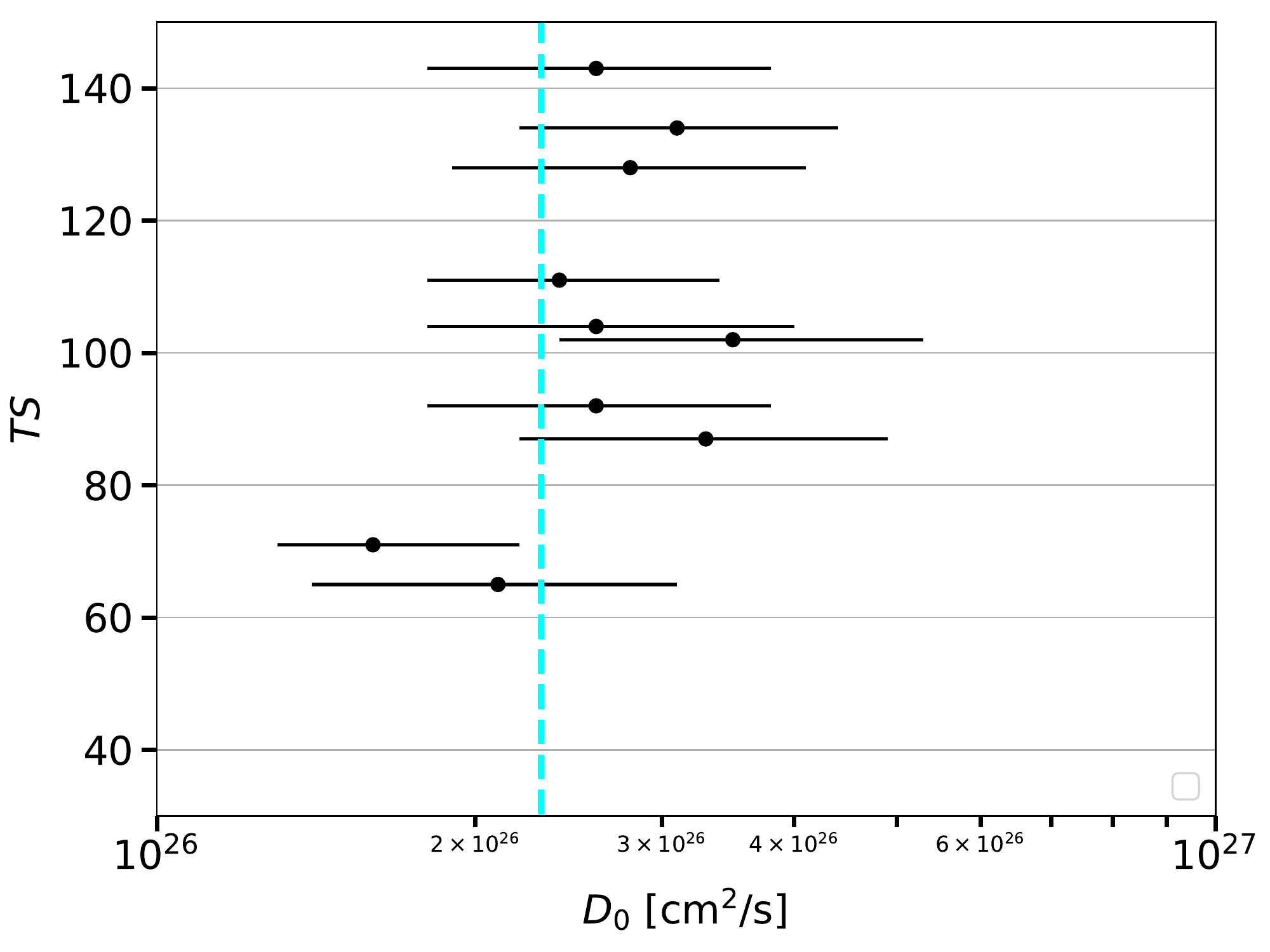}
\caption{$TS$, best fit and $1\sigma$ error for $D_0$ for the search of Geminga ICS halo with the official IEM, with the eight models used in \cite{Acero:2015prw} and with the model used in the \cite{TheFermi-LAT:2017vmf}. The cyan vertical line represents the weighted average (see the text for further details).}  
\label{fig:testsTS}
\end{figure*}

The Monogem ICS halo is detected with $TS$ values between 3 and 26 depending on the IEM considered. 
Since the detection of Monogem ICS halo is not significant regardless the choice of the IEM, we place $95\%$ lower limits on the value of the diffusion coefficient.
Our analysis is not designed to place upper limits on $D_0$ because for values larger than about $10^{28}$ cm$^2$/s the ICS $\gamma$-ray emission starts to be very extended and almost isotropic. Therefore, our analysis does not have enough sensitivity to constrain very large values of $D_0$ for such nearby pulsars.
The $95\%$ lower limits for the diffusion coefficient are between $D_0 > 1-10 \cdot 10^{26}$ cm$^2$/s and are compatible with the value reported by the HAWC Collaboration in HAWC2017 for this source.
The values of these lower limits might change if one would assume a two-zone diffusion model with different values for $r_b$ and considering particular radial shapes for the change of $D$ between the low and high-diffusion zones.


The Geminga ICS halo is detected with $TS=65-143$ and $D_0=1.6-3.5 \cdot 10^{26}$ cm$^2$/s depending on the IEM considered.
In Tab.~\ref{tab:altmodel} we report the results of the analysis for Geminga and Monogem for each IEM.
In Fig.~\ref{fig:testsTS} we show the $TS$ versus the best fit values and $1\sigma$ errors for $D_0$ as derived in our analysis for the different IEMs.
The weighted average\footnote{We calculate the weighted average by considering the following equation: $\bar{D}_0 = \sum_j\frac{1}{\sigma_j^2} \cdot \sum_i \frac{D_{0,i}}{\sigma_i^2} $, where $\sigma_i$ and $\bar{D}_{0,i}$ are the best fit and $1\sigma$ error for the measurement of $D_0$ for each IEM model.} of the diffusion coefficient is $D_0 = 2.3 \cdot 10^{26}$ cm$^2$/s that corresponds, in the energy range of our {\it Fermi}-LAT analysis, to $D (100 \,\rm{GeV}) = 1.1 \cdot 10^{27}$ cm$^2$/s.
The value we find for $D_0$ is compatible within $2\sigma$ errors with the result from the HAWC Collaboration (see HAWC2017, $D_0=6.9^{+3.0}_{-2.2} \cdot 10^{25}$ cm$^2$/s
\footnote{This number is obtained by rescaling their diffusion coefficient for electrons at 100 TeV, $D_{100},{\rm to} \; D_0$.}).
The discrepancy in the rescaling of the diffusion coefficient from our analysis ($E>10$ GeV) and from HAWC2017 ($E>5$ TeV) to 1 GeV might be the due to a slope of the diffusion coefficient which is different from $\delta=1/3$.
Indeed, the best-fit value for the diffuse coefficient that that we find in {\it Fermi}-LAT energy range and the one found by the HAWC Collaboration at 100 TeV, are compatible if we consider a slope of the diffusion coefficient of $\delta=0.21$. However, considering that the discrepancy between our results and the one reported in HAWC2017 are only at the level of $1-2\sigma$, we conclude there are not strong evidence for a $\delta$ value different from $1/3$ so we decide to use this value in our analysis.
We also note that at energies of about 10 GeV other effects may cause a discrepancy from a delta value of 1/3, i.e.~cosmic-ray diffusive
re-acceleration, the possible presence of convection, as well as un-modeled uncertainties in the energy losses due to synchrotron radiation or
bremsstrahlung emission.

In order to see if our analysis is able to detect the effect of the known Geminga proper motion on the ICS $\gamma$-ray morphology, we run the analysis also for $v_T=0$. We thus fix the Geminga pulsar velocity to zero and the spatial template for the ICS emission is spherically symmetric around the source (see Fig.~\ref{fig:mapcube_Geminga}). The analysis pipeline is the same as before. The result is reported in Tab.~\ref{tab:altmodel} where we put the $TS$ for the proper motion ($TS^{\rm{motion}}$). This is calculated as twice the difference between the likelihood found including the proper motion ($\mathcal{L}_{\rm{motion}}$) and the likelihood with pulsar velocity equal to zero ($\mathcal{L}_{0}$): $TS^{\rm{motion}} = -2 \log{ (\mathcal{L}_{\rm{motion}} - \mathcal{L}_{0} ) }$.
The $TS$ for the motion is between 22 and 51. Considering one degree of freedom that is the transverse velocity, these $TS$ values correspond to a significance in the range between $4.7-7.1 \sigma$. Our analysis thus significantly detects the motion of Geminga pulsar by fitting its ICS halo.

\begin{figure*}[t]
\centering\includegraphics[width=0.49\textwidth]{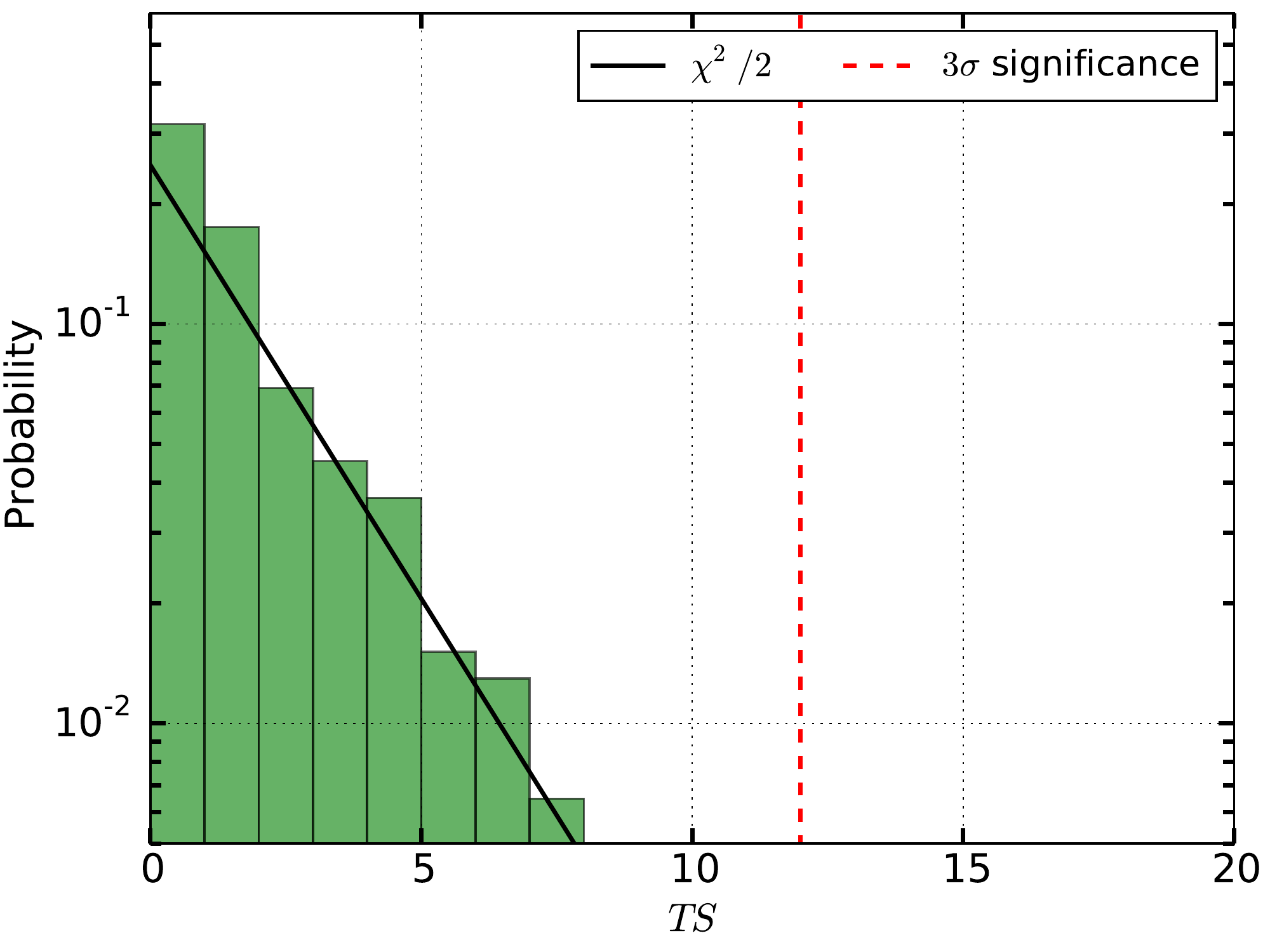}
\centering\includegraphics[width=0.49\textwidth]{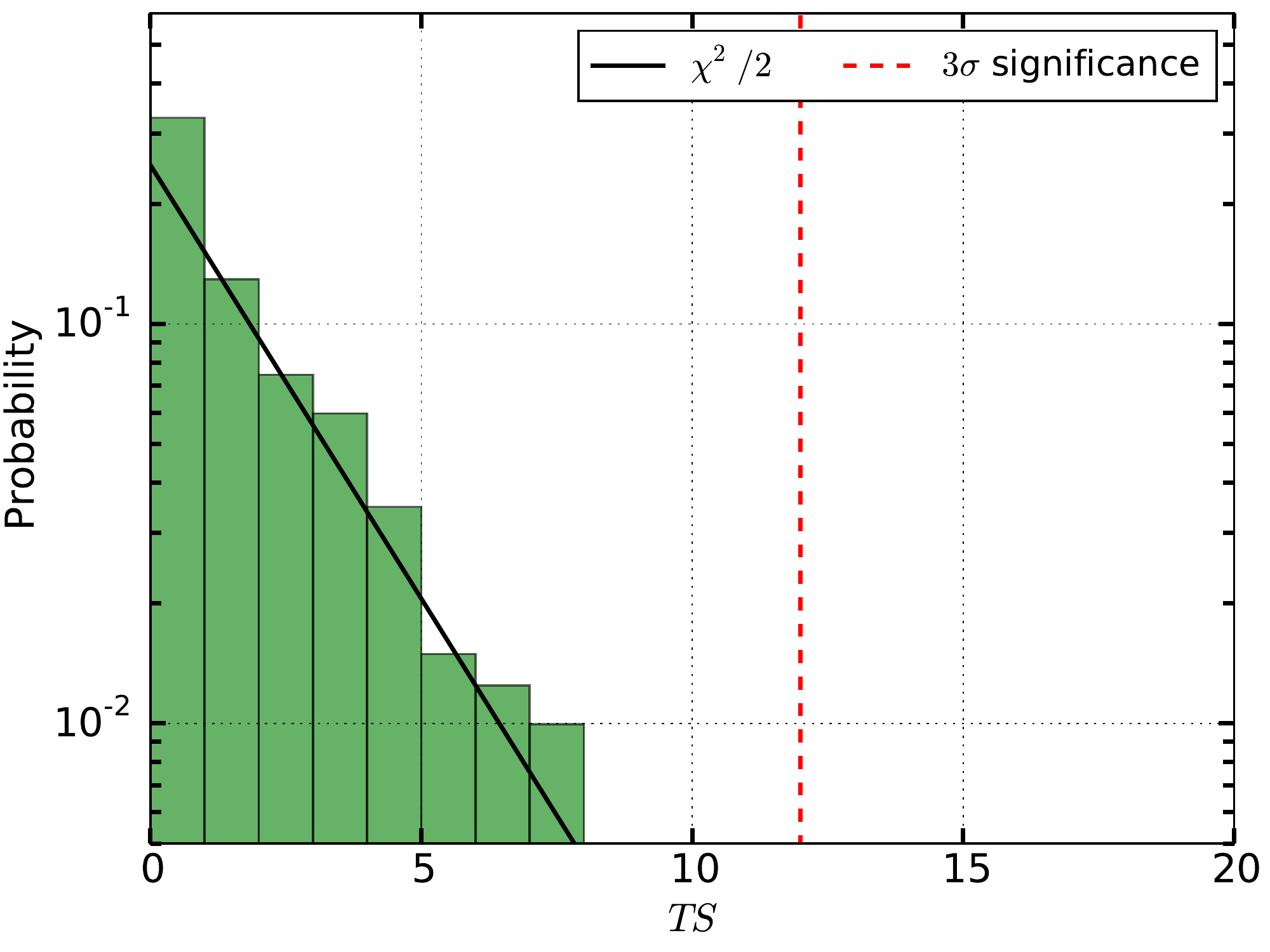}
\caption{Validation tests for the detection of Geminga (left panel) and Monogem (right panel) ICS halo. The green histogram represents the Probability for the detection of a ICS halo in simulations of our ROI as a function of the $TS$. We also show the $\chi^2$ distribution (black solid line) and the corresponding $3\sigma$ significance detection level (red dashed line).}  
\label{fig:testsnull}
\end{figure*}

The values of $TS$ we find for Geminga ICS halo with the different IEMs correspond to a significance in the range $7.8-11.8 \sigma$ if the null hypothesis $TS$ is distributed as a $\chi^2/2$ with 2 degrees of freedom. 
In order to demonstrate that the $TS$ distribution follows the probability distribution function (PDF) of $\chi^2/2$ with 2 degrees of freedom, we run simulations with the null signal.
We perform 1000 simulations of the ROI without including the Geminga and Monogem ICS halos using {\tt Fermipy}.
Then, we run the same analysis applied to the real data: we search for the Geminga and Monogem ICS halos varying $D_0$ in the range $10^{25}-10^{29}$ cm$^2$/s.
We thus compute the significance of the Geminga and Monogem ICS halos for each simulations. 
We show the results of this analysis in Fig.~\ref{fig:testsnull}.
The $TS$ distribution is highly peaked at $TS\approx 0$, as expected for the null signal and it is compatible with the $\chi^2$ PDF.
This justifies our conversion of the $TS$ into a significance as done above.

\begin{figure*}[t]
\centering\includegraphics[width=0.49\textwidth]{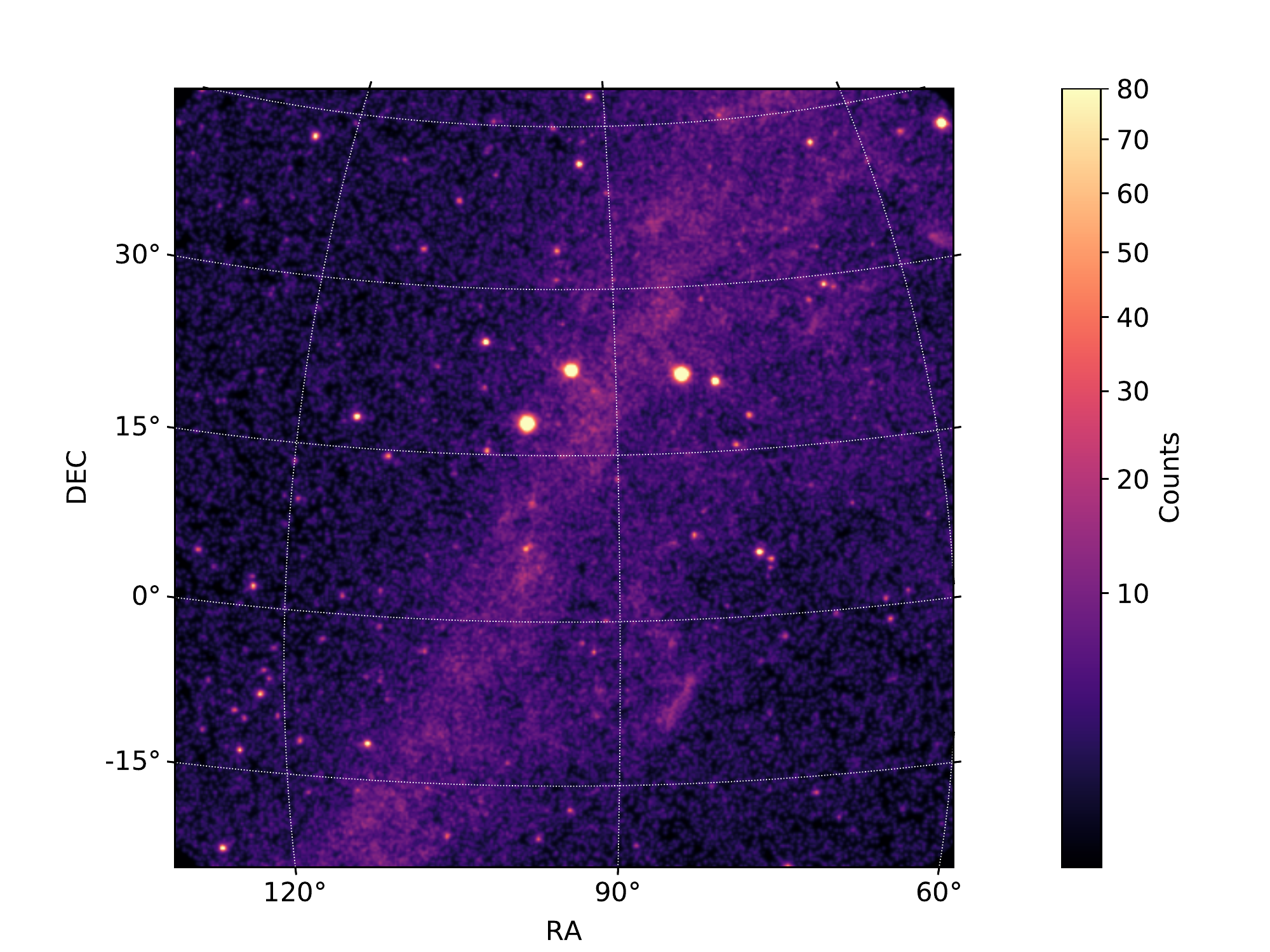}
\centering\includegraphics[width=0.49\textwidth]{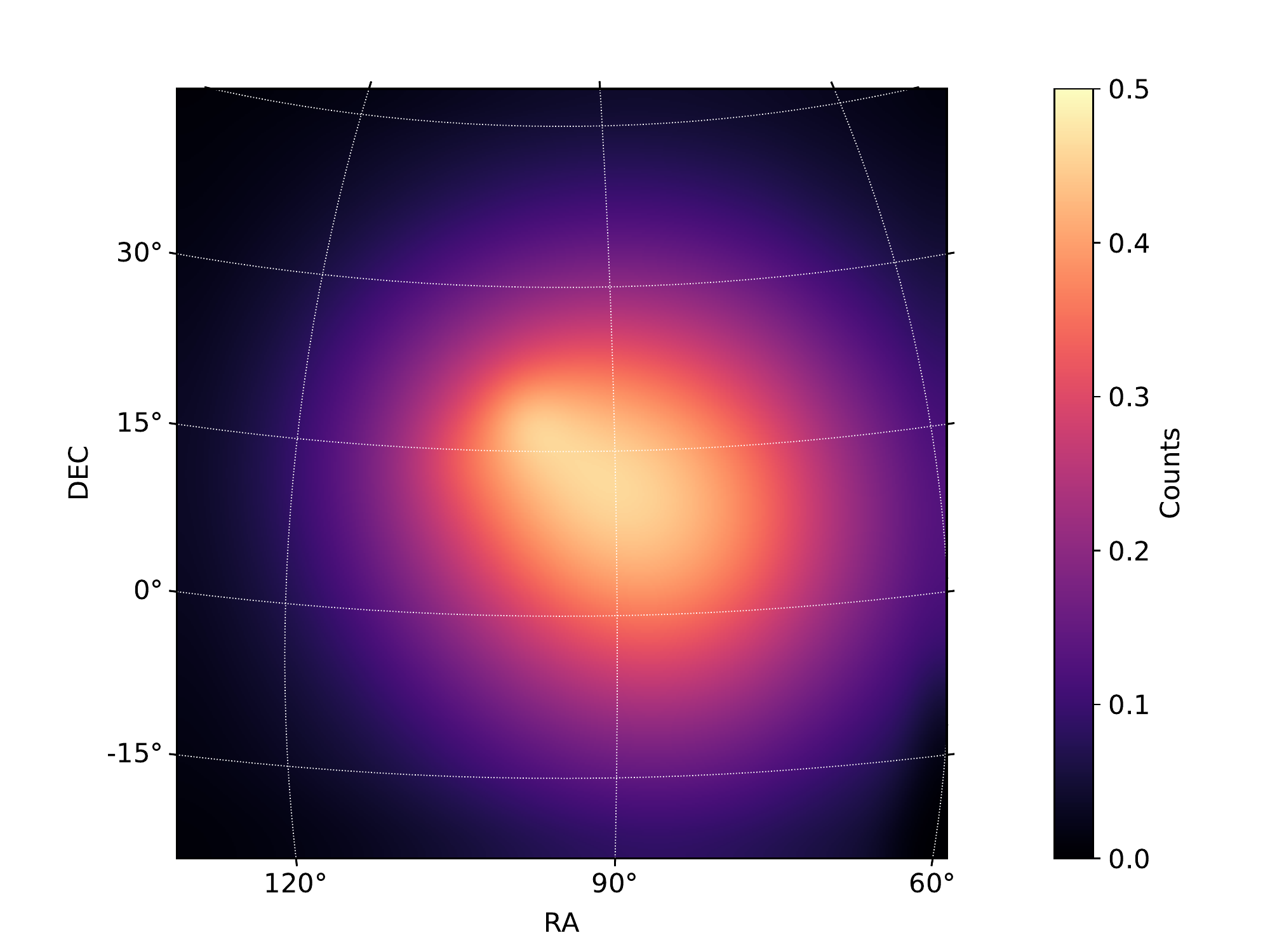}
\centering\includegraphics[width=0.49\textwidth]{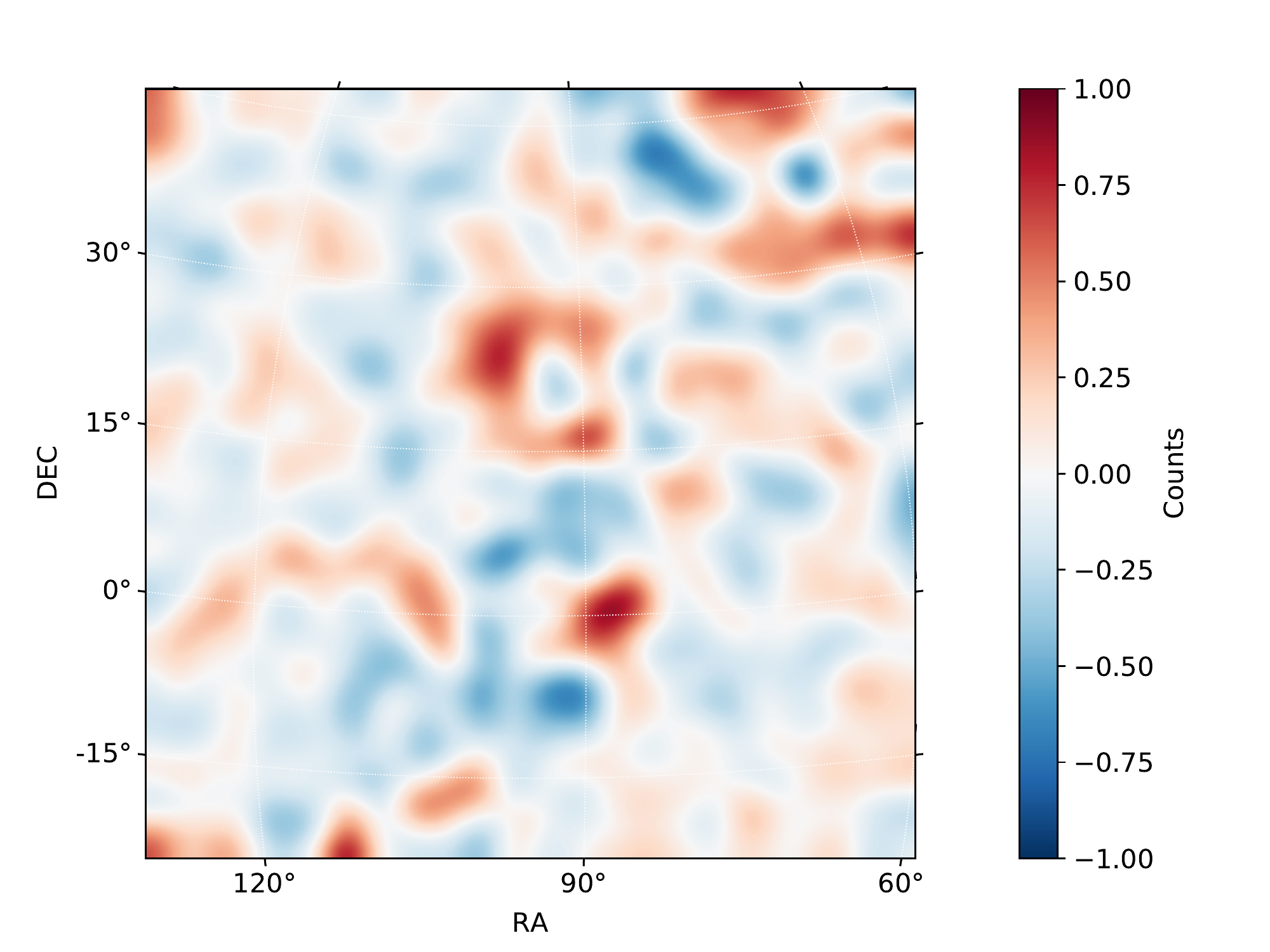}
\centering\includegraphics[width=0.49\textwidth]{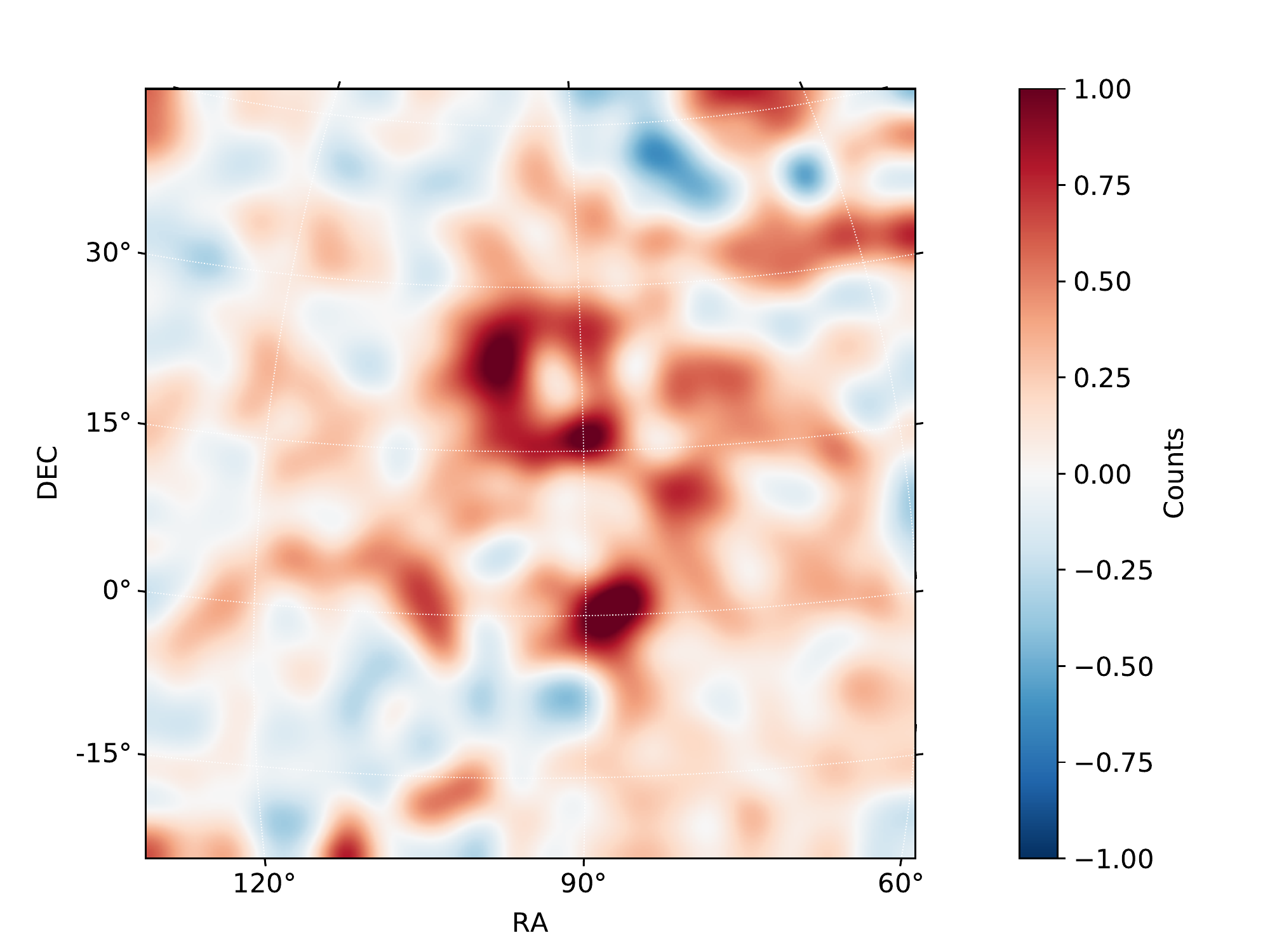}
\caption{Count maps  of the ROI centered at the Geminga pulsar for our analysis of Fermi-LAT data. The color bar represents the number of counts per pixel.
Top left: the total count map. 
Top right: map of our model for the Geminga ICS halo. 
Bottom left: residual map with all the components subtracted. 
Bottom right panel: residual map with all the components subtracted except the Geminga ICS halo. All the maps have been created for a pixel size of $0.4^{\circ}$, smoothed with a Gaussian function.
The positions of Geminga and Monogem pulsars are respectively $RA=98.48^{\circ}$ and $DEC=17.77^{\circ}$ and $RA=104.95^{\circ}$ and $DEC=14.24^{\circ}$}.
\label{fig:maps}  
\end{figure*}

In Fig.~\ref{fig:maps} we show the following count maps of the ROI: 
the total count map, the model map for the Geminga ICS halo, the residual map after all the components of our model are subtracted and the residual map without the Geminga halo subtraction.
We perform a fit after we remove the Geminga halo.
All the maps have been produced with a pixel size of $0.4^{\circ}$, applying a Gaussian filter with standard deviation of $4\sigma$ and using the Python function {\tt scipy.ndimage.gaussian\_filter}.
In the count map we clearly see the emission from the IEM, the flux of Geminga pulsar in the center of the map, together with other very bright sources such as: the blazars RX J0648.7+1516, TXS 0518+211, 1ES 0647+250, and PKS 0735+17, the SNR IC 443 and the Crab PWN.
The residual map with the ICS halo included do not contain bright residuals, meaning that our model represents the data well. 
The largest residuals in this map contain at most 1 photon per pixel.
The residual map derived with the Geminga ICS halo not included in the model is brighter than the one generated with all the components included.
As expected from our model, the Geminga ICS halo contribution is brighter in the center, where it contributes with 0.5 photon per pixel, and fainter at the edges, where its contribution is a factor of 5 smaller.

\begin{figure*}[t]
\centering\includegraphics[width=0.49\textwidth]{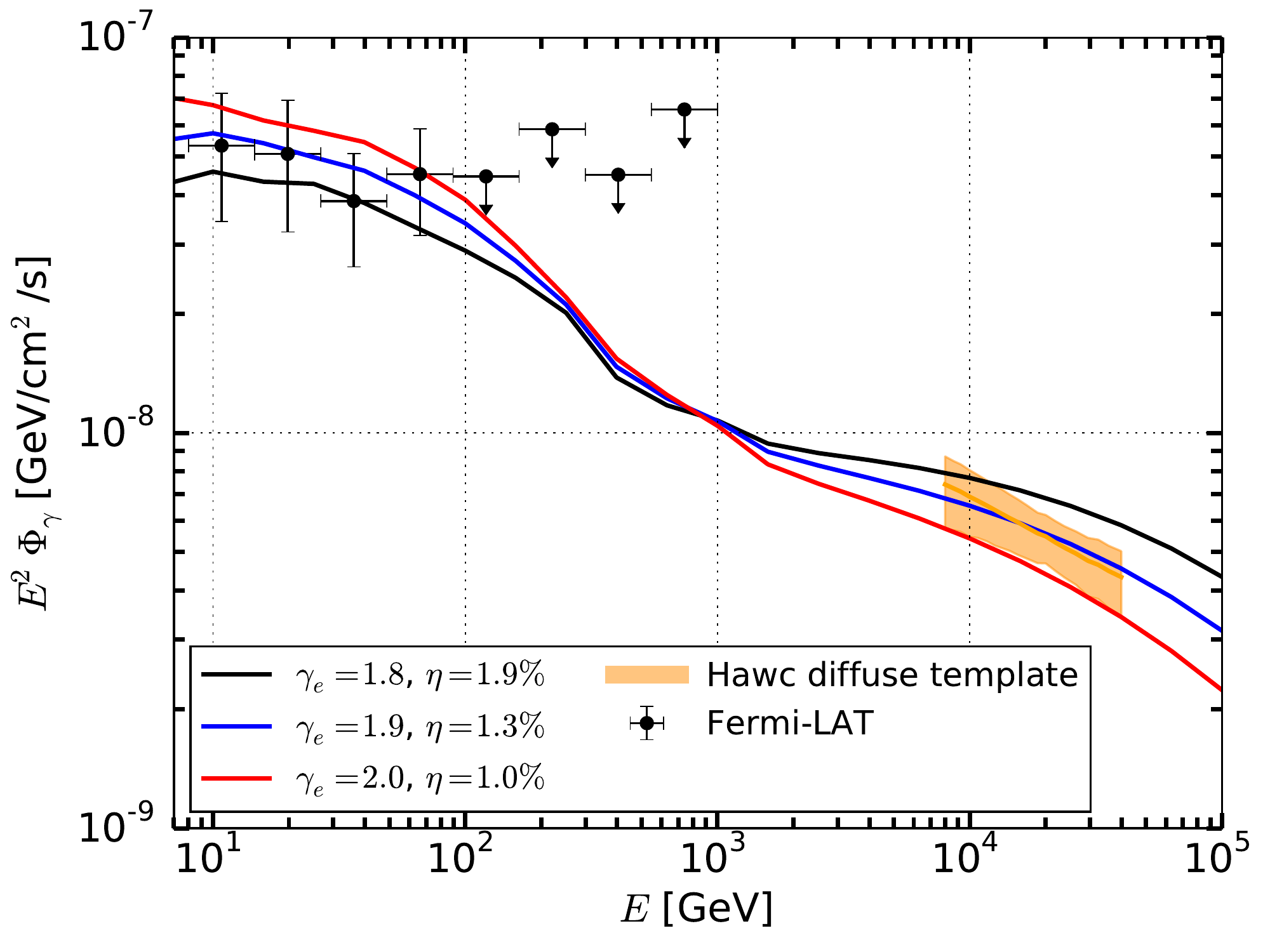}
\centering\includegraphics[width=0.49\textwidth]{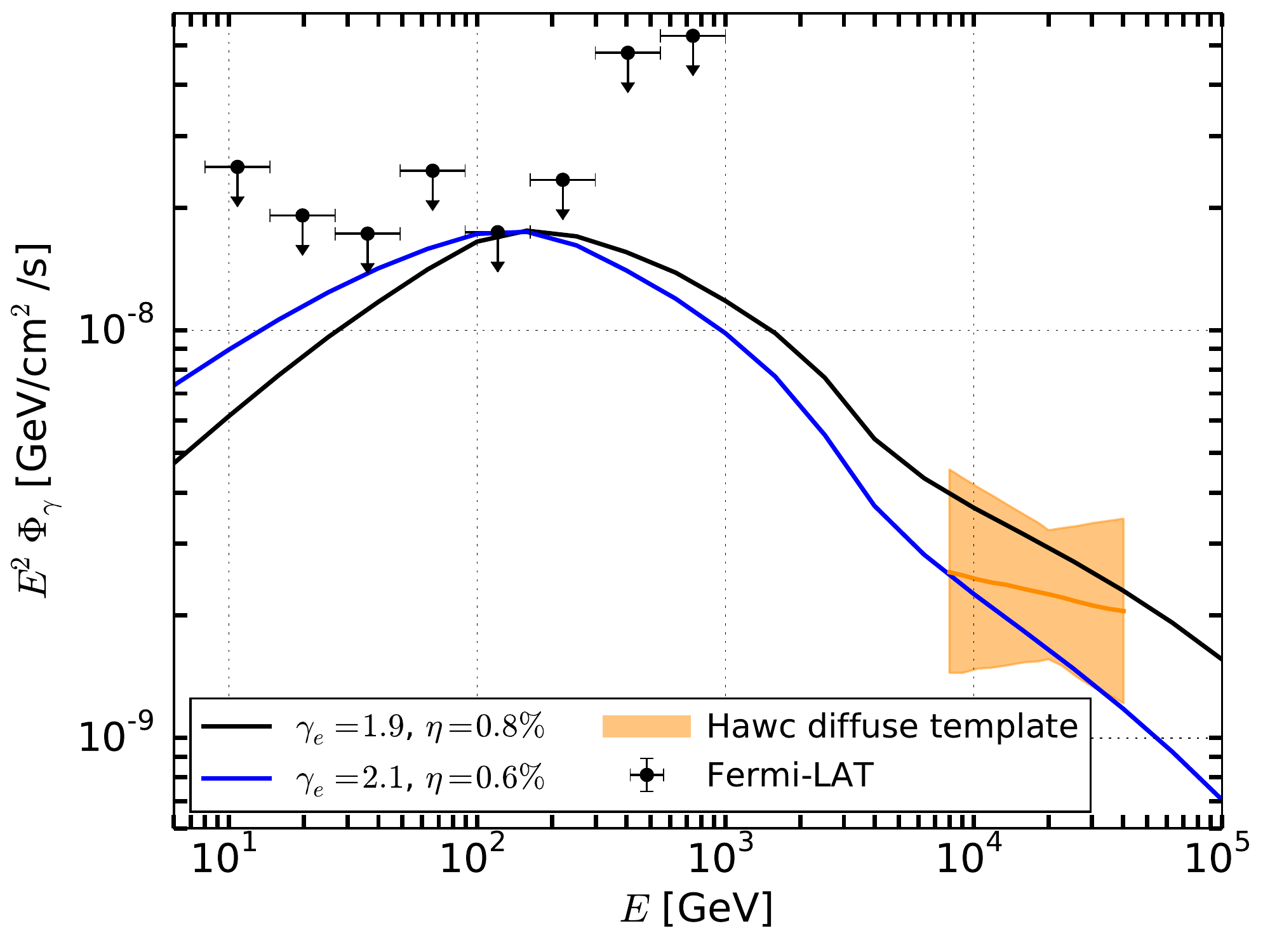}
\caption{The $\gamma$-ray flux for ICS from Geminga (left panel) and Monogem (right panel). 
The {\it Fermi}-LAT data are shown as black dots. We report the HAWC data (obtained using a diffuse template) as an orange band. 
The curves are the flux predictions obtained for different values of $\gamma_e$. }
\label{fig:ICSED_refined}  
\end{figure*}

\begin{figure*}[t]
\centering\includegraphics[width=0.49\textwidth]{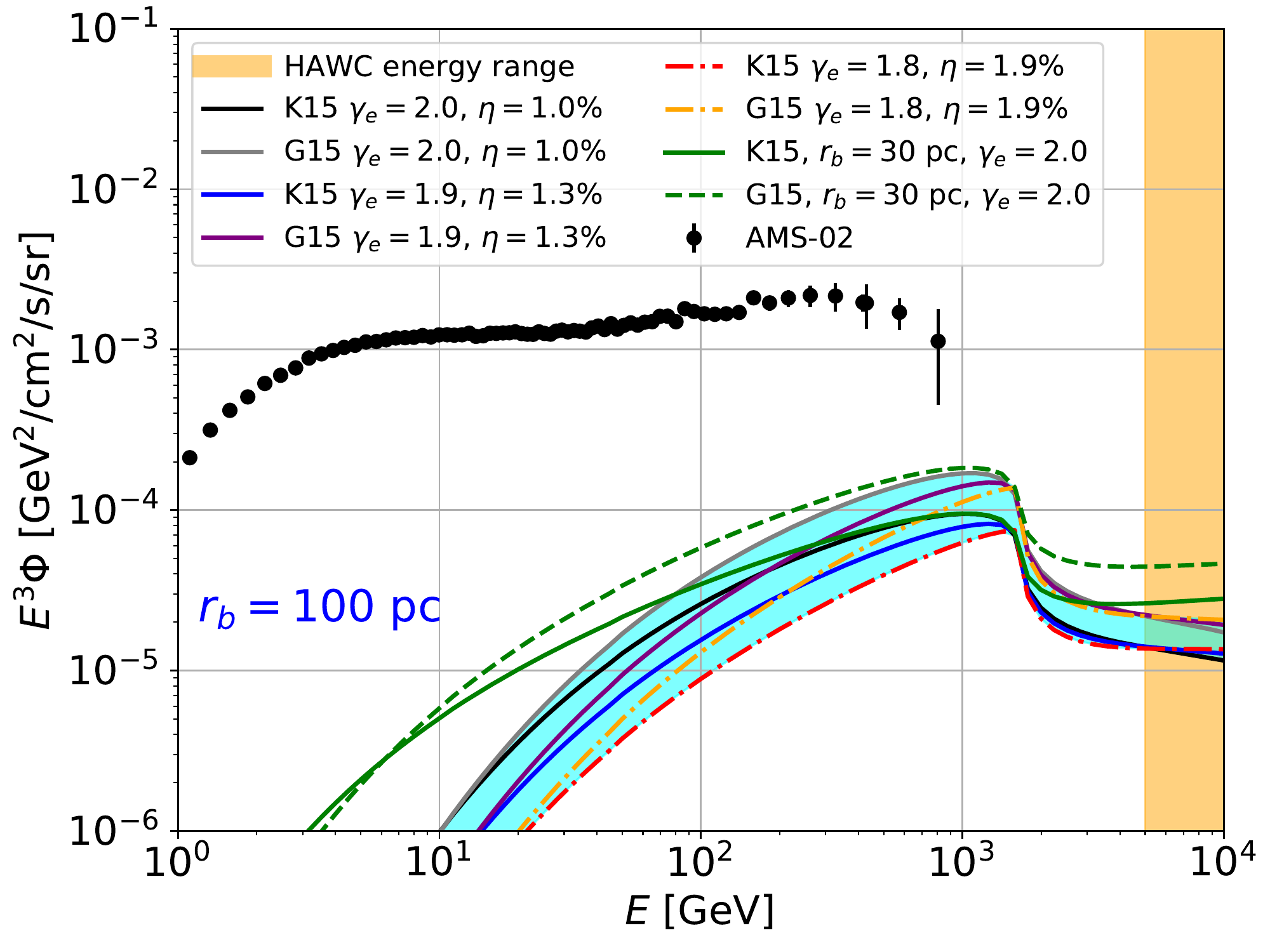}
\centering\includegraphics[width=0.49\textwidth]{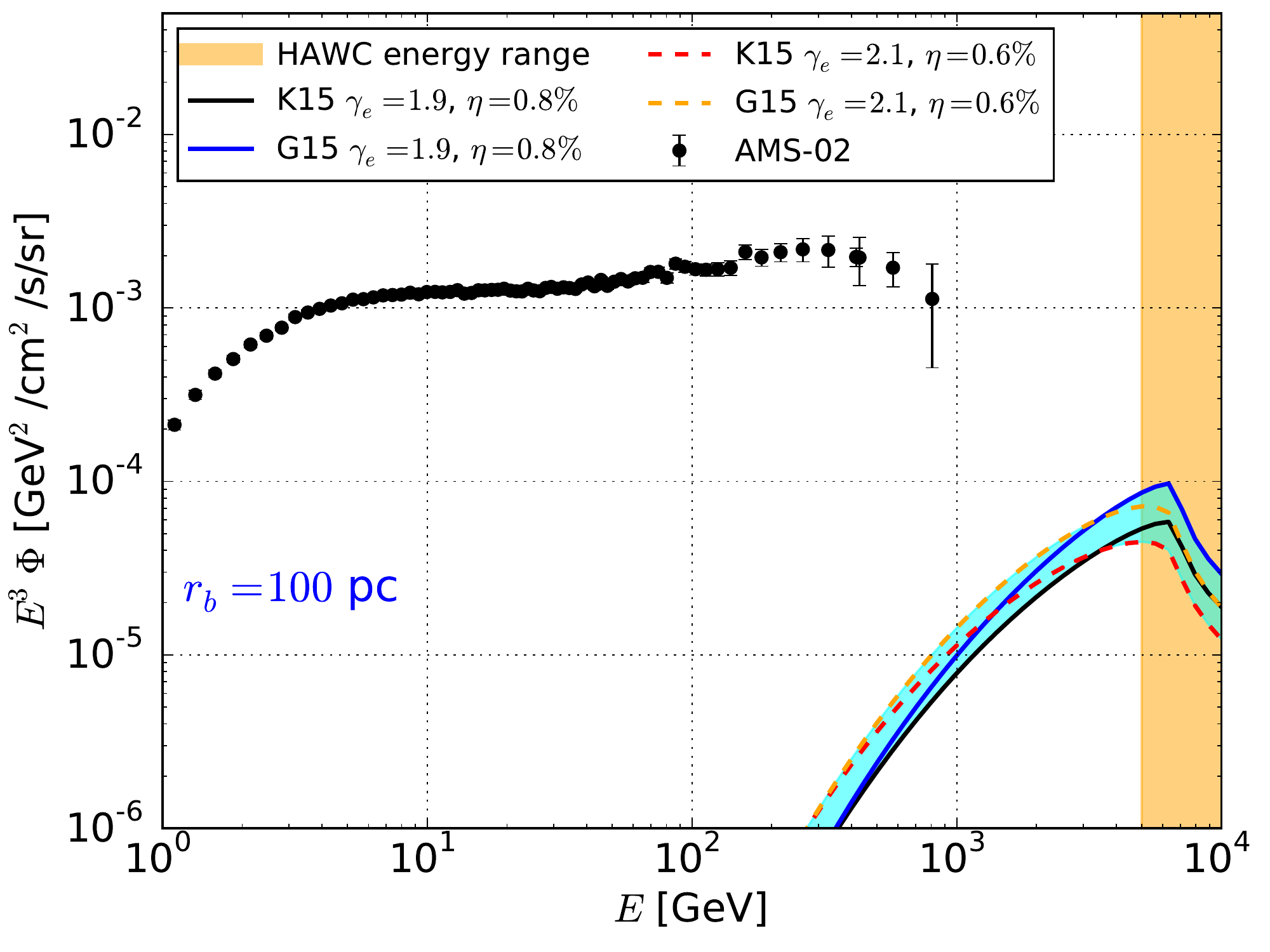}
\centering\includegraphics[width=0.49\textwidth]{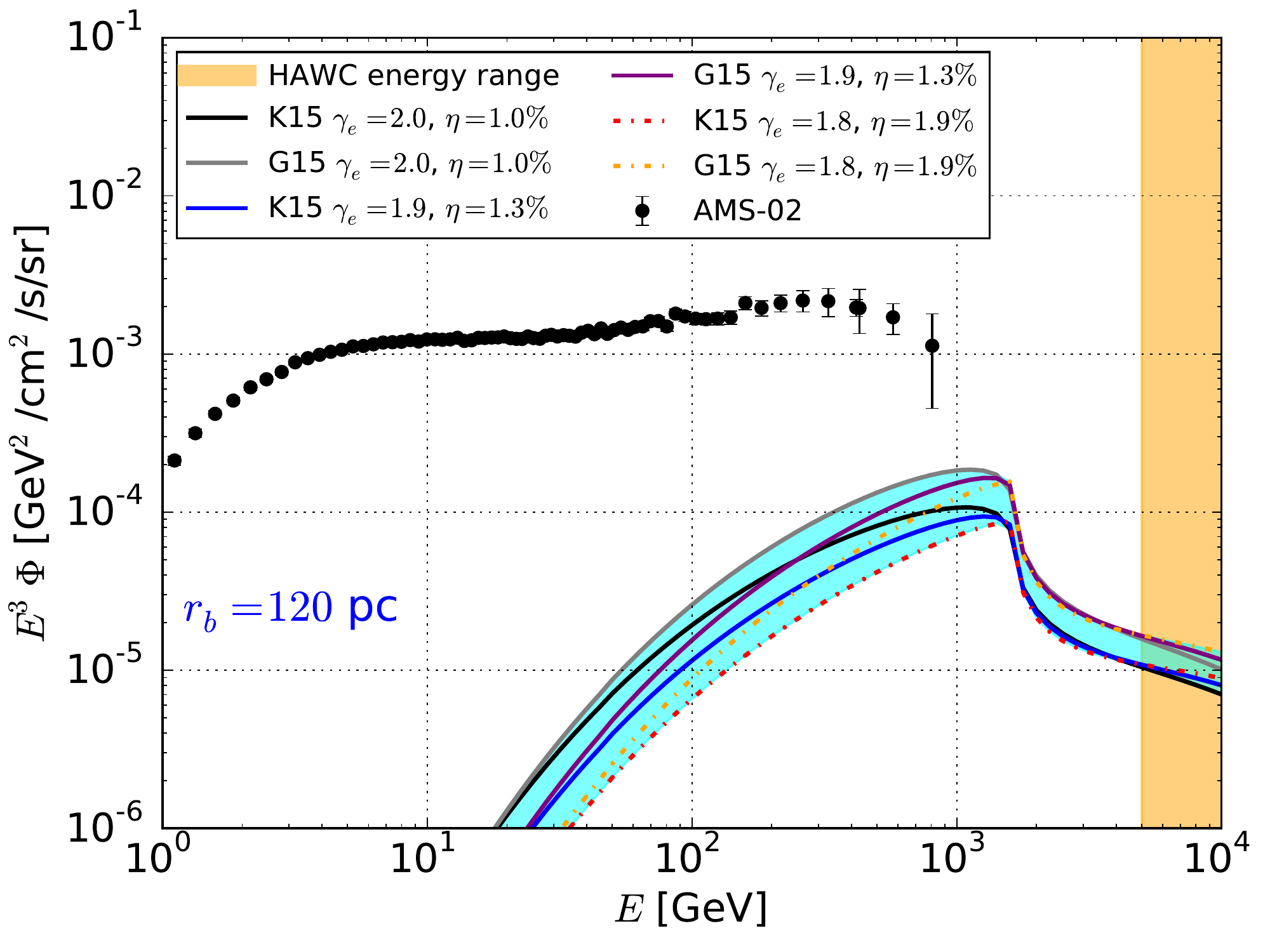}
\centering\includegraphics[width=0.49\textwidth]{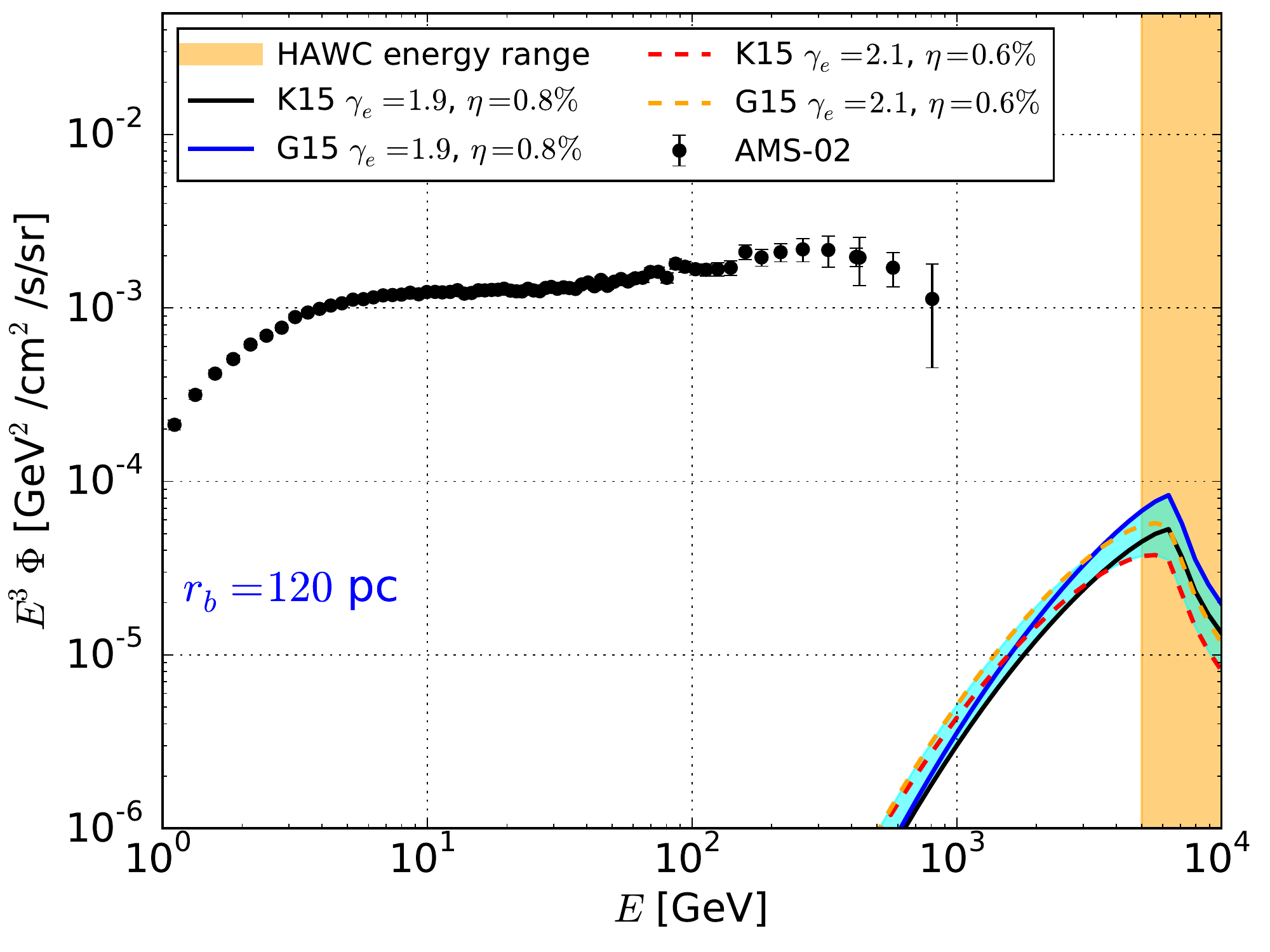}
\centering\includegraphics[width=0.49\textwidth]{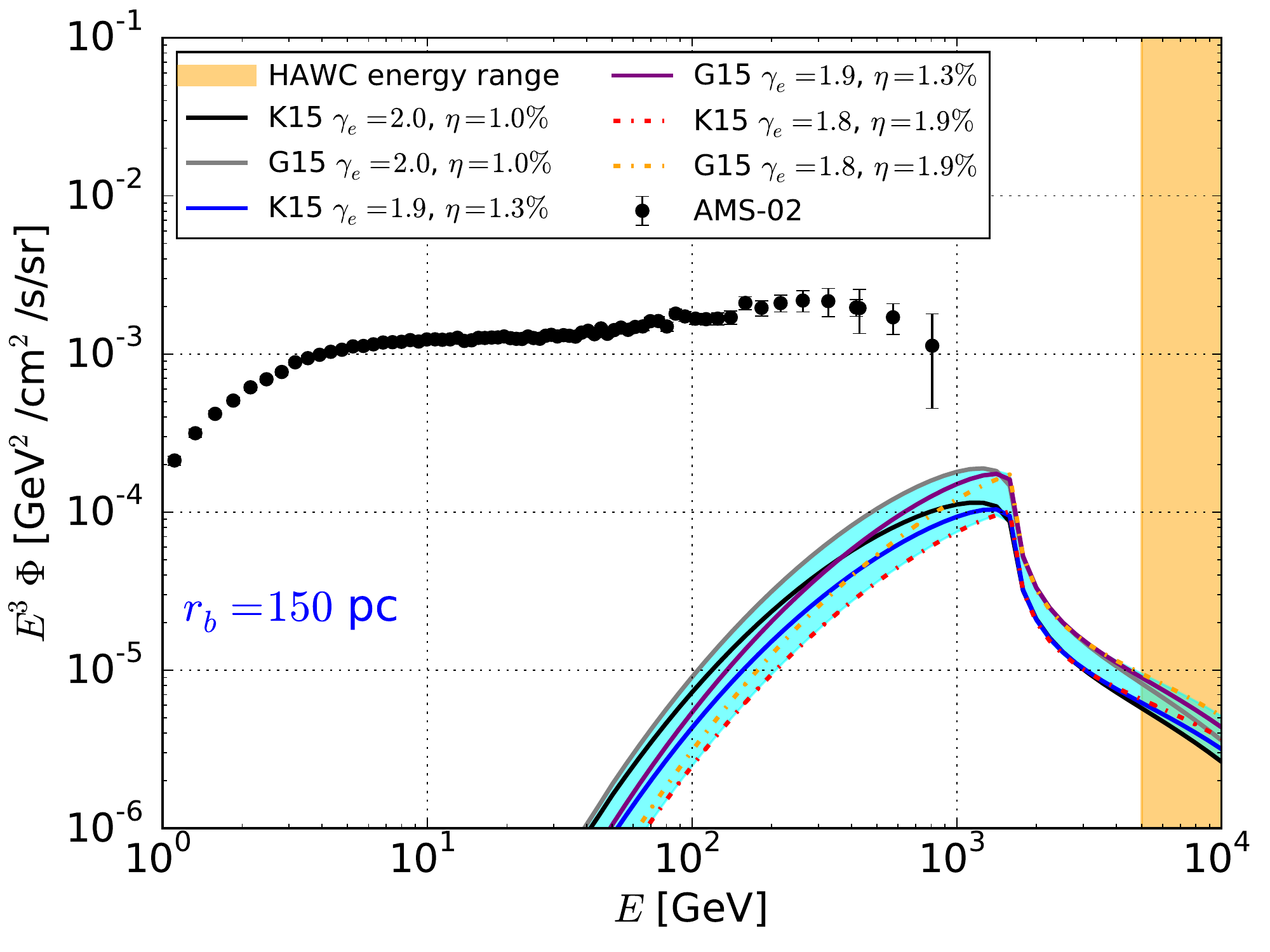}
\centering\includegraphics[width=0.49\textwidth]{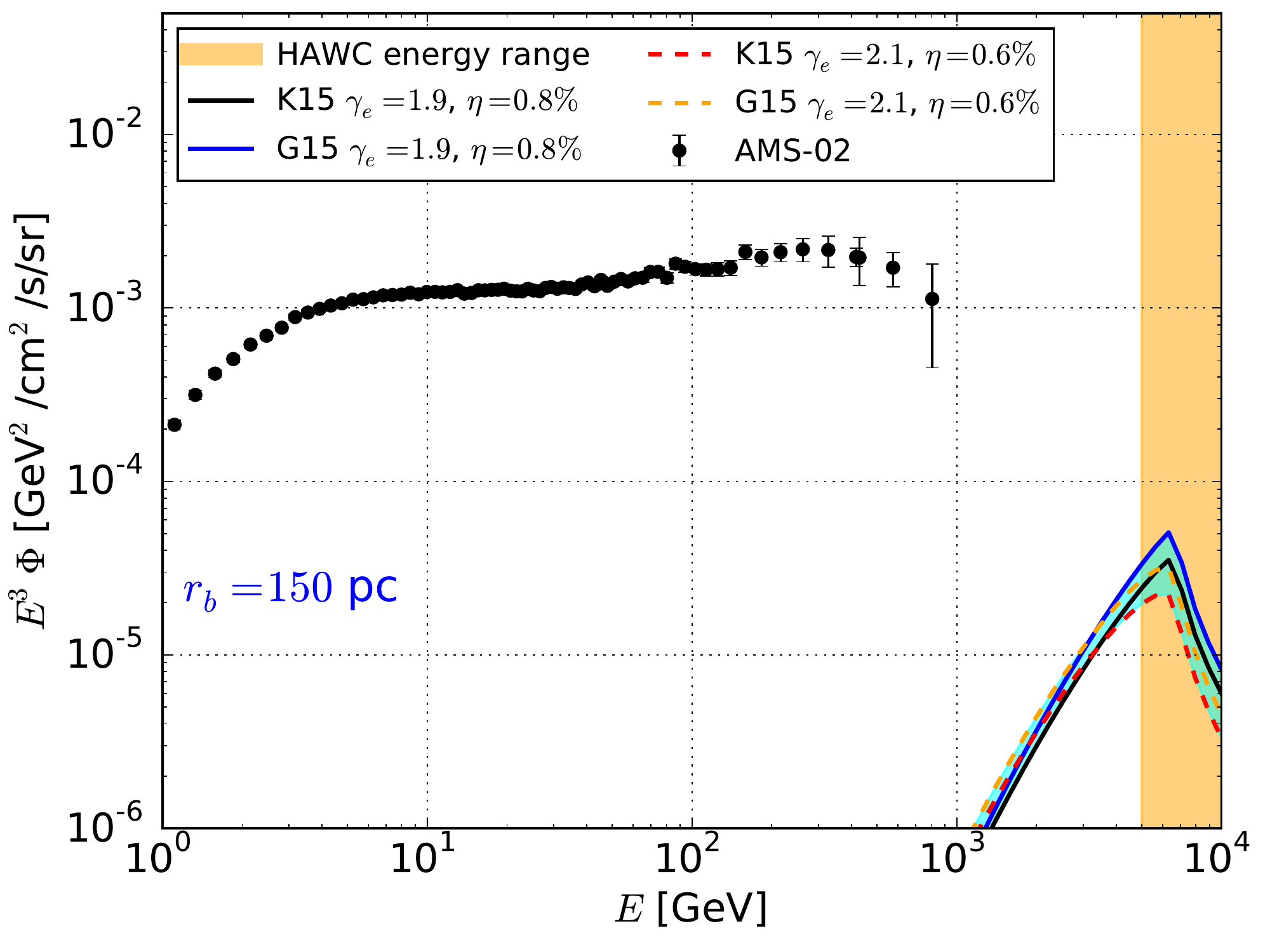}
\caption{$e^+$ flux at Earth from Geminga (left panel) and Monogem (right panel). Blue (purple) curves are for G15 (K15) propagation model and for $r_b=100$ pc, $120$ pc and $150$ pc. The cyan band embeds the differences in the results considering these two propagation parameters and the choice of $\gamma_e$. The results for Monogem are upper limits.
We also report with a green line (solid for K15 and dashed for G15) in the top left panel the case with $r_b=30$ pc, $\gamma_e=2.0$ and with $\eta$ found from a fit to the HAWC surface brightness (see Sec.~\ref{sec:hawcresults} and Fig.~\ref{fig:SB_initial_Geminga_2p0}).}
\label{fig:positron_refined_20pc}
\end{figure*}

\begin{figure*}[t]
\centering\includegraphics[width=0.47\textwidth]{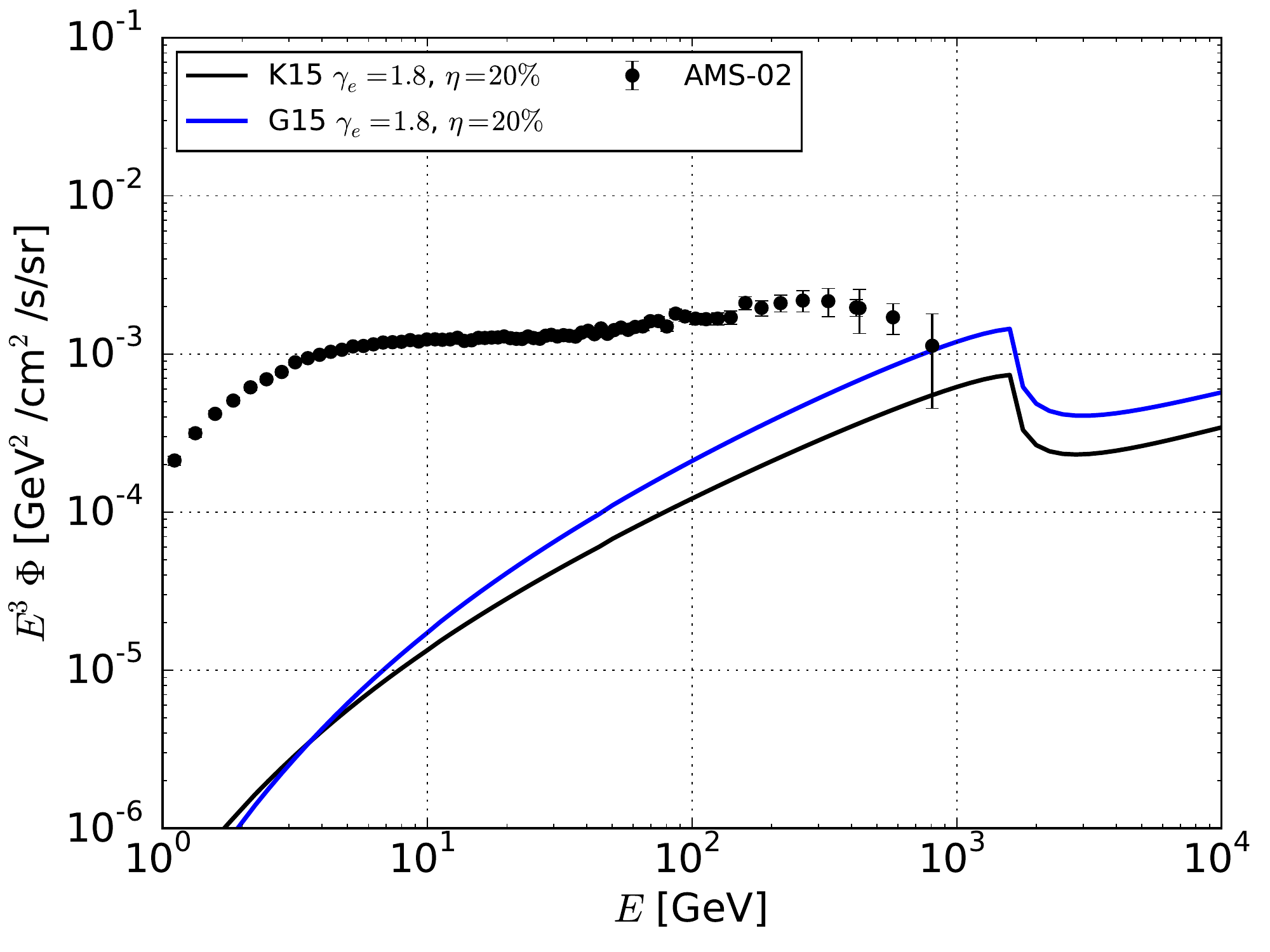}
\centering\includegraphics[width=0.52\textwidth]{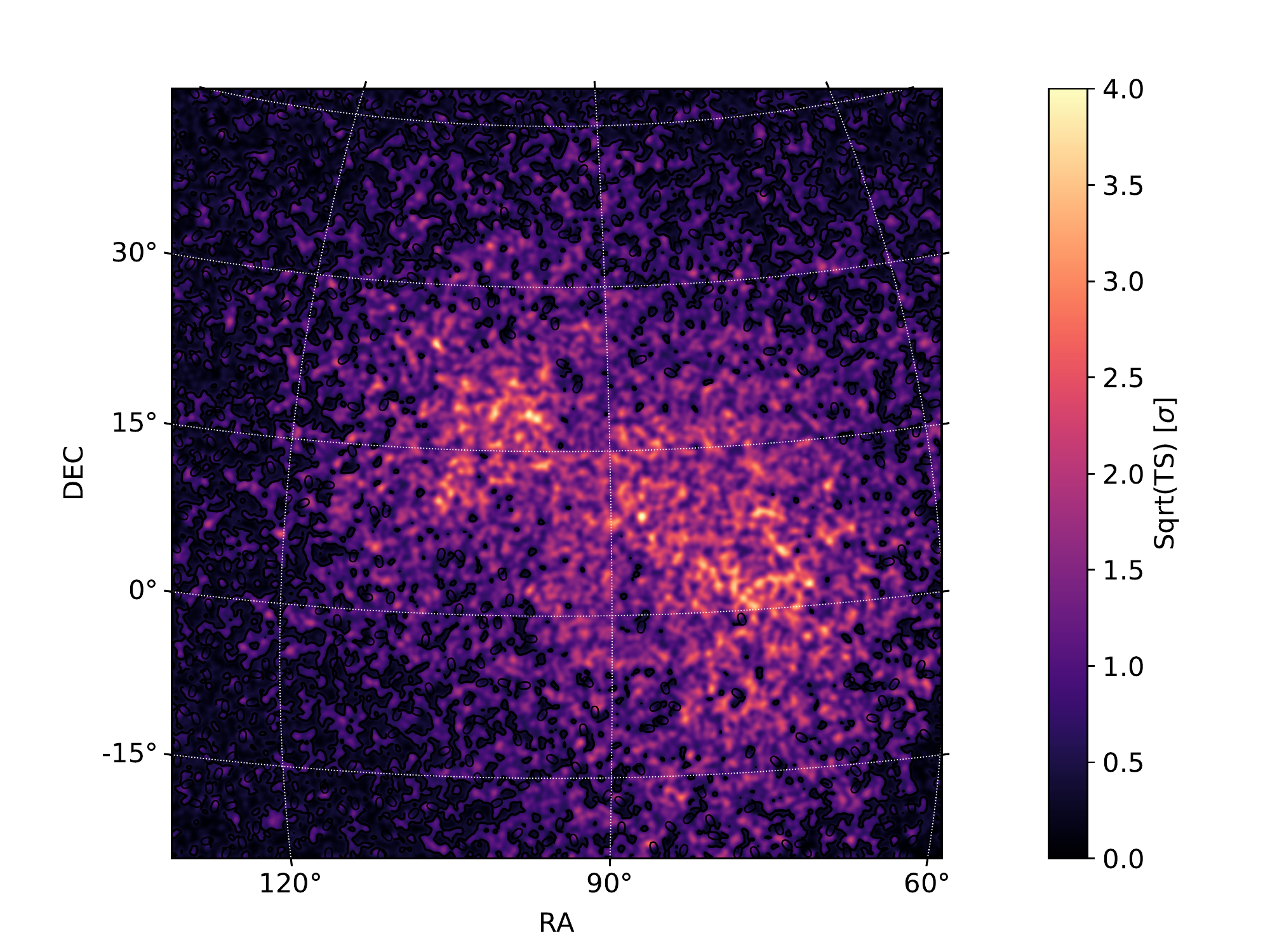}
\caption{Effect of an artificial high efficiency Geminga source. 
Left Panel: $e^+$ flux at Earth from Geminga setting  $\eta W_0=2.4\cdot10^{48}$ erg (that corresponds to an efficiency of about $\eta=20\%$) and $\gamma_e=1.80$. Blue (purple) lines are for G15 (K15) propagation model. 
Right Panel: significance map ($\sigma = \sqrt{TS}$ of the residuals for the artificially high efficiency Geminga. The color bar reports the significance of the residuals.}  
\label{fig:sim}  
\end{figure*}

Since the Geminga ICS halo is quite extended we check the correlation coefficient between the normalization of the isotropic template and the normalization and the slope of the IEM with respect to the normalization and the slope of the Geminga ICS halo. 
We find that there is a weak anticorrelation between the ICS halo SED parameters and the one of the isotropic and diffuse template with values that are always smaller than of $-0.30$. 
The normalization of the isotropic template is 0.58 for the Off. IEM and we checked that this value is not affected by the presence or absence of the Geminga ICS halo.
As an additional test we run the analysis for a ROI $60^{\circ}\times60^{\circ}$, i.e. $10^{\circ}$ smaller than before, and we find results for the significance of the Geminga ICS halo and $D_0$ that are perfectly compatible with the one reported above.
We will also show in Sec.~\ref{sec:tests} further tests that validate our results.

The flux values evaluated independently in different energy bins for the Geminga (Monogem) ICS halo are reported in the left (right) panel of Fig.~\ref{fig:ICSED_refined}.
We calculate these fluxes leaving free to vary the SED parameters of the sources in the model and the IEM and isotropic templates.
The {\it Fermi}-LAT measures the Geminga ICS halo from 8 GeV up to 100 GeV with a precision of about 30\%. For the remaining explored energies we obtain upper limits. 
We also report our predictions for the SED derived using the template presented in Sec.~\ref{sec:template}.
For $\gamma_e=[1.8,1.9,2.0]$ (Geminga analysis), we derive $\eta$ by fitting the {\it Fermi}-LAT  data  and find $\eta=[0.019,0.013,0.010]$. 
The chosen  $\gamma_e$ values bracket the HAWC measurements. 
The analogous analysis for Monogem at $\gamma_e= 1.9$ and 2.1 results in $\eta\leq $ 0.008 and 0.006, respectively. 
The values that we use here for $\gamma_e$ are harder than the one reported in HAWC2017. 
However, we checked that by using $\gamma_e = 2.0$, $D_0=2.3 \cdot 10^{26}$ cm$^2$/s and $\delta=0.21$ for the slope of the diffusion coefficient, we still find a surface brightness that is perfectly compatible ({\it i.e.} reduced $\chi^2$ smaller than 1) with the HAWC data.

Ref.~\cite{Shao-Qiang:2018zla} analyzed {\it Fermi}-LAT data to search for an extended emission from Geminga and Monogem PWNe. They do not find any significant emission and calculate upper limits in the energy range $10-500$ GeV. They use an ROI of size $22^{\circ}\times22^{\circ}$ centered on Geminga PSR. Their non-detection is explained by the fact the ICS halo emission is very extended and the effect of the Geminga proper motion makes the morphology elongated by about $20^{\circ}$ from the actual position of the pulsar. Therefore, the ROI considered in \cite{Shao-Qiang:2018zla} is not large enough to detect the Geminga ICS halo.
 
We then use our findings to predict the contribution of Geminga and Monogem to the $e^+$ flux at Earth. 
The latter is computed implementing the efficiencies fitted on the {\it Fermi}-LAT data, for the different $e^+$ spectral index.  
Since, positrons emitted from the Geminga and Monogem PWNe have to travel in both the low and high-diffusion zones before reaching the Earth, we take into account the two-zone propagation model. As in Sec.~\ref{sec:hawcresults}, we repeat the calculation using for $r>r_b$ the K15 and G15 Galactic propagation models which are suitable to model the average propagation of CRs in the Galaxy.
The results are shown in Fig.~\ref{fig:positron_refined_20pc} for $r_b=100$ pc, $120$ pc and $150$ pc.
These values for $r_b$ are consistent with the size of the ICS halo, which is about 110 pc with the Off IEM.
We have estimated this by calculating the $68\%$ containment radius, i.e.~the angle within which $68\%$ of the flux is contained. 
Considering the variation of $D_0$ found in Tab.~\ref{tab:altmodel} with all the 10 IEMs used in the analysis, the ICS halo size varies between $100-120$ pc.

The size of the ICS halo measured by HAWC is of the order of 25 pc, see HAWC2017. Therefore, using HAWC data alone the value of $r_b$ could be much smaller than the values used above.
We also show in Fig.~\ref{fig:positron_refined_20pc} the case with $r_b=30$ pc, which is not ruled out by HAWC observations. We consider for this case $\gamma_e=2.0$ and $\eta$ found from a fit to the HAWC surface brightness (see Sec.~\ref{sec:hawcresults} and Fig.~\ref{fig:SB_initial_Geminga_2p0}). Even with such a small size for the low-diffusion bubble, which is however inconsistent with the ICS halos we detect in {\it Fermi}-LAT data, the contribution of Geminga PWN to the $e^+$ excess is at most about 10\%.

The different Galactic propagation parameters act as a normalization of $e^+$ flux, specifically a factor of 3 with a negligible energy dependence. 
It is clearly visible that the different positron injection spectra and conversion efficiencies  give very similar predictions at hundreds of GeV up to TeV energies, 
where the  {\it Fermi}-LAT $\gamma$ rays calibrate the progenitor leptons. Therefore, at lower $e^+$ energies softer injection spectra give  higher $e^+$ flux. 
The Geminga PWN contributes at few per-cent level to the positron flux at 100 GeV. The highest contribution from Geminga is about 10\% of the last AMS-02 energy data point at around 800 GeV. 
As for Monogem, our predictions are derived from {\it Fermi}-LAT upper bounds. Similar considerations as done for Geminga hold here. Monogem can at most produce 3\% of the highest energy measured $e^+$ flux. 
The exact spatial distribution of the diffusion at the edge of the low and high-diffusion zone is not well know. 
In order to account for this uncertainty we calculate the $e^+$ flux for three different values of $r_b$ that are compatible with our observations.
We have verified that for Geminga the effect of decreasing (increasing) the boundary radius from $r_b=120$ pc to $r_b=100$ pc ($r_b=150$ pc) leads to about $15\%$ smaller ($15\%$ larger) contribution to the highest energy $e^+$ AMS-02 data and to about a factor of 40\% larger (2.5 times smaller) contribution at 100 GeV where, however, its contribution is at the \% level.
Including in the calculation of the $\gamma$-ray ICS flux the two-zone diffusion model with $r_b$ compatible with the observed Geminga halo can produce changes in the results. In particular we check that assuming $r_b=100$ pc would produce a $\gamma$-ray flux  lower than the one-zone model by a factor of about $40\%$ at 10 GeV. Also the spatial distribution of photons for ICS would change. 
At the same energy the flux becomes smaller by a factor of $20\%$ at the direction of the source and by a factor of $45\%$ at $\theta=15^{\circ}$. 
These differences decrease at higher energies and become negligible for $E_{\gamma}>1$ TeV. 
This will slightly change the best-fit values of $D_0$ and of the efficiency for Geminga.
However, the $\gamma$-ray flux will be still compatible with the one reported in Fig.~\ref{fig:ICSED_refined} so the contribution of Geminga to the $e^+$ flux could increase at most by a factor of about $50\%$.
On the other hand no significant differences are present for $r_b>150$ pc.
In Fig.~\ref{fig:contburst} we have shown how the positron flux changes with different parameter choices, such as the ones linked to the propagation model, the Galactic magnetic field strength, the local ISRF model, the choice of $\tau_0$ and the injection type. Considering all these parameters together with the value of $r_b$, we estimate that  the $e^+$ flux can vary at most by a factor of about 2 in the AMS-02 energy range.
As we noted before, the typical size of the Geminga ICS halo is around 80~pc. Thus, using a two-zone diffusion model for the ICS $\gamma$-ray emission with $r_b$ larger than 100~pc would not significantly change the results.
This implies that these sources alone, as bound now by {\it Fermi}-LAT data, cannot be the major contributors to the  $e^+$ excess.

\subsection{Additional tests for the detection of Geminga ICS halo}
\label{sec:tests}

We perform additional tests for validating the detection of the Geminga ICS halo in {\it Fermi}-LAT data.

The signal that we detect in the direction of Geminga should also be tested against systematic uncertainties in the {\it Fermi}-LAT instrument response functions (IRFs). The primary IRF uncertainty relevant for our study is the point spread function (PSF) for which the systematic uncertainties are of the order of $5\%$ at 10 GeV and increase to $25\%$ at 1 TeV\footnote{\url{https://fermi.gsfc.nasa.gov/ssc/data/analysis/LAT_caveats.html}}. 
These systematics can leave residuals when performing an analysis in the direction of the brightest $\gamma$-ray sources.
In order to check this effect we perform our analysis using the ICS templates derived for Geminga to analyze some of the brightest {\it Fermi}-LAT $\gamma$-ray sources.
We use ROIs centered in the direction of the Vela pulsar, 3C454.3, PSR J1836+5925, PSR J1709-4429 which, together with the Geminga pulsar, are the brightest sources detected above 100 MeV, and for PG 1553+113 and Mkn 421, which are among the brightest above 10 GeV. 
We run the analysis varying $D_0$ and $\eta$ exactly as done for the Geminga ICS halo.
We find for all these sources a maximum $TS$ of 2, implying that our result is not affected by the systematics of the PSF.

As presented in the previous section we detect with a significance of at least 4.7$\sigma$ the proper motion of Geminga pulsar. In order to test this result we re-run the analysis with the template for Geminga rotated by 90, 180 and 270 degrees with respect to the templates in Fig.~\ref{fig:mapcube_motionGeminga}. 
If our analysis would prefer the template with the proper motion with respect to the spherically symmetric model (velocity equal to zero) and we really detect the pulsar proper motion, we should find that the likelihood of the fit with the template rotated should be worse than the case used in the previous section.
We perform this check with the Off. IEM and we do not anticipate any significant change in the results using any of the other IEMs. The result of this check is that the model with the correct direction for the pulsar proper motion is preferred with a $TS$ of 32, 46 and 34 with respect to the model with the Geminga template rotated by 90, 180 and 270 degrees, respectively. 

In addition we test whether we find similar results by using Pass~8 {\tt CLEAN} and Pass~8 {\tt ULTRACLEANVETO} event classes. The data selected with these two classes have a lower contamination of falsely classified cosmic rays at the cost of a reduced effective area and consequently fewer number of $\gamma$-ray counts. Pass~8 {\tt CLEAN} and Pass~8 {\tt ULTRACLEANVETO} are generally used to perform $\gamma$-ray analyses of diffusive emission, which require a low level of cosmic-ray contamination. Indeed, they have between $20-50\%$ lower cosmic-ray background rate than {\tt SOURCE}\footnote{\url{https://fermi.gsfc.nasa.gov/ssc/data/analysis/documentation/Cicerone/Cicerone_Data/LAT_DP.html}}. We find that the best fit values for $D_0$ are $2.5^{+1.1}_{-0.8}$ and $2.7^{+1.2}_{-0.9}$, respectively for {\tt CLEAN} and Pass~8 {\tt ULTRACLEANVETO}. These values are compatible within $1\sigma$ errors with the ones found with the {\tt SOURCE} class (see Tab.~\ref{tab:altmodel}). Finally, we detect an ICS halos around Geminga with a $TS$ of 55 and 48, i.e.~with a slightly lower significance with respect to the {\tt SOURCE} class since we have a lower statistics of $\gamma$-ray counts.

The two tests reported above together with what is presented in the previous section demonstrate that the found $\gamma$-ray halo is robust, not compatible with the null signal, that the value of $D_0$ is not affected by the choice of the IEM and that we are detecting the proper motion of Geminga pulsar.

Additional effects might change the results of this paper to some extent. For example, the diffusion around Geminga could be anisotropic and the turbulent magnetic fields (generated by the
pulsar) may be pulled along with the pulsar (see, e.g., \cite{Liu:2019zyj}). Including in our model these mechanisms to check if they provide a better representation of the data is beyond the scope of this paper.
However, we can test whether a model with a different pulsar velocity is preferred in the model. 
This could provide us hints that our model with the correct pulsar proper motion is preferred by the data and that the effects reported above are probably second order.
We test the model with half of the velocity of the Geminga pulsar, i.e.~using $105$ km/s. We create the templates using this velocity and we perform a fit to {\it Fermi}-LAT data as done in Sec.~\ref{sec:results} and using the Off IEM. This model gives a best fit for the diffusion coefficient of the order of $3\cdot 10^{26}$ cm$^2$/s, i.e. slightly larger than the value found for the Geminga pulsar velocity, and a slightly worse fit with respect to the case with $211$ km/s (the model with 211 km/s is preferred with a $TS=8$).

In the previous section we found that the contribution of Geminga and Monogem to the positron excess
is at the per-cent level.
As a final exercise, we make the unrealistic assumption that instead the contribution of Geminga is about at the same level of the highest energy $e^+$ data point. 
We artificially increase the flux by setting $\eta W_0=2.4\cdot10^{48}$ erg and $\gamma_e=1.80$, uplifting the efficiency to $20\%$.
This scenario is similar to what have been published in \cite{Profumo:2018fmz,Hooper:2017gtd}.
The corresponding artificial $e^+$ flux is reported in Fig.~\ref{fig:sim} (left panel) for the two representative Galactic propagation models. 
We simulate the corresponding $\gamma$-ray halo emission using the {\tt Fermipy} tools. 
This source would be detected with {\it Fermi}-LAT with a $TS=24000$, which implies a detection at about $150\sigma$ significance.
The signature of this source in {\it Fermi}-LAT data would be given by very large residuals up to $20^{\circ}$ from the center of source. This is clearly shown from Fig.~\ref{fig:sim} (right panel), where we plot 
the square root of $TS$ (that is approximately equal to the significance) in the ROI around the source. 
This plot maps the residuals in the ROI without the Geminga ICS halo in the source model.
The result of this exercise demonstrates once more that if the Geminga PWN produces most of the contribution to the $e^+$ excess, the LAT would have detected an overwhelming number of events in a $10^{\circ}$ square around it.
Therefore, the results presented in \cite{Profumo:2018fmz,Hooper:2017gtd} for the contribution of Geminga PWN to the positron excess are strongly disfavored by {\it Fermi}-LAT data.
The results presented here are valid for a one-zone diffusion model and they might change assuming a two-zone diffusion model and depending on the value of $r_b$.
In particular for a two-zone diffusion model with $r_b$ of the size of the ICS halo of $\sim 100$ pc size, as we detect for Geminga, the conclusions we have drawn before are still true: the $\gamma$-ray flux would not change significantly with respect to the case reported above and an extremely significant signal from the Geminga ICS halo would be present in {\it Fermi}-LAT data. This statement is supported by \cite{Johannesson:2019jlk} where they predict similar ICS $\gamma$-ray and $e^+$ flux assuming a two-zone diffusion model with $r_b \sim 70$ pc. 
On the other hand, a smaller $r_b$ ($r_b\leq 50$ pc) would give a lower ICS $\gamma$-ray flux but a similar $e^+$ flux at the $\sim 100$ GeV energies, see \cite{Johannesson:2019jlk}. In this case Geminga PWN could contribute most of the $e^+$ excess as we report in Fig.~\ref{fig:sim} but with an ICS $\gamma$-ray flux that is a factor of at least 5 smaller than the case we show in the same figure. However, this small size of the low-diffusion bubble is disfavored by our detection of a very extended ICS halo in {\it Fermi}-LAT data.

Finally, we note that if the observed $\gamma$-ray emission originates from the ICS of $e^\pm$ with the ambient radiation, a diffuse emission originating from synchrotron emission should be present with a similar spatial extension. 
The synchrotron emission peaks near a critical frequency $\nu_c$ which is connected to the energy of the  $e^\pm$  through the typical relation in Eq.~\ref{eq:nnuc}. Thus, depending on the electron energy, an emission from radio up to the X-ray band is expected. 
In particular, in a  magnetic field of the order of few $\mu$G,  the same  $e^\pm$ which produce the observed ICS emission at 10 TeV (10 GeV) should radiate  at energies peaked at roughly $1.2$ keV ($1.2$ eV). 
Since the extension of Geminga is at least a few degrees, the detection of the synchrotron halo would be particularly prohibitive at those energies. However, if the presence of ICS halos around pulsars would be confirmed by the observation of other systems, a synchrotron counterpart of ICS halos in other wavelengths could be detectable for more distant and luminous sources, for which the angular size would be smaller.

\section{Conclusions}
\label{sec:conclusions}
The HAWC detection of a multi TeV $\gamma$-ray halo around two close PWNe has a natural interpretation in terms of ICS by more energetic $e^\pm$. In HAWC2017 it is shown that the contribution of Geminga and Monogem PWNe to the $e^+$ excess, measured firstly by Pamela and then confirmed with higher significance by AMS-02 at energies 
from tens of GeV up few hundreds of  GeV, is below the \% level. 
We build a model for predicting the $e^+$ flux at Earth from PWNe, which is based on a continuous injection from the source and on two diffusive regimes - one in the PWN halo region, the other in the ISM. 
The calibration of our model to the HAWC data  
leads to predictions for the $e^+$ flux which are variable by an order of magnitude at AMS-02 energies, 
contributing from a few \% up to 30 \% of the $e^+$ excess. 
\\
In order to obtain a more robust prediction for the $e^+$ flux at the excess energies, we have analyzed almost 10 years of {\it Fermi}-LAT data above 8 GeV. 
We have demonstrated that at these energies the proper motion of the Geminga pulsar is particularly relevant for the ICS $\gamma$-ray flux so we have included this effect in our analysis.
We report here the detection at $7.8-11.8\sigma$ significance of an extended emission around the Geminga PWN, depending on the IEM considered in the analysis.
Moreover, we detect the proper motion of Geminga pulsar through the ICS halo with $TS\in[20,51]$. 
This signal is straightforwardly interpreted with $\gamma$ rays produced via ICS off the photon fields located within a distance of about 100 pc from the pulsar, where the diffusion coefficient is estimated to be in the range of $1.6-3.5 \cdot 10^{26}$ cm$^2$/s at 1 GeV depending on the IEM and with a weighted average of $2.3 \cdot 10^{26}$ cm$^2$/s.

With an efficiency of about $0.01$ for the conversion of the PWN released energy into $e^\pm$ escaping the nebula, we find that the flux for the {\it Fermi}-LAT Geminga halo is compatible with the HAWC data.
The inferred contribution of Geminga to the $e^+$ flux at Earth is at most 20\% at the high-energy AMS-02 data.
We have also derived an upper limit for an halo of very high-energy $\gamma$ rays around Monogem PWN that translates into an upper limit on the efficiency of $\eta=0.008$ and into a contribution to the positron flux of a few $\%$.
Recently, the authors of \cite{Cholis:2018izy} showed that a Galactic population of pulsars with efficiency in the range of $1-3\%$ and physical spin-down properties can explain the $e^+$ flux excess. 
This result, together with the results discussed in \cite{2014JCAP...04..006D} for cataloged pulsars, 
indicate that  the cumulative positron emission from Galactic PWNe remains a viable interpretation for the positron excess.


\begin{acknowledgments}
The {\it Fermi} LAT Collaboration acknowledges generous ongoing support from a number of agencies and institutes that have supported both the development and the operation of the LAT as well as scientific data analysis. These include the National Aeronautics and Space Administration and the Department of Energy in the United States, the Commissariat\'a l'Energie Atomique and the Centre National de la Recherche Scientifique / Institut National de Physique Nucl\'eaire et de Physique des Particules in France, the Agenzia Spaziale Italiana and the Istituto Nazionale di Fisica Nucleare in Italy, the Ministry of Education, Culture, Sports, Science and Technology (MEXT), High Energy Accelerator Research Organization (KEK) and Japan Aerospace Exploration Agency (JAXA) in Japan, and the K. A. Wallenberg Foundation, the Swedish Research Council and the Swedish National Space Board in Sweden.
Additional support for science analysis during the operations phase is gratefully acknowledged from the Istituto Nazionale di Astrofisica in Italy and the Centre National d'Etudes Spatiales in France. This work performed in part under DOE Contract DE- AC02-76SF00515.

The authors thank J. Beacom, E. Bottacini, J. I. R. Cordova, D. Hooper, G. J\'ohannesson, T. Linden, R. L\'opez-Coto, A. Manfreda, P. Martin, F. Massaro, M. Meyer, S. Profumo, P. Salati, F. Salesa Greus, L. Tibaldo and H. Zhou for insightful discussions.
MDM acknowledges support by the NASA Fermi Guest Investigator Program 2014 through the Fermi 
multi-year Large Program N. 81303 (P.I. E.~Charles). 
The work of FD and SM  is supported by the "Departments of Excellence 2018 - 2022" Grant awarded by
the Italian Ministry of Education, University and Research (MIUR) (L. 232/2016).
FD and SM acknowledge financial contribution from the agreement ASI-INAF
n.2017-14-H.0 and the Fondazione CRT for the grant 2017/58675.
\end{acknowledgments}

\bibliography{paper}

\end{document}